\documentclass[a4paper,11pt]{article}
\usepackage{jheppub} % for details on the use of the package, please see the JINST-author-manual
\usepackage{amsmath,amssymb,slashed}
\usepackage[utf8]{inputenc}
\usepackage{graphicx}
\usepackage{hyperref}
\usepackage{orcidlink}
\usepackage{bm,bbold}
\usepackage{accents}
\usepackage{ulem}
\usepackage{slashed}
\def\!{\mskip-\thinmuskip}

\newcommand{\mean}[1]{\langle #1 \rangle}

\newcommand{\di}{{\rm d}}

\newcommand{\wj}{\widehat{j}}
\newcommand{\wT}{\widehat{T}}
\newcommand{\wP}{\widehat{P}}

\newcommand{\wrho}{\widehat{\rho}}

\newcommand{\h}[1]{\widehat{#1}}
\newcommand{\tr}{{\rm tr}}  
  
\newcommand{\e}{{\rm e}}
\newcommand{\I}{{\rm i}}

\newcommand{\de}{\partial}

\renewcommand{\tilde}[1]{{\widetilde{#1}}}

\newcommand{\group}[1]{\relax\ifmmode\mathsf{#1}\else\textsf{#1}\fi}

\newcommand{\SP}{\mathfrak{S}}

\newcommand{\corrLTE}[2]{\left( #1,\, #2\right)_{\rm LTE}}
\newcommand{\corrD}[2]{\left( #1,\, #2\right)_{\rm D}}
\newcommand{\corrALTE}[1]{\left( \hWFA^\mu,\, #1\right)_{\rm LTE}}
\newcommand{\corrAD}[1]{\left( \hWFA^\mu,\, #1\right)_{\rm D}}
\newcommand{\mcU}{\mathcal{U}}
\newcommand{\ortu}[1]{\langle#1\rangle} %An index orthogonal to u
\newcommand{\WFA}{\mathcal{A}}
\newcommand{\WFV}{\mathcal{V}}
\newcommand{\hWFA}{\h{\mathcal{A}}}
\newcommand{\hWFV}{\h{\mathcal{V}}}
\newcommand{\mf}[1]{\mathfrak{#1}}

\arxivnumber{2502.15520}
\title{Kubo formulas for spin polarization in dissipative relativistic spin hydrodynamics: a first-order gradient expansion approach}

\author{Matteo Buzzegoli\orcidlink{0000-0002-2114-5431}}
\affiliation{Department of Physics, West University of Timișoara,\\
Bd. Vasile Pârvan 4, Timișoara 300223, Romania}
\emailAdd{matteo.buzzegoli@e-uvt.ro}

\abstract{We use linear response theory to derive both the non-dissipative and dissipative effects of spin polarization for massive and massless interacting spin 1/2 particles in a relativistic fluid.
We list and classify all the possible contributions up to first order in gradients of hydrodynamic fields including the axial chemical potential and the spin potential, and we obtain the corresponding Kubo formulas.
We find that all the possible dissipative contributions, except those coming from the gradients of spin potential, require a chiral imbalance or parity violating interactions.
In a fluid with chiral imbalance we find a chiral version of the spin Hall effect, i.e. a spin polarization is induced by the gradients of temperature and of axial chemical potential in the direction orthogonal to the momentum of the particle and to the gradients.
Moreover, we identify several other new non-dissipative contributions that are not present for free fields. 
}
\begin{document}
\maketitle
\flushbottom

%*******************************************************************************************************
%***********************************************************************************************************
%***********************************************************************************************************
\section{Introduction}
Spin polarization in a relativistic fluid is an important aspect of the relativistic heavy-ion collision phenomenology where measurements of spin polarization of particles are used to reveal properties of the quark gluon plasma (QGP) formed during the collisions. Based on the analysis of partonic angular momentum transfer~\cite{Liang:2004ph,Gao:2007bc,Huang:2011ru} and calculations using statistical local thermal equilibrium~\cite{Becattini:2007sr,Becattini:2013fla}, it was initially realized that thermal vorticity tensor (i.e., the rotation and acceleration fields in the fluid) should induce spin polarization in a fluid. Consequently, for the QGP, spin polarization should be detected in the direction of the large angular momentum generated in non-central collisions. The measurements of the global spin polarization of $\Lambda$ and $\bar{\Lambda}$~\cite{STAR:2017ckg,STAR:2018gyt,STAR:2021beb}, and later of $\Xi$ and $\Omega$ hyperons~\cite{STAR:2020xbm} carried out by the STAR collaboration at RHIC and of $\Lambda$ and $\bar{\Lambda}$ hyperons~\cite{ALICE:2019onw} by ALICE collaboration at LHC, confirmed the predictions of the vorticity-induced spin polarization based on hydrodynamics and statistical local equilibrium~\cite{Becattini:2015ska,Karpenko:2016jyx,Xie:2017upb,Wu:2019eyi,Ivanov:2020udj,Fu:2020oxj}.

Subsequent measurements of the spin polarization as a function of momentum~\cite{Niida:2018hfw,STAR:2019erd,ALICE:2021pzu}, termed ``local spin polarization'', differed drastically from the theoretical predictions. In particular, the longitudinal component of the spin polarization vector was found to have an opposite sign compared to the theoretical calculations, giving rise to many theoretical investigations, see for instance the review~\cite{Becattini:2020ngo}. These studies lead to the finding that other hydrodynamic fields in an out-of-equilibrium fluid can also induce spin polarization, such as the thermal shear tensor~\cite{Liu:2021uhn,Becattini:2021suc} and the gradient of chemical potential~\cite{Liu:2020dxg}. On the one hand, predictions that include these additional effects~\cite{Fu:2021pok,Becattini:2021iol,Yi:2021ryh,Wu:2022mkr,Alzhrani:2022dpi,Palermo:2024tza} can reproduce most of the experimental data. However, a strong sensitivity to transport coefficients and initial hydrodynamic conditions has been observed~\cite{Yi:2021ryh,Palermo:2024tza}. On the other hand, local spin polarization data raise the question of whether the spin degrees of freedom in the QGP are actually equilibrated~\cite{Kapusta:2019sad,Ayala:2019iin,Hongo:2022izs,Wagner:2024fhf}. To address this question it has been necessary to develop spin hydrodynamics, i.e., a framework that includes the evolution of a spin  tensor~\cite{Florkowski:2017ruc,Weickgenannt:2019dks}.

Evidence of spin alignment of vector mesons (a phenomenon analogous to spin polarization) has also been detected in heavy-ion collisions~\cite{STAR:2022fan,ALICE:2019aid}. These observations present a challenging problem because they depend drastically on the collision energy and the type of meson considered. Furthermore, the effect of spin alignment is larger than expected from hydrodynamic models, giving a strong indication that there is a strong spin correlation between the quark and anti-quark that
combine into the vector meson~\cite{Yang:2017sdk,Sheng:2019kmk,Sheng:2020ghv,Sheng:2022wsy,Xia:2020tyd,Wei:2023pdf,Kumar:2023ghs,Sheng:2023urn,Lv:2024uev}. For recent reviews on spin polarization, spin hydrodynamics and spin alignment we refer to~\cite{Becattini:2024uha,Niida:2024ntm}. 

Therefore, studying all mechanisms that can polarize the spin of a quark in the QGP is central to understanding spin polarization and spin alignment in heavy-ion collisions. Relativistic hydrodynamics has proven to be a very effective tool for studying the properties of the QGP, including spin, because it drastically reduces the effective degrees of freedom from interacting quantum fields to a few classical fields. Taking advantage of the large separation of scales between macroscopic fields and microscopic interactions, hydrodynamics and relativistic statistical mechanics allow for approximating a physical observable with a gradient expansion, which is a series based on the gradients of the hydrodynamic fields~\cite{Romatschke:2017ejr}. These terms can be further classified in two categories: the one stemming from the local equilibrium part of the statistical operator, which are not associated to dissipative or dynamical processes and which are proportional to what we call thermal coefficients, and those stemming from the dissipative part of the statistical operator and that are proportional to transport coefficients~\cite{Becattini:2019dxo}. Most of the studies so far focused their attention on the local equilibrium part and obtained the spin polarization of non-interacting particles, and very few attempted to address dissipative contributions~\cite{Shi:2020htn,Weickgenannt:2022qvh,Li:2025pef,Wagner:2024fry}.

The scope of this work is to classify all possible contributions to spin polarization in a dissipative fluid, considering terms up to the first order in the gradient expansion. This work considers a general case including gradients of the fluid velocity (vorticity and shear), vector and axial chemical potentials (the latter describing a fluid with a chiral imbalance, i.e., a difference between right-handed and left-handed particles), and the spin potential (which describes a situation where the spin degrees of freedom have not relaxed to equilibrium). The underlying microscopic theory is not specified; instead, the effects of the hydrodynamic fields will be obtained using linear response theory, and their strength (the thermal and transport coefficients) would have to be calculated using Kubo formulas.
%These formulas are correlators between quantum operators obtained tracing them with the the statistical operator in the grand canonical equilibrium form.
With these formulas, the coefficients are obtained by tracing quantum operators with the statistical operator in the grand canonical equilibrium form.
The symmetries of the equilibrium statistical operator and of the spin polarization vector will be used to classify and discuss the effects obtained with this procedure. It is worth noting that all out-of-equilibrium effects depend on the specific form of the quantum operators used, notably the energy-momentum tensor and the spin tensor. The out-of-equilibrium statistical operator~\cite{Becattini:2018duy} and, consequently, the spin polarization~\cite{Buzzegoli:2021wlg}, depend on this choice, which is known as the pseudo-gauge choice. We do not specify this choice, because it will not affect the form of the Kubo formulas but only the result of the calculation.

To study all first order contribution to spin polarization, we use the Zubarev non-equilibrium density operator method~\cite{Zubarev:1979,vanWeert1982,Becattini:2014yxa,Hayata:2015lga}, which allows to clearly distinguish the dissipative contributions from the local equilibrium (non-dissipative) ones~\cite{Becattini:2019dxo}. Assuming that the system reaches local thermal equilibrium at some time, the method consist of deriving the covariant form of the statistical operator by maximazing the entropy  while constraining the energy-momentum densities and other relevant quantities. It is important to stress that this procedure and the notion of local thermal equilibrium are valid for any underlying quantum field theory. Because of this property and of the premises, the Zubarev method is particularly useful for studying the QGP formed in heavy-ion collisions, especially the quantum effects induced by gradients, including the thermal vorticity. Indeed, this method was used to derive, for the first time, the Kubo formula for the shear viscosity of a quantum relativistic fluid~\cite{Hosoya:1983id}. Recently, it has been successfully used to study the transport coefficients associated with the gradient expansion in a relativistic fluid~\cite{Huang:2011dc,Harutyunyan:2018cmm,Harutyunyan:2021rmb}, including those associated to the spin potential~\cite{Hu:2021lnx,Montenegro:2020paq,Torrieri:2022ogj,Becattini:2023ouz,Tiwari:2024trl,Dey:2024cwo,She:2024rnx,Buzzegoli:2024mra}, and to obtain the thermal coefficients related to thermal vorticity~\cite{Buzzegoli:2017cqy,Buzzegoli:2018wpy,Buzzegoli:2020ycf,Becattini:2020qol,Palermo:2021hlf}, to connect the hydrodynamic fields to the spin polarization~\cite{Buzzegoli:2021wlg,Liu:2021nyg,Buzzegoli:2022qrr,Sheng:2024pbw,Li:2025pef} and to the spin alignment~\cite{Li:2022vmb,Yang:2024fkn,Zhang:2024mhs,Li:2025pef}.

The spin polarization of a particle with a definite momentum is obtained from the axial and vector parts of the Wigner function of the corresponding  field~\cite{Becattini:2020sww,Palermo:2023cup,Becattini:2024uha}. The classification of all possible contributions to spin polarization of a spin-{\textonehalf} fermion then boils down to the gradient expansion of the Dirac Wigner function. We perform this expansion up to first order in gradients. Differently from usual physical observable considered in hydrodynamics, like the energy-momentum tensor, the Wigner function is a distribution depending on the point $x$ and on momentum $k$. In order to account for the momentum dependence we follow the methods of~\cite{Liu:2021uhn,Liu:2020dxg} and we decompose the Wigner function in terms construed with the momentum $k$ and the hydrodynamic fields. As a result, differently from the usual Kubo formulas, where thermal and transport coefficients depends only on coordinates through equilibrium thermodynamic fields, the thermal and transport coefficients appearing in the Wigner function will also depend on the momentum of the particle state where the spin polarization has been evaluated.

The paper is organized as follows: section~\ref{sec:Zubarev} reviews the Zubarev method and how to obtain a gradient expansion of a generic local observable. In section~\ref{sec:SpinPol}, the gradient expansion is applied to the spin polarization of a Dirac field, demonstrating that this is obtained from the gradient expansion of the axial and vector parts of the Wigner function. Section~\ref{sec:KuboGeneral} provides the general methods for deriving Kubo formulas related to the linear response theory of different types of hydrodynamic fields, classified by their irreducible representation under rotation: scalar, vector, symmetric and antisymmetric tensor and rank-3 spin-like tensor. In section~\ref{sec:KuboA}, this method is applied to obtain the linear response of the axial and vector parts of the Wigner function up to the first order in the gradient expansion, to classify the types of coefficients obtained and to determine their Kubo formulas. The main results and their discussion are given in section~\ref{sec:Results} and we conclude in section~\ref{sec:Conclusions}.

\subsection{Notation}
In this paper we use the natural units with $\hbar = c = k_B = 1$, the Minkowskian metric is diag$(1,-1,-1,-1)$; for the Levi-Civita symbol we use the convention $\epsilon^{0123} = 1$. Operators in Hilbert space are denoted by a wide upper hat, e.g. $\h{O}$. We will use the relativistic notation with repeated indices assumed to be saturated. To describe hydrodynamic quantities in a relativistic fluid with fluid velocity $u$, where $u^2=1$, we define
\begin{equation}
\label{eq:ProjDef}
\Delta^{\mu\nu} = \eta^{\mu\nu} - u^\mu u^\nu,\quad
V^{\ortu\rho} = V^\rho_\perp = \Delta^\rho_\lambda V^\lambda,\quad
D=u^\mu \de_\mu,\quad \theta=\de_\mu u^\mu,
\end{equation}
that are, respectively, the projector to the space orthogonal to $u(x)$, a four-vector orthogonal to $u(x)$, the derivative along the fluid velocity and the rate of expansion. The quantity
\begin{equation}
\label{eq:Delta4Def}
\Delta_{\mu\nu\rho\sigma}= \frac{1}{2}\left(\Delta_{\mu\rho}\Delta_{\nu\sigma}
	+\Delta_{\mu\sigma}\Delta_{\nu\rho}\right)
	-\frac{1}{3}\Delta_{\mu\nu}\Delta_{\rho\sigma}
\end{equation}
is the  projection of a rank-2 tensor to its symmetric traceless part orthogonal to the fluid velocity.
We use the subscripts $S$ and $A$ in a rank-2 tensor to denote its symmetric and antisymmetric part respectively, for instance
\begin{equation}
T^{\mu\nu}_S = \frac{1}{2}\left( T^{\mu\nu} + T^{\nu\mu} \right),\quad
T^{\mu\nu}_A = \frac{1}{2}\left( T^{\mu\nu} - T^{\nu\mu} \right).
\end{equation}
The dual of a rank-2 tensor is denoted with a tilde and is defined as follows:
\begin{equation}
\tilde\varpi^{\mu\nu} = \frac{1}{2} \epsilon^{\mu\nu\alpha\beta}\varpi_{\alpha\beta} .
\end{equation}
To describe the Wigner function, that is a distribution function that depends both on coordinates $x$ and momentum $k$, we also define the projector
\begin{equation}
\label{eq:QDef}
Q^{\mu\nu} = \frac{\Delta^{\mu\nu}}{3} - \frac{k_\perp^\mu k_\perp^\nu}{k_\perp^2},
\end{equation}
that satisfy the identities
\begin{equation}
\Delta_{\mu\nu}Q^{\mu\nu} = \eta_{\mu\nu}Q^{\mu\nu} = 0,\quad
Q_{\mu\nu} Q^{\mu\nu} = \frac{2}{3}.
\end{equation}
%
%The Wigner function of the fluid is expressed using a derivative expansion and linear response theory, and
The coefficients related to the response of a hydrodynamic field $\mcU$ are classified as follows.
A response coefficient can be thermal (with time-reversal \group{T} symmetry) or dissipative (breaking time-reversal symmetry), non-chiral or chiral (regarding parity transformation \group{P}), not charged or charged (regarding charge conjugation \group{C}). If the coefficient is dissipative, it will be denoted with a bar, if it is chiral it will be written with the fraktur font, and if it is charged it will have a superscript ``c'', as illustrated in the following table
\begin{equation}
\label{tablecoeff}
\begin{array}{|l|cccccccc|}
\hline
    & a_\mcU,\, v_\mcU & a_\mcU^c,\, v_\mcU^c & \bar{a}_\mcU,\, \bar{v}_\mcU & \mf{a}_\mcU,\, \mf{v}_\mcU
    & \bar{a}_\mcU^c,\, \bar{v}_\mcU^c & \mf{a}_\mcU^c,\, \mf{v}_\mcU^c &
    \bar{\mf{a}}_\mcU,\, \bar{\mf{v}}_\mcU & \bar{\mf{a}}_\mcU^c,\, \bar{\mf{v}}_\mcU^c \\
\hline
\group{P} & + & + & + & - & + & - & - & - \\
\group{T} & + & + & - & + & - & + & - & - \\
\group{C} & + & - & + & + & - & - & + & - \\
\hline
\end{array}
\end{equation}

%***********************************************************************************************************
%***********************************************************************************************************
%***********************************************************************************************************
\section{Gradient expansion for an out-of-equilibrium fluid}
\label{sec:Zubarev}
\begin{figure}[t!bph]
\centering
\includegraphics[width=0.5\textwidth]{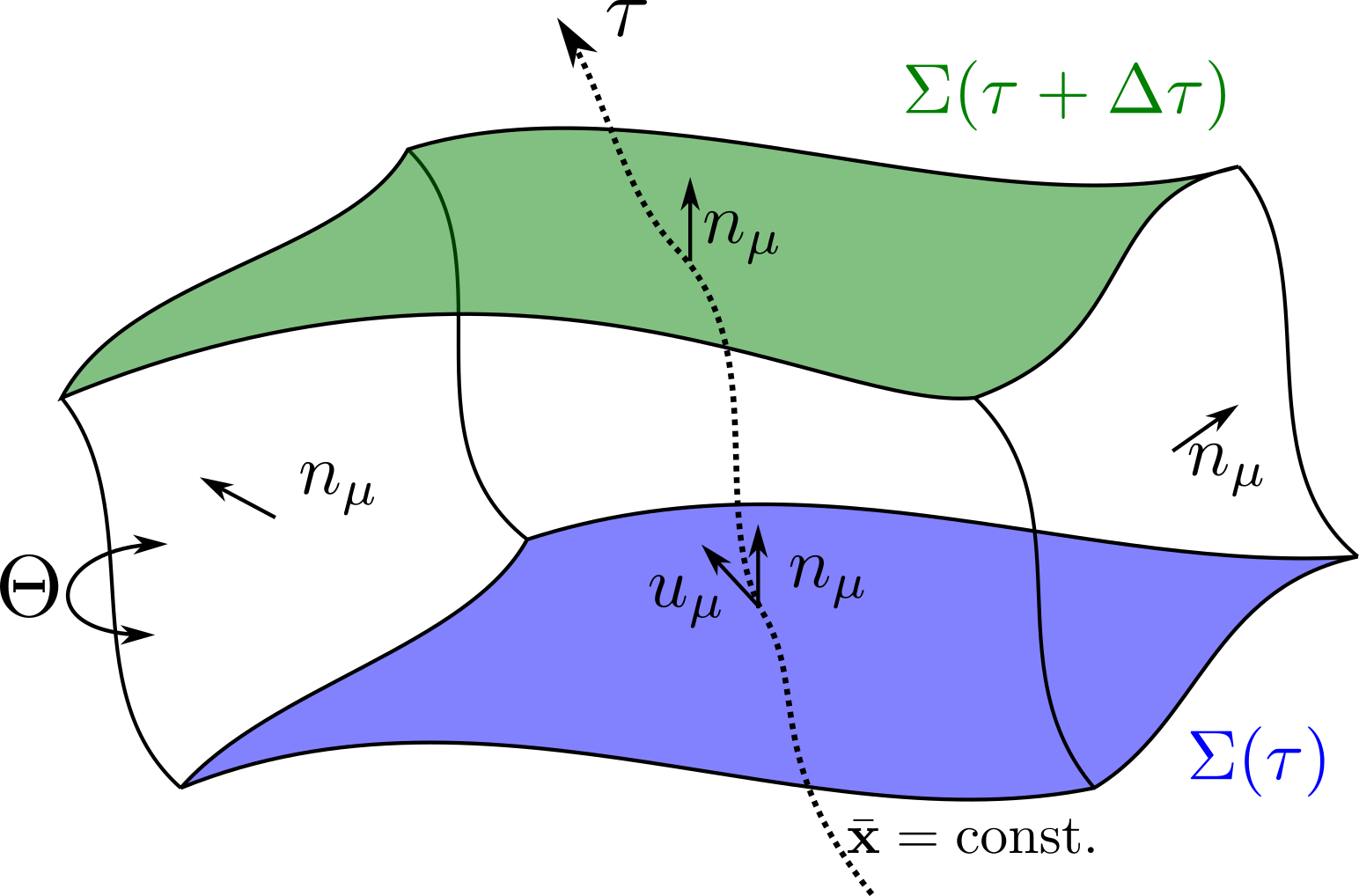}
\caption{The foliation of spacetime with $\Sigma(\tau)$ a spacelike hypersurface parametrized by $\tau(x)=$const. The normal unit vector to the hypersurfaces is denoted by $n^\mu$. The whole region represents the spacetime volume $\Omega$ enclosed by two spacelike hypersurface $\Sigma(\tau)$ and $\Sigma(\tau+\Delta\tau)$ and a timelike boundary $\Theta$.}
\label{fig:foliation} 
\end{figure}
The description of a relativistic fluid using relativistic quantum statistical mechanics, obtained from the Zubarev method, begins by considering a system that has reached local thermal equilibrium on a 3D spacelike hypersurface $\Sigma(\tau)$ at ``time'' $\tau$, defining a foliation of spacetime (see figure~\ref{fig:foliation}). The system is described by an energy-momentum tensor operator $\h{T}^{\mu\nu}$, a spin tensor operator $\h{S}^{\lambda,\mu\nu}$, a conserved current $\h{j}^\mu$ and an axial current $\h{j}_A^\mu$ obtained from a quantum field theory that we do not specify. At equilibration the system is instead described effectively by the classical tensor fields $T^{\mu\nu}$ and $S^{\lambda,\mu\nu}$ giving the energy-momentum density and the orbital and spin angular momentum densities, as well as by the currents $j^\mu$ and $j_A^\mu$. The statistical operator $\wrho$ describing the system is then obtained by maximizing the entropy $S=-\tr\left[\wrho\log\wrho\right]$ with the constraints:
\begin{subequations}\begin{align}
n_\mu \tr\left[\wrho\, \h{T}^{\mu\nu}  \right] &= n_\mu T^{\mu\nu}, &
n_\lambda \tr\left[\wrho\, \h{S}^{\lambda,\mu\nu}  \right] &= n_\lambda S^{\lambda,\mu\nu},\\
n_\mu \tr\left[\wrho\, \h{j}^{\mu}  \right] &= n_\mu j^{\mu}, &
n_\mu \tr\left[\wrho\, \h{j}_A^{\mu}  \right] &= n_\mu j_A^{\mu},
\end{align}\end{subequations}
where $n$ is the unit vector orthogonal to $\Sigma(\tau)$. The resulting operator is~\cite{Zubarev:1979,vanWeert1982,Becattini:2014yxa,Hayata:2015lga,Buzzegoli:2018wpy,Becattini:2018duy}
\begin{equation}
\label{eq:NEDO}
\h{\rho} = \frac{1}{Z} \exp\left\{ 
-\int_{\Sigma(\tau)} {\rm d}\Sigma_\mu\left(\h{T}^{\mu\nu}\beta_\nu
 -\zeta\,\h{j}^\mu -\zeta_A\,\h{j}_A^\mu-\frac{1}{2}\mf{S}_{\lambda\nu}\h{S}^{\mu\lambda\nu}\right) \right\},
\end{equation}
where $\beta$, $\zeta$, $\zeta_A$ and $\mf{S}$ are the relevant Lagrange multiplier functions, and correspond respectively to the four-temperature vector, the ratio between local chemical and axial chemical potentials and temperature and the spin potential~\cite{Becattini:2018duy}. Since $\beta$ is a timelike vector, it is convenient to identify the fluid velocity as
\begin{equation}
u = \frac{\beta}{\sqrt{\beta^2}} = T\,\beta;
\end{equation}
this choice is known as $\beta$ or thermodynamic frame~\cite{Becattini:2014yxa}. It is worth noting that, despite the operators being time-dependent, the non-equilibrium statistical operator (\ref{eq:NEDO}) is actually stationary, simply because the time $\tau$ at which it is evaluated is fixed. However, it is more convenient to express the statistical operator using a later time 3D hypersurface. For instance, to predict spin polarization in heavy-ion collisions, the statistical operator is evaluated on the freezeout hypersurface~\cite{Becattini:2024uha}. This can be achieved by applying Gauss's theorem: the integral in eq.~(\ref{eq:NEDO}) over the hypersurface $\Sigma(\tau)$ can be expressed as the same integral over $\Sigma(\tau+\Delta\tau)$ plus the integral of the divergence of the integrand over the region $\Omega$ marked by the two hypersurfaces, provided that the flux of the integrand vanishes at the timelike boundary $\Theta$, as depicted in figure~\ref{fig:foliation}. Therefore, choosing a later time hypersurface $\Sigma$
we obtain the statistical operator
\begin{multline}
\label{eq:StatOper}
\h{\rho} = \frac{1}{Z} \exp\Bigg\{ 
-\int_{\Sigma} {\rm d}\Sigma_\mu(y)\left(\h{T}^{\mu\nu}(y)\beta_\nu(y)
 -\zeta(y)\,\h{j}^\mu(y) -\zeta_A(y)\,\h{j}_A^\mu(y)
 -\frac{1}{2}\mf{S}_{\lambda\nu}(y)\h{S}^{\mu\lambda\nu}(y)\right) \\ +
\int_\Omega {\rm d}\Omega\left[\h{T}_S^{\mu\nu}\xi_{\mu\nu} + \h{T}_A^{\mu\nu}\left(\mf{S}_{\mu\nu}-\varpi_{\mu\nu}\right)
 -\h{j}^\mu \nabla_\mu \zeta - \nabla_\mu\left(\zeta_A \h{j}_A^\mu \right) -\frac{1}{2}\h{S}^{\mu\lambda\nu} \nabla_\mu \mf{S}_{\lambda\nu} \right]\Bigg\},
\end{multline}
where we made use of the equations of motion
\begin{equation}
\nabla_\mu \h{T}^{\mu\nu} = 0,\quad
\nabla_\lambda \h{S}^{\lambda,\mu\nu} = -2 \h{T}^{\mu\nu}_A,\quad
\nabla_\mu \h{j}^\mu = 0.
\end{equation}
The part expressed as an integral over $\Sigma$ is the local equilibrium part and does not generate entropy, while the other part gives rise to entropy production~\cite{Becattini:2019dxo}. For this reason the contributions to physical observables obtained from the latter part are related to dissipative processes, see also~\cite{Hayata:2015lga}.

Consider now the mean value of an operator $\h{O}$ at the point $x$ (and at momentum $k$ when considering the Wigner function), which is $\langle\h{O}(x)\rangle = \tr[\,\wrho\,\h{O}(x)]$. The operator $\h{O}$ can be a tensor with a set of indices ${(\mu)}$ that we omit: $\h{O}(x)=\h{O}^{(\mu)}(x)$. Taking advantage of the hydrodynamic regime, the mean value can be worked out perturbatively with a gradient expansion of the hydrodynamic fields.  For this purpose, the hydrodynamic fields in the local equilibrium part of the statistical operator are expanded in a Taylor series around $x$, that is
\begin{subequations}
\begin{align}
\beta_\nu(y) = & \beta_\nu(x) + (y-x)^\rho \de_\rho\beta_\nu(x) + \cdots\\
    =& \beta_\nu(x) - (y-x)^\rho \varpi_{\rho\nu}(x)  + (y-x)^\rho \xi_{\rho\nu}(x)+ \cdots,\nonumber\\
\zeta(y) = & \zeta(x) + \de_\rho\zeta(x) (y-x)^\rho + \cdots,\quad
    \zeta_A(y) = \zeta_A(x) +  \de_\rho\zeta_A(x) (y-x)^\rho + \cdots,\\
\mf{S}_{\lambda\nu}(y) = & \mf{S}_{\lambda\nu}(x) + (y-x)^\mu \de_\mu \mf{S}_{\lambda\nu}(x) + \cdots,
\end{align}
\end{subequations}
and where we defined the thermal vorticity tensor
\begin{equation}
\varpi_{\rho\sigma}(x) = -\frac{1}{2} \left[\de_\rho\beta_\sigma(x) 
    - \de_\sigma\beta_\rho(x) \right],
\end{equation}
and the thermal shear tensor
\begin{equation}
\xi_{\rho\sigma}(x) = \frac{1}{2} \left[\de_\rho\beta_\sigma(x) 
    + \de_\sigma\beta_\rho(x) \right].
\end{equation}
In this work we keep all the terms that are of first order in derivative, including the derivatives of the axial chemical potential and the spin potential, and we neglect higher derivatives. Even if one expect the gradients of the axial chemical potential and of spin potential to be small, it will be shown that these induce the leading contributions for the dissipative part of the spin polarization in a fluid without parity violating effects and therefore must be included in order to derive the dissipative part of the spin polarization formula. The inclusion of the higher order terms are left for future studies.
Keeping only these terms, the statistical operator~(\ref{eq:StatOper}) is approximated as
\begin{equation}
\label{eq:statopABC}
\h{\rho} \simeq \frac{1}{Z}\exp\left\{\h{A}+\h{B}_{\rm LTE} +\h{C}_{\rm D} \right\},\quad
\h{B}_{\rm LTE}=\h{B}_\varpi+\h{B}_\xi+\h{B}_{\zeta_A}
+\h{B}_{\mf{S}}+\h{B}_{\partial\mf{S}}
+\h{B}_{\partial\zeta}+\h{B}_{\partial\zeta_A},
\end{equation}
where the leading term is
\begin{equation}
\label{eq:leadingSO}
\wrho_{{\rm eq}(x)} =\frac{\e^{\h{A}}}{\tr\left[e^{\h{A}}\right]}
    =\frac{\e^{-\beta(x)\cdot\h{P}+\zeta(x)\h{Q}} }{\tr\left[\e^{-\beta(x)\cdot\h{P}+\zeta(x)\h{Q}}\right]},\quad
\h{A} = -\beta(x)\cdot\h{P}+\zeta(x)\h{Q},
\end{equation}
where $\h{P}^\nu = \int_\Sigma \di\Sigma_\mu\,\h{T}^{\mu\nu}$ and $\h{Q} = \int_\Sigma \di\Sigma_\mu\,\h{j}^\mu$ are the four-momentum operator and the total charge, and $\wrho_{{\rm eq}(x)}$ has the grand canonical form with the hydrodynamic fields evaluated in~$x$. The first order local equilibrium corrections are
\begin{subequations}
\begin{align}
\label{eq:BLTEa}
\h{B}_\varpi =& \frac{1}{2}\varpi_{\rho\sigma}(x)\h{J}_x^{\rho\sigma}, &
\h{B}_\xi =& -\frac{1}{2}\xi_{\rho\sigma}(x)\h{\Xi}_x^{\rho\sigma},\quad
\h{B}_{\zeta_A} = \zeta_{A}(x)\h{Q}_A,\\
\label{eq:BLTEb}
\h{B}_{\mf{S}} =& \frac{1}{2}(\mf{S}_{\rho\sigma}(x)-\varpi_{\rho\sigma}(x))\h{S}^{\rho\sigma}, &
\h{B}_{\partial\mf{S}} =& \de_\mu\mf{S}_{\lambda\nu}(x)\int_\Sigma\di\Sigma_\rho(y)\,(y-x)^\mu\h{S}^{\rho,\lambda\nu}(y),\\
\label{eq:BLTEc}
\h{B}_{\partial\zeta} = & \de_\rho\zeta(x)\int_\Sigma\di\Sigma_\lambda(y)\,(y-x)^\rho\h{j}^\lambda(y), &
\h{B}_{\partial\zeta_A} =& \de_\rho\zeta_A(x)\int_\Sigma\di\Sigma_\lambda(y)\,(y-x)^\rho\h{j}_A^\lambda(y),
\end{align}
\end{subequations}
with the angular-boost momentum operator around $x$ given by
\begin{subequations}
\begin{equation}
\h{J}_x^{\rho\sigma} = \int_\Sigma{\rm d}\Sigma_\lambda(y)\,\h{\mathcal{J}}^{\lambda,\rho\sigma}
= \int_\Sigma{\rm d}\Sigma_\lambda(y)
    \left[(y-x)^\rho\wT^{\lambda\sigma}(y) - (y-x)^\sigma\wT^{\lambda\rho}(y) + \h{S}^{\lambda,\rho\sigma}(y)\right],
\end{equation}
and the other global operators are defined by
\begin{align}
\h{\Xi}_x^{\rho\sigma}(\Sigma)= &\int_\Sigma{\rm d}\Sigma_\lambda(y)
    \left[(y-x)^\rho\wT^{\lambda\sigma}(y) + (y-x)^\sigma\wT^{\lambda\rho}(y)\right],\\
\h{Q}_A(\Sigma) =& \int_\Sigma\di\Sigma_\lambda(y)\,\h{j}_A^\lambda(y),\quad
\h{S}^{\rho\sigma}(\Sigma) =\int_\Sigma\di\Sigma_\lambda(y)\,\h{S}^{\lambda,\rho\sigma}(y).
\end{align}
\end{subequations}
Since $\nabla_\lambda\h{\mathcal{J}}^{\lambda,\rho\sigma}=0$, the operator $\h{J}$ is independent of $\Sigma$, whereas the other operators are not conserved and depend on the hypersurface $\Sigma$. The dissipative part is included in
\begin{equation}
\h{C}_{\rm D} = \int_\Omega {\rm d}\Omega\left[\h{T}_S^{\mu\nu}\xi_{\mu\nu} + \h{T}_A^{\mu\nu}\left(\mf{S}_{\mu\nu}-\varpi_{\mu\nu}\right)
 -\h{j}^\mu \nabla_\mu \zeta  - \h{j}_A^\mu\nabla_\mu\zeta_A - \zeta_A\nabla_\mu \h{j}_A^\mu -\frac{1}{2}\h{S}^{\mu\lambda\nu} \nabla_\mu \mf{S}_{\lambda\nu} \right],
\end{equation}
which we later expand in gradients.
Because the system is assumed to be in a hydrodynamic regime with small correlation lengths, the operator $\h{A}$ is dominant compared to all the other operators that are collected in $\h{B}=\h{B}_{\rm LTE} +\h{C}_{\rm D}$. Consequently, these corrections can be accounted for with linear response theory.

Taking advantage of the Kubo identity
\begin{equation}
\e^{\h{A}+\h{B}} \simeq \e^{\h{A}} + \int_0^1 {\rm d}x\, \e^{-z\h{A}}\, \h{B}\, \e^{z\h{A}}
    +\mathcal{O}\left(\h{B}^2 \right),
\end{equation}
and including the partition function $Z=\tr[\h{A}+\h{B}]$, the first-order expansion of the expectation value of an operator $\h{O}(x)$, see for instance for this  formalism~\cite{Buzzegoli:2017cqy,Buzzegoli:2018wpy}, is
\begin{equation}
\langle\h{O}(x)\rangle = \tr\left[\,\wrho\,\h{O}(x)\right]\simeq \langle\h{O}(x)\rangle_\beta + \langle\h{B}_{[1]}\, \h{O}(x)\rangle_{\beta,c}+\cdots
\end{equation}
where the subscript ``c'' denotes the connected correlator
\begin{equation}
\langle\h{B}\, \h{O} \rangle_c =  \langle\h{B}\, \h{O} \rangle
    - \langle\h{B} \rangle\langle\h{O} \rangle,
\end{equation}
and the subscript $\beta$ denotes thermal averages with the leading order statistical operator, that is
\begin{equation}
\label{eq:MeanLO}
\langle\h{O}(x)\rangle_\beta
=\tr\left[\wrho_{{\rm eq}(x)} \h{O}(x)\right]
=\frac{\tr\left[\e^{\h{A}} \h{O}(x)\right]}{\tr\left[\e^{\h{A}}\right]}
=\frac{\tr\left[\e^{-\beta(x)\cdot\h{P}+\zeta(x)\h{Q}} \h{O}(x)\right]}{\tr\left[\e^{-\beta(x)\cdot\h{P}+\zeta(x)\h{Q}}\right]},
\end{equation}
and we defined
\begin{equation}
\h{B}_{[1]} = \int_0^1\!{\rm d}z\, \h{B}_{[z]}
= \int_0^1\!{\rm d}z\, \e^{-z\h{A}} \h{B} \e^{z\h{A}},
\end{equation}
with $\h{B}_{[z]}\equiv \e^{-z\h{A}} \h{B} \e^{z\h{A}}$. Note that the operators $\h{B}_{\rm LTE}$ and $\h{C}_{\rm D}$ in the statistical operator (\ref{eq:statopABC}) moved from the exponent to linear terms; the total contribution is simply obtained by summing all the terms from each operator.

The different contributions contained in $\h{B}=\h{B}_{\rm LTE} +\h{C}_{\rm D}$ will now be analyzed. In general, $\mcU$ denotes a hydrodynamic field or its derivative, that is
\begin{equation}
\label{eq:mcUReducible}
\mcU_{(\alpha)} = \varpi_{\rho\sigma},\,
\xi_{\rho\sigma},\, \de_\rho\zeta,\,
\zeta_{A},\, \de_\rho\zeta_A,\,
\mf{S}_{\rho\sigma},\, \de_\tau\mf{S}_{\rho\sigma},
\end{equation}
where $(\alpha)$ represents a set of indices. 
First, consider the dissipative part. Using this notation, each term in $\h{C}_{\rm D}$ can be written as
\begin{equation}
\h{C}_{\mcU} = c_{\mcU}\int_\Omega \di\Omega\,\mcU_{(\alpha)}(x_2)\,
    \h{\mathcal{C}}_{\mcU}^{(\alpha)}(x_2),
\end{equation}
with $c_{\mcU}$ a number.
The dissipative contribution of $\mcU$ to the mean value $O$ is then
\begin{equation}
\label{eq:LRDiss0}
\Delta_{\mcU,\,{\rm D}} O(x) = \frac{\tr\left[\e^{\h{A}+\h{C}_{\mcU}} \h{O}(x)\right]}{\tr\left[\e^{\h{A}+\h{C}_{\mcU}}\right]} - \langle\h{O}(x)\rangle_\beta
    = \langle\h{C}_{\mcU\,[1]}\, \h{O}(x)\rangle_{\beta,c}.
\end{equation}
The correlator on the RHS of (\ref{eq:LRDiss0}) can be worked out and written as
\begin{equation}
\label{eq:LRDiss}
\Delta_{\mcU,\,{\rm D}} O(x) = c_{\mcU} \frac{\I}{|\beta(x)|}
    \int_{-\infty}^t \di^4 x_2 \int_{-\infty}^{t_2} \di s \,
    \left\langle\left[\h{O}(x),\, \h{\mathcal{C}}_{\mcU}^{(\alpha)}(s,\, \bm{x}_2) \right] \right\rangle_{\beta(x)}
    \, \mcU_{(\alpha)}(x_2)\, ,
\end{equation}
where $|\beta(x)|=\sqrt{\beta(x)^2}=T(x)^{-1}$. The derivation of  eq.~(\ref{eq:LRDiss}) from eq.~(\ref{eq:LRDiss0}) is a known result~\cite{Hosoya:1983id,Huang:2011dc,Harutyunyan:2018cmm,Becattini:2019dxo,Harutyunyan:2021rmb} and will not be repeated here. However, the important properties and approximations used are listed:
\begin{itemize}
    \item The integral over the volume $\Omega$ between the initial time $\tau$ and the time at the point $x$ has been approximated as the volume within the two hyperplanes with normal $u^\mu$ at the point $x$ and at the initial time hypersurface $\tau$.\footnote{In heavy-ion collisions, where the freezeout hypersurface is usually chosen as the final time hypersurface, it is questionable whether this remains a good approximation. However, similarly to what is done in appendix~\ref{app:Kubo} for the local equilibrium corrections, expressing the linear response correction with the retarded Green's function allows the volume integral to be factored into $\int_\Omega \di^4 x_2 \e^{-\I q(y-x)}$ that can be modified accordingly.}
    \item The operators $\h{\mathcal{C}}_{\mcU}^{(\alpha)}(x_2)$ commute with $\h{Q}$, such that
    \begin{equation}
        \e^{z\left(\beta(x)\cdot\h{P}-\zeta(x)\h{Q}\right)}
            \h{\mathcal{C}}_{\mcU}^{(\alpha)}(x_2)
            \e^{-z\left(\beta(x)\cdot\h{P}-\zeta(x)\h{Q}\right)} =
        \h{\mathcal{C}}_{\mcU}^{(\alpha)}\left(x_2 - \I z \sqrt{\beta(x)^2}\right).
    \end{equation}
    \item The correlation between the operator $\h{O}$ at time $t$ and $\h{\mathcal{C}}_{\mcU}^{(\alpha)}$ at the (infinitely) remote past is vanishing:
    \begin{equation}\begin{split}
        \lim_{t_0\to\infty} \mean{\h{O}(x)\h{\mathcal{C}}_{\mcU}^{(\alpha)}(t_0,\bm{x}_2) }_{\beta(x)}
        \simeq& \lim_{t_0\to\infty} \mean{\h{O}(x)}_{\beta(x)}\mean{\h{\mathcal{C}}_{\mcU}^{(\alpha)}(t_0,\bm{x}_2) }_{\beta(x)}\\
        =& \mean{\h{O}(x)}_{\beta(x)}\mean{\h{\mathcal{C}}_{\mcU}^{(\alpha)}(x_2) }_{\beta(x)}\,.
    \end{split}\end{equation}
\end{itemize}
From eq.~(\ref{eq:LRDiss}) we see that in principle the dissipative effects on the observable $\h{O}$ depends on the whole history of the hydrodynamic fields $\mcU$. However, in the hydrodynamic regime considered in this work, the correlation length between $\h{O}(x)$ and $\h{\mathcal{C}}_{\mcU}(x_2)$ is typically much smaller than the distance over which the gradients of $\mcU$ have significant variation. Therefore, in the hydrodynamic limit, we obtain a good approximation by replacing the hydrodynamic field $\mcU(x_2)$ with its Taylor expansion around the point $x$ where we are evaluating our observable.
The dissipative part $\h{C}_{\rm D}$ is then decomposed in
\begin{equation}
\h{C}_{\rm D}\simeq \h{C}_\xi +\h{C}_{\partial\zeta}
    +\h{C}_{\zeta_A} +\h{C}_{\partial\zeta_A}
    +\h{C}_{\mf{S}} +\h{C}_{\partial\mf{S}},
\end{equation}
where
\begin{subequations}
\begin{align}
\label{eq:CDissa}
\h{C}_\xi = & \xi_{\rho\sigma}(x) \int\di^4 x\, \h{T}^{\rho\sigma}_S(x_2),\quad
\h{C}_{\partial\zeta} =  -\de_\rho\zeta(x) \int\di^4 x\, \h{j}^{\rho}(x_2),\\
\label{eq:CDissb}
\h{C}_{\zeta_A} = & -\zeta_A(x)\int\di^4 x\, \de_\lambda\h{j}_A^\lambda(x_2),\,
\h{C}_{\de\zeta_A} =  -\de_\rho\zeta_A(x)\int\di^4 x\, \left[\h{j}_A^\rho(x_2)
    + (x_2-x)^\rho\de_\lambda\h{j}^\lambda_A \right],\\
\label{eq:CDissc}
\h{C}_{\mf{S}} =& (\mf{S}_{\rho\sigma}(x) - \varpi_{\rho\sigma}(x))
    \int\di^4 x\, \h{T}^{\rho\sigma}_A(x_2),\\
\label{eq:CDissd}
\h{C}_{\de\mf{S}} =& -\frac{1}{2}\de_\tau\mf{S}_{\rho\sigma}(x)\int\di^4 x\left[ \h{S}^{\tau,\rho\sigma}(x_2)
    -2(x_2-x)^\tau\h{T}^{\rho\sigma}_A(x_2)\right].
\end{align}
\end{subequations}
The linear response theory of these terms result in
\begin{equation}
\label{eq:DissResponse}
\Delta_{\mcU,\,{\rm D}} O(x) = \mcU_{(\alpha)}(x)\,c_{\mcU}\,
    \left( \h{O},\, \h{\mathcal{C}}_{\mcU}^{(\alpha)}\right)_{\rm D}\, ,
\end{equation}
where we defined
\begin{equation}
\label{eq:CorrDCoord}
\boxed{
\left( \h{X},\, \h{Y}\right)_{\rm D} =  \frac{\I}{|\beta(x)|}
    \int_{-\infty}^t \di^4 x_2 \int_{-\infty}^{t_2} \di s \,
    \left\langle\left[\h{X}(x),\, \h{Y}(s,\, \bm{x}_2) \right] \right\rangle_{\beta(x)}.}
\end{equation}
All the transport coefficients will be expressed with this thermal correlator. More commonly Kubo formulas are expressed in momentum space. Indeed, the correlator (\ref{eq:CorrDCoord}) can be expressed with the retarded Green's function in momentum space defined by
\begin{align}
G^R_{\h{X}\h{Y}}(x-x_2) = & -\I\theta(x-x_2)
    \left\langle\left[\h{X}(x),\, \h{Y}(x_2) \right] \right\rangle_{\beta(x)},\\
G^R_{\h{X}\h{Y}}(x) = & \int \frac{\di^4 p}{(2\pi)^3}
    \e^{-\I p \cdot x} G^R_{\h{X}\h{Y}}(p).
\end{align}
Working out the eq.~(\ref{eq:CorrDCoord}), see \cite{Hosoya:1983id,Huang:2011dc,Harutyunyan:2018cmm,Becattini:2019dxo,Harutyunyan:2021rmb} and appendix~\ref{app:Kubo}, we eventually obtain
\begin{equation}
\label{eq:CorrDMom}
\left( \h{X},\, \h{Y}\right)_{\rm D} = 
    -\frac{1}{|\beta(x)|}u^\lambda \lim_{p\cdot u\to 0} \lim_{p_\perp \to 0} 
    \frac{\de}{\de p^\lambda} {\rm Im}\, G^R_{\h{X}\h{Y}}(p) .
\end{equation}
This exact method and these assumptions were used to obtained for the first time the Kubo formula for the shear viscosity in a relativistic fluid with an underlying quantum field theory as microscopic interactions \cite{Hosoya:1983id}. In the first order approximation adopted here (where past correlation of the hydrodynamic field and microscopic interactions are neglected) we obtained the linear response as a static (zero-frequency) retarded Green's function. The study of higher orders can be understood as including non-local (memory) effects and it leads to consider retarded Green's functions at finite frequency, or at least its higher derivative~\cite{Harutyunyan:2021rmb}. The study of these higher corrections in the derivative expansion naturally generates finite relaxation time scales in the transport equations, see for instance~\cite{Zubarev:1972285,Koide:2006ef,Koide:2008nw,Harutyunyan:2021rmb}. We conclude the discussion of the dissipative part by considering the second terms inside the integral of $\h{C}_{\de\zeta_A}$ and in $\h{C}_{\de\mf{S}}$ of eqs.~(\ref{eq:CDissb}) and (\ref{eq:CDissd}). In these cases it could be convenient to split the coordinate from the operator, obtaining (see appendix~\ref{app:Kubo})
\begin{equation}
\left( \h{O},\, \int\di^4 x_2 (x_2-x)^\rho\, \de_\lambda\h{j}^\lambda_A\right)_{\rm D} =
-\frac{1}{|\beta(x)|}u^\sigma \lim_{p\cdot u\to 0} \lim_{p_\perp \to 0}
\frac{\de}{\de p^\sigma} \frac{\de}{\de p^\rho} {\rm Re}\, G^R_{\h{O}\de_\lambda\h{j}^\lambda_A}(p),
\end{equation}
and
\begin{equation}
\left( \h{O},\, \int\di^4 x_2 (x_2-x)^\tau \,\h{T}^{\rho\sigma}_A\right)_{\rm D} =
-\frac{1}{|\beta(x)|}u^\sigma \lim_{p\cdot u\to 0} \lim_{p_\perp \to 0}
 \frac{\de}{\de p^\sigma} \frac{\de}{\de p^\tau} {\rm Re}\, G^R_{\h{O}\h{T}^{\rho\sigma}_A}(p).
\end{equation}

For the non-dissipative part, each term in the operator $\h{B}_{\rm LTE}$ has the form
\begin{equation}
\h{B}_{\mcU}
= b_{\mcU}\,\mcU_{(\alpha)}(x)\, \h{\mathcal{B}}_{\mcU}^{(\alpha)},
\end{equation}
resulting in the following contribution to the mean value of the operator $\h{O}$:
\begin{equation}
\label{eq:LTEResponse}
\Delta_{\mcU,\,{\rm LTE}} O(x)  = \frac{\tr\left[\e^{\h{A}+\h{B}_{\mcU}} \h{O}(x)\right]}{\tr\left[\e^{\h{A}+\h{B}_{\mcU}}\right]} - \langle\h{O}(x)\rangle_\beta
= \mcU_{(\alpha)}(x)\,b_{\mcU}\,
    \left( \h{O},\, \h{\mathcal{B}}_{\mcU}^{(\alpha)}\right)_{\rm LTE}
\end{equation}
where
\begin{equation}
\label{eq:CorrLTECoord}
\boxed{
\left( \h{X},\, \h{Y}\right)_{\rm LTE}
    = \int_0^{|\beta|} \frac{\di\tau}{|\beta(x)|}
        \mean{\h{Y}_{[\tau/|\beta|]}\h{X}(x)}_{\beta(x),\,{\rm c}}\, ,
}
\end{equation}
and we remind
\begin{equation}
\h{Y}_{[\tau/|\beta|]} =
\e^{\frac{\tau}{|\beta|}\left(\beta(x)\cdot\h{P}-\zeta(x)\h{Q}\right)}
            \h{Y}
            \e^{-\frac{\tau}{|\beta|}\left(\beta(x)\cdot\h{P}-\zeta(x)\h{Q}\right)}.
\end{equation}
To write the correlator (\ref{eq:CorrLTECoord}) in terms of retarded Green's function in momentum space it is convenient to consider three different cases. The first is the response to thermal vorticity $\h{B}_\varpi$ which is given in terms of the angular-boost generators $\h{J}$ that are conserved operators and are present even at global thermodynamic equilibrium. In this case we have:
\begin{equation}
\h{B}_\varpi = \frac{1}{2}\varpi_{\rho\sigma}(x)\h{J}_x^{\rho\sigma},\quad
\Delta_{\varpi,\,{\rm LTE}} O(x) = \frac{1}{2}\varpi_{\rho\sigma}(x)\,
    \left( \h{O},\, \h{J}_x^{\rho\sigma}\right)_{\rm LTE},
\end{equation}
where
\begin{equation}
\begin{split}
\left( \h{O},\, \h{J}_x^{\rho\sigma}\right)_{\rm LTE}
    = & \int_0^{|\beta|} \frac{\di\tau}{|\beta(x)|}
        \mean{\h{J}^{\rho\sigma}_{x[\tau/|\beta|]}\h{O}(x)}_{\beta(x),\,{\rm c}}\\
=& \frac{2\I}{|\beta(x)|} \int_{-\infty}^t\di t_2 \int_\Sigma \di\Sigma_\lambda(y) (y-x)^{\rho} 
        \left\langle\left[\h{O}(x),\, \h{T}_B^{\lambda\sigma}(t_2,\, \bm{y}_2) \right] \right\rangle_{\beta(x)}.
\end{split}
\end{equation}
The operator $\h{J}$ is a conserved quantity and it is independent of the pseudo-gauge used to compute it; therefore it is more convenient to use the symmetric Belinfante energy-momentum tensor $\h{T}_B$ as done above, such that we do not have to include an explicit spin part. The factor $2$ already consider that the correlator must be contracted with the antisymmetric tensor $\varpi$. Being a conserved operator, $\h{J}$ is also independent of the hypersurface and without loss of generality, we can choose $\Sigma$ as the hyperplane with normal $u(x)$, obtaining
\begin{equation}
\begin{split}
\left( \h{O},\, \h{J}_x^{\rho\sigma}\right)_{\rm LTE}
=& \frac{2\I}{|\beta(x)|} u_\lambda(x) \int_{-\infty}^t\di t_2 \int \di^3 y\, (y-x)^\rho
        \left\langle\left[\h{O}(x),\, \h{T}_B^{\lambda\sigma}(t_2,\, \bm{y}_2) \right] \right\rangle_{\beta(x)}\\
=& \frac{2\I}{|\beta(x)|} u_\lambda(x) \int_{-\infty}^0\di t_2 \int \di^3 x_2\, x_2^\rho
        \left\langle\left[\h{O}(0),\, \h{T}_B^{\lambda\sigma}(t_2,\, \bm{x}_2) \right] \right\rangle_{\beta(x)},
\end{split}
\end{equation}
which can be written as (see appendix~\ref{app:Kubo})
\begin{equation}
\left( \h{O},\, \h{J}_x^{\rho\sigma}\right)_{\rm LTE} = 
    \frac{1}{|\beta(x)|}u_\lambda(x)  \lim_{u\cdot p \to 0} \lim_{p_\perp \to 0}
    \frac{\de}{\de p_\rho}  {\rm Im}\,G^R_{\h{T}_B^{\lambda\sigma}\h{O}}(p) .
\end{equation}
All the other terms resulting from the linear response of the local thermal equilibrium operators $\h{B}_{\rm LTE}$ depend on the specific form of the hypersurface $\Sigma$. This is a consequence of being out-of-equilibrium contributions, yet non-dissipative, and consequently of being derived from non conserved quantum operators. Consider first $\h{B}_{\zeta_A}$ and $\h{B}_{\mf{S}}$; they can be written as
\begin{equation}
\h{B}_{\mcU,0}
= b_{\mcU}\,\mcU_{(\alpha)}(x)\, \h{\mathcal{B}}_{\mcU,0}^{(\alpha)}
= b_{\mcU}\,\mcU_{(\alpha)}(x)\int_\Sigma\di\Sigma_\lambda(y)
    \, \h{\mathcal{B}}_{\mcU}^{\lambda(\alpha)}(y).
\end{equation}
Their linear contribution to the mean value of an observable is
\begin{equation}
\Delta_{\mcU,0,\,{\rm LTE}} O(x) = \mcU_{(\alpha)}(x)\,b_{\mcU}\,
    \left( \h{O},\, \h{\mathcal{B}}_{\mcU,0}^{(\alpha)}\right)_{\rm LTE}. 
\end{equation}
If the hypersurface has the same form as the freezeout in heavy-ion collisions, the correlator can be written as (see appendix~\ref{app:Kubo})
\begin{equation}
\label{eq:CorrGRU0FO}
\left( \h{O},\, \h{\mathcal{B}}_{\mcU,0}^{(\alpha)}\right)_{\rm LTE}
= \frac{\hat{t}_\lambda}{|\beta(x)|} \lim_{p\cdot u\to 0} \lim_{p_T\to 0}
    {\rm Re}\,G^{R}_{\h{O}\h{\mathcal{B}}^{\lambda(\alpha)}_{\mcU}}(p),
\end{equation}
with $\h{t}$ is the time direction in the center of mass frame; instead, approximating the hypersurface with an hyperplane orthogonal to $u$ (see appendix~\ref{app:Kubo}):
\begin{equation}
\begin{split}
\left( \h{O},\, \h{\mathcal{B}}_{\mcU,0}^{(\alpha)}\right)_{\rm LTE}
\simeq& \frac{ u_\lambda}{|\beta(x)|} \lim_{p\cdot u\to 0} \lim_{p_T\to 0}
    {\rm Re}\,G^{R}_{\h{O}\h{\mathcal{B}}^{\lambda(\alpha)}_{\mcU}}(p).
\end{split}
\end{equation}
Lastly, the operators $\h{B}_\xi,\, \h{B}_{\partial\mf{S}},\, \h{B}_{\partial\zeta},\, \h{B}_{\partial\zeta_A}$ all have the form
\begin{equation}
\h{B}_{\mcU,1}
= b_{\mcU}\,\mcU_{\rho(\alpha_2)}(x)\, \h{\mathcal{B}}_{\mcU,1}^{\rho(\alpha_2)}
= b_{\mcU}\,\mcU_{\rho(\alpha_2)}(x)\int_\Sigma\di\Sigma_\lambda(y)\,(y-x)^\rho
    \, \h{\mathcal{B}}_{\mcU}^{\lambda(\alpha_2)}(y),
\end{equation}
and their contribution is expressed as
\begin{equation}
\Delta_{\mcU,1,\,{\rm LTE}} O(x) = \mcU_{\rho(\alpha_2)}(x)\,b_{\mcU}\,
    \left( \h{O},\, \h{\mathcal{B}}_{\mcU,1}^{\rho(\alpha_2)}\right)_{\rm LTE}.
\end{equation}
In momentum space, we can show that (see appendix~\ref{app:Kubo}) the correlator becomes
\begin{equation}
\label{eq:CorrGRU1FO}
\left( \h{O},\, \h{\mathcal{B}}_{\mcU,1}^{\rho(\alpha_2)}\right)_{\rm LTE}
= \frac{\hat{t}_\lambda}{|\beta(x)|} \lim_{p\cdot u\to 0} \lim_{p_T\to 0}
     \Delta^\rho_\sigma \frac{\de}{\de p_\sigma}
     {\rm Im}\,G^{R}_{\h{O}\h{\mathcal{B}}^{\lambda(\alpha_2)}_{\mcU}}(p)
\end{equation}
when $\Sigma$ is the freezeout hypersurface, or, approximating the hypersurface with an hyperplane orthogonal to $u$ (see appendix~\ref{app:Kubo}):
\begin{equation}
\begin{split}
\left( \h{O},\, \h{\mathcal{B}}_{\mcU,1}^{\rho(\alpha_2)}\right)_{\rm LTE}
\simeq& \frac{u_\lambda}{|\beta(x)|} \lim_{p\cdot u\to 0} \lim_{p_T\to 0}
     \frac{\de}{\de p_\rho}{\rm Im}\,G^{R}_{\h{O}\h{\mathcal{B}}^{\lambda(\alpha_2)}_{\mcU}}(p).
\end{split}
\end{equation}
%

%******************************************************************************************************
\subsection{Chiral coefficients}
Many dissipative corrections to the spin polarization are given by chiral transport coefficients, meaning that they are obtained with thermal correlators of two operators that transform with opposite sign under a parity transformation $\h{\group{P}}$; that is, if
\begin{equation}
    \h{\group{P}}\, \h{O}\, \h{\group{P}}^{-1} = \eta_O \h{O},\quad
    \h{\group{P}}\, \h{B}\, \h{\group{P}}^{-1} = \eta_B \h{B},
\end{equation}
the coefficient given by $\mean{\h{O}\,\h{B}}_\beta$ is chiral if $\eta_O\,\eta_B=-1$.
Notable examples of chiral coefficients are those related to the chiral magnetic effect and to the chiral vortical effect. Chiral coefficients are not vanishing only if the system has interactions that breaks parity or if there is a chiral imbalance $\zeta_A$. Because otherwise $\h{\group{P}}\, \wrho_{{\rm eq}(x)}\, \h{\group{P}}^{-1} = \wrho_{{\rm eq}(x)}$ and we have
\begin{equation}
\begin{split}
\mean{\h{O}\,\h{B}}_\beta =& \tr\left[ \wrho_{{\rm eq}(x)}\, \h{O}\, \h{B} \right]
    = \tr\left[ \wrho_{{\rm eq}(x)}\, \h{\group{P}}\, \h{\group{P}}^{-1}\, \h{O}\, \h{\group{P}}\, \h{\group{P}}^{-1}\, \h{B} \, \h{\group{P}}\, \h{\group{P}}^{-1}\right]\\
=& \eta_O\,\eta_B \tr\left[ \h{\group{P}}^{-1} \wrho_{{\rm eq}(x)}\, \h{\group{P}} \h{O}\, \h{B} \right]
    = - \mean{\h{O}\,\h{B}}_\beta =0.
\end{split}
\end{equation}
In the presence of a chiral imbalance, we can obtain a chiral coefficient including the axial chemical potential in the equilibrium statistical operator
\begin{equation}
\label{eq:SOeqmuA}
\wrho_{{\rm eq}(x)} = \frac{1}{\mathcal{Z}_{\rm eq}}\exp\left\{-\beta(x)\cdot\wP + \zeta(x)\h{Q} + \zeta_A(x)\h{Q}_A\right\} 
\end{equation}
as done in~\cite{Buzzegoli:2018wpy} for free chiral fermions. Another option to estimate them is to include a higher-order correction in $\zeta_A$.
This is done by including higher order of $\zeta_A(x) \h{Q}_A$ in the linear response expansion. Consider for instance a chiral coefficient resulting from an operator $\h{B}_X$; we expand as before the statistical operator as follows
\begin{equation}
\h{\rho} \simeq \frac{1}{Z}\exp\left\{\h{A} +\zeta_A(x) \h{Q}_A + \h{B}_{X} + \dots\right\}
\end{equation}
and the average of an operator is obtained from the linear response theory at second order in the axial charge:
\begin{equation}
\langle\h{O}(x)\rangle = \tr\left[\,\wrho\,\h{O}(x)\right]\simeq \langle\h{O}(x)\rangle_\beta + \langle\h{B}_{[1]}\, \h{O}(x)\rangle_{\beta,c}
+ \frac{1}{2}\langle\h{B}_{[2]}\, \h{O}(x)\rangle_{\beta,c}  + \cdots
\end{equation}
where
\begin{equation}
\h{B}_{[2]} = \frac{1}{2}\int_0^1\di\lambda_1 \int_0^1\di\lambda_2
    T_\lambda \left\{ \h{B}_{[\lambda_1]} \h{B}_{[\lambda_2]}\right\}
\end{equation}
where $T_\lambda$ is the time-ordering operator with respect to the variables $\lambda_1$
and $\lambda_2$, and it is straightforward to see that
\begin{equation}
\label{eq:ConnectedAndTimeOrdered}
\frac{1}{2}\mean{\h{B}_{[2]} \h{O}(x)}_{\beta,{\rm c}} = \left(\h{B},\,\h{B},\,\h{O}(x) \right),
\end{equation}
where we defined the three-point correlation function as
\begin{equation}
\label{eq:threepoint}
\begin{split}
\big(\h{Y},\,\h{Z},\,\h{X}\big) \equiv& \frac{1}{2}\int_0^1\di\lambda_1\int_0^1\di\lambda_2 T_\lambda
	\left\{ \mean{\h{Y}_{[\lambda_1]} \h{Z}_{[\lambda_2]} \h{X} }_\beta\right.
- \mean{\h{Y}_{[\lambda_1]} \h{Z}_{[\lambda_2]}}_\beta \mean{\h{X} }_\beta - \mean{\h{Y}_{[\lambda_1]}}_\beta \mean{\h{Z}_{[\lambda_2]} \h{X} }_\beta\\
&\left. -\mean{\h{Z}_{[\lambda_2]}}_\beta \mean{\h{Y}_{[\lambda_1]} \h{X} }_\beta
	+ 2 \mean{\h{Y}_{[\lambda_1]}}_\beta \mean{\h{Z}_{[\lambda_2]}}_\beta \mean{\h{X} }_\beta \right\}.
\end{split}
\end{equation}
It is also straightforward to find the identities:
\begin{equation}
\label{eq:ThreeCorrIdentities}
\begin{split}
\left(\h{A} + \h{B},\, \h{C},\, \h{O} \right) = & \left(\h{A},\, \h{C},\, \h{O} \right) + \left(\h{B},\, \h{C},\, \h{O} \right)\\
\left(\h{A},\, \h{B},\, \h{O} \right) = & \left(\h{B},\, \h{A},\, \h{O} \right).
\end{split}
\end{equation}
The order $\mathcal{O}(\delta_A\de^1)$ is then only contained in:
\begin{multline*}
\left(\zeta_A(x) \h{Q}_A + \h{B}_{X},\,\zeta_A(x) \h{Q}_A + \h{B}_{X},\,\h{O}(x) \right) =
    \zeta_A(x)^2\left(\h{Q}_A,\,\h{Q}_A,\,\h{O}(x) \right)\\
	+2 \zeta_A(x)\left(\h{Q}_A,\,\h{B}_X,\,\h{O}(x) \right) + \left(\h{B}_X,\,\h{B}_X,\,\h{O}(x) \right)
\simeq  2 \zeta_A(x)\left(\h{Q}_A,\,\h{B}_X,\,\h{O}(x) \right).
\end{multline*}
In conclusion, a chiral coefficient obtained at first order in $\h{B}_X$ is approximated as:
\begin{equation}
\left( \h{B}_X,\,\h{O}(x) \right) \simeq 2 \zeta_A(x)\left(\h{Q}_A,\,\h{B}_X,\,\h{O}(x) \right).
\end{equation}
%

%*******************************************************************************************************
\subsection{Spin polarization of fermions in a dissipative fluid}
\label{sec:SpinPol}
The main purpose of this work is to determine the spin polarization of a Dirac field in a fluid, considering terms up to the first order in the derivative expansion. To that aim, the covariant Wigner operator of a Dirac field must be introduced. The Wigner operator is defined as the Wigner transform of the two-point function
\begin{equation}
\h{W}(x,\,k)_{AB} = \frac{1}{(2\pi)^4} \! \int \!{\rm d}^4 y\, {\rm e}^{-{\rm i} k \cdot y}
	: \h{\bar{\Psi}}_B \left(x +\frac{y}{2}\right) \h{\Psi}_A \left(x-\frac{y}{2}\right) :\, ,
\end{equation}
where $\h{\Psi}$ is the Dirac field, and $:\,:$ denotes the normal ordering of creation and destruction operators. As being constructed with the spinors $\h{\Psi}$, the Wigner operator is a $4\times 4$ spinoral matrix. With the Wigner transform the non-local two-point function has become an operator depending on the coordinate $x$ and on the off-shell momentum $k$. We instead refer to the mean value of the Wigner operator as the Wigner function:
\begin{equation}
W(x,\,k) = \tr\left[ \wrho\, \h{W}(x,\,k)\right]= \frac{1}{(2\pi)^4} \! \int \!{\rm d}^4 y\, {\rm e}^{-{\rm i} k \cdot y} 	\left\langle: \h{\bar{\Psi}}_B \left(x +\frac{y}{2}\right)
    \h{\Psi}_A \left(x-\frac{y}{2}\right):\right\rangle.
\end{equation}
Without loss of generality, the Wigner function can be expanded on the basis of the 16 independent generators of Clifford algebra as
\begin{equation}
W(x,\,k) = \frac{1}{4} \left[ \mathcal{F}(x,\,k) + \I\gamma^5 \mathcal{P}(x,\,k)
    +\gamma^\mu \mathcal{V}_\mu(x,\,k) + \gamma^5\gamma^\mu \mathcal{A}_\mu(x,\,k)
    +\frac{\I}{4}[\gamma^\mu,\,\gamma^\nu]\mathcal{S}_{\mu\nu}(x,\,k)\right].
\end{equation}
The component functions can be extracted from the Wigner function by tracing over spinoral indices, for example:
\begin{equation}
\mathcal{F}(x,\,k) = \tr_4\left[W(x,\,k)\right],\quad
\mathcal{V}^\mu(x,\,k) = \tr_4\left[\gamma^\mu \, W(x,\,k)\right],\quad
\mathcal{A}^\mu(x,\,k) = \tr_4\left[\gamma^\mu \gamma^5 \, W(x,\,k)\right].
\end{equation}
The Wigner operator can be further decomposed into three different terms that correspond to the particle (future timelike), antiparticle (past timelike), and spacelike parts:
\begin{equation}
\begin{split}
\h{W}(x,\,k) =& \h{W}(x,\,k)\theta(k^2)\theta(k\cdot u) + \h{W}(x,\,k)\theta(k^2)\theta(-k\cdot u)
    + \h{W}(x,\,k)\theta(-k^2)\\
=& \h{W}_+(x,\,k) + \h{W}_-(x,\,k) + \h{W}_S(x,\,k),
\end{split}
\end{equation}
and correspondingly for the Wigner function.

The spin polarization vector of the Dirac field particles with momentum $k$ is obtained directly from the particle part of the Wigner function. Choosing a spacelike 3D hypersurface $\Sigma$, the spin polarization of a massive field is obtained from the formula~\cite{Becattini:2020sww,Becattini:2024uha}
\begin{equation}
\label{eq:SMass}
S^\mu(k)= \frac{1}{2}\frac{\int_\Sigma {\rm d}\Sigma\cdot k\; \WFA_+^\mu(x,k)}
 {\int_\Sigma {\rm d}\Sigma\cdot k\; \mathcal{F}_+(x,k)},
\end{equation}
where $\mathcal{F}_+$ is the scalar particle part of the Wigner function and $\WFA_+^\mu$ is the axial particle part of the Wigner function. If instead we are considering a massless fermion, its spin polarization vector is given by~\cite{Palermo:2023cup}
\begin{equation}
\label{eq:SMassless}
S^\mu(|\bm{k}|,\bm{k})= \frac{k^\mu}{2}\frac{\int_\Sigma {\rm d}\Sigma\cdot k\; q_\alpha \WFA_+^\alpha(x,k)}
 {\int_\Sigma {\rm d}\Sigma\cdot k\; q_\alpha \WFV_+^\alpha(x,k)},
\end{equation}
with $\WFV_+^\mu$ the vector particle part of the Wigner function, and $q$ is an arbitrary lightlike vector non orthogonal to $k$, that is $k\cdot q \neq 0$. The result is ultimately independent of the choice of $q$~\cite{Palermo:2023cup}.

It is now straightforward to obtain the first-order gradient expansion of the spin polarization in a fluid described by the statistical operator (\ref{eq:StatOper}): one only needs to apply the gradient expansion developed above for an operator $\h{O}$ to the Wigner operator. Consider the massive case first. Since the equilibrium operator at point $x$ does not have any breaking of parity, the leading order mean value of the axial Wigner operator is simply vanishing: $\mean{\h{\mathcal{A}}^\mu(x,\,k)}_\beta=0$. The numerator of Eq~(\ref{eq:SMass}) being already of the first order, we do not need to expand the denominator, obtaining:
\begin{equation}
\label{eq:SMassExp}
S^\mu(k)\simeq \frac{1}{2}\frac{\int_\Sigma {\rm d}\Sigma\cdot k\; \Delta\WFA_+^\mu(x,k)}
 {\int_\Sigma {\rm d}\Sigma\cdot k\; \mean{\h{\mathcal{F}}_+(x,k)}_\beta}.
\end{equation}
For the massless case we write the more general expansion where we allow the axial part of the Wigner function to have a non-vanishing value at leading order, for instance by including a chiral imbalance in the equilibrium statistical operator as discussed above, and we obtain
\begin{equation}
\label{eq:SMasslessExp}
\begin{split}
S^\mu(|\bm{k}|,\,\bm{k})\simeq \frac{k^\mu}{2}&\left\{\frac{\int_\Sigma {\rm d}\Sigma\cdot k\; q_\alpha\mean{\h{\mathcal{A}}^\alpha_+(x,k)}_\beta}
 {\int_\Sigma {\rm d}\Sigma\cdot k\; q_\alpha \mean{\h{\mathcal{V}}^\alpha_+(x,k)}_\beta}
\left[1 -\frac{\int_\Sigma {\rm d}\Sigma\cdot k\; q_\alpha \Delta \WFV_+^\alpha(x,k)}
 {\int_\Sigma {\rm d}\Sigma\cdot k\; q_\alpha \mean{\h{\mathcal{V}}^\alpha_+(x,k)}_\beta} \right]\right. \\
 &\left.+\frac{\int_\Sigma {\rm d}\Sigma\cdot k\; q_\alpha \Delta\WFA_+^\alpha(x,k)}
 {\int_\Sigma {\rm d}\Sigma\cdot k\; q_\alpha \mean{\h{\mathcal{V}}^\alpha_+(x,k)}_\beta} \right\}.
\end{split}
\end{equation}
In the next sections, all the possible contributions in $\Delta\WFA_+^\mu$ and $\Delta\WFV_+^\mu$ induced by first-order hydrodynamic fields will be derived, and the corresponding Kubo formulas will be given. The coefficients will be classified based on the properties of their corresponding correlator under discrete transformations: parity $\group{P}$, time reversal $\group{T}$ and charge conjuration $\group{C}$. If the parity of the coefficient under that transformation is even (odd), it also means that the coefficient describes an effect in which a term in our mean value operator (the axial or vector part of the Wigner function) is induced by a hydrodynamic quantity that has the same (opposite) parity under that transformation. As it is expected, all terms that breaks the time reversal symmetry come from the dissipative part of the statistical operator. Coefficients with even (odd) parity under time reversal $\group{T}$ are then named non-dissipative (dissipative). Coefficients with even (odd) parity under a parity transformation $\group{P}$ are then named non-chiral (chiral), as discussed above. Lastly, coefficients are called charged or not-charged based on charge conjugation $\group{C}$. For charged coefficients, it is important to stress a difference between spin polarization and usual mean values of local operators. The terms in a local operator, such as an electric current $\mean{\h{j}^\mu}$, that are proportional to charged coefficients are usually proportional to the chemical potential $\zeta$. Otherwise, the contribution of the particle part will be canceled by the antiparticle part. However, the spin polarization is related only to the particle part of the Wigner function, and a term proportional to the charged coefficient can be different from zero even at zero chemical potential because there is no antiparticle part to cancel the particle contribution. This is, for instance, the case for the spin Hall effect in spin polarization. An electric field or a gradient of chemical potential is still needed to source the effect, but the coefficient itself, despite being charged, does not vanish at zero chemical potential. The same does not apply to the chiral coefficients, except when the antiparticles have the opposite chirality of the particle.

%***********************************************************************************************************
%***********************************************************************************************************
%***********************************************************************************************************
\section{Momentum dependent Kubo formulas}
\label{sec:KuboGeneral}
In this section we describe the strategy used to identify and classify all the possible first order contributions to the axial and vector parts of the Wigner function. Consider the linear response of a hydrodynamic field $\mcU_{(\alpha)}$ of eq.~(\ref{eq:mcUReducible}). One of the advantages of linear response theory is that all thermal expectation values are reduced to thermal averages obtained using the equilibrium statistical operator~(\ref{eq:leadingSO}). This statistical operator has full $\group{T}^4$ spacetime translation invariance but not full Lorentz group invariance because the symmetry is broken by the four timelike vector fluid velocity $u(x)$. Nevertheless, the statistical operator has a residual $SO(3)$ symmetry, coming from the rotation in the three-dimensional space that leaves invariant the four vector $u$. As a first step, $\mcU_{(\alpha)}$ is decomposed into its irreducible components of this $SO(3)$ group. Each of the hydrodynamic fields $\mcU_{(\alpha)}$ generally has a scalar $X$, vector $V^\rho$, symmetric $\Sigma^{\rho\sigma}$ and anti-symmetric $\phi^{\rho\sigma}$ rank-2, and rank-3 spin-like $\Phi^{\lambda,\mu\nu}$ components. We will also generically denote one of these components with $\mcU_{(\alpha)}$.

Consider one such decomposed hydrodynamic field $\mcU_{(\alpha)}$. The axial and vector parts of the Wigner function are vectors under Lorentz transformations, see appendix~\ref{App:WigVectors}, and depend on the four-vector $k$ and on $x$ through the hydrodynamic fields. We then write down all possible vectors that we can build with the fluid velocity $u^\mu$, the metric tensor $\eta^{\mu\nu}$, the four-momentum $k^\mu$, the Levi-Civita symbol $\epsilon^{\mu\nu\rho\sigma}$ and that are linear in $\mcU_{(\alpha)}$. These terms correspond to all the possible linear contributions induced by $\mcU_{(\alpha)}$. We assign to each of them a coefficient that is also a function of $k$ and of $x$, through the equilibrium hydrodynamic fields: $\beta(x)$ and $\zeta(x)$. The Kubo formulas for these coefficients are obtained by matching all these terms with the local equilibrium (\ref{eq:LTEResponse}) and dissipative (\ref{eq:DissResponse}) response proportional to $\mcU_{(\alpha)}$ using the appropriate projector. These coefficients are Lorentz scalars and differ from the usual coefficients related to local operators as they also depend on the momentum $k$. Reminding the general properties under discrete transformations of the main quantities:
\begin{equation}
\label{tab:PTCTransf}
\begin{array}{|l|cc|cc|}
\hline
& \, & \, & \, & \, \\[-1em]
 & \hWFA^\mu & \hWFV^\mu & \partial^\mu & u^\mu,\,k^\mu  \\
\hline
\group{P} &  (-,+) &  (+,-) & (+,-) &  (+,-) \\
\group{T} & (+,-) &  (+,-) &  (-,+) &  (+,-) \\
\group{C} & (+,+) &  (-,-) &  (+,+) &  (+,+) \\
\hline
\end{array}
\end{equation}
where the first term in the bracket correspond to the component along $u$ and the second to the orthogonal part, we can then classify these coefficients based on their properties under these transformations.

Since the identification of all the possible terms contributing to a vector quantity (such as $\WFA^\mu$ and $\WFV^\mu$) and their corresponding Kubo formulas can be obtained independently of the physical meaning of the hydrodynamic field $\mcU_{(\alpha)}$, the decomposition and the identification of the Kubo formulas for the general case of a scalar $X$, vector $V^\rho$, symmetric $\Sigma^{\rho\sigma}$ and anti-symmetric $\phi^{\rho\sigma}$ rank-2, and rank-3 spin-like $\Phi^{\lambda,\mu\nu}$ quantities will be obtained. The results are written for $\WFA^\mu$ but since the only properties of $\WFA^\mu$ used is that is a vector, the results are valid for any vector, including $\WFV^\mu$. These results will be then used in section~\ref{sec:KuboA} to obtain all the possible terms for the axial and vector parts of the Wigner function resulting from the first order gradient expansion in a general relativistic fluid described with the statistical operator (\ref{eq:StatOper}) and will be classified by their properties. With the exception of including the momentum $k$, the procedure follows the standard method used for gradient expansions in hydrodynamics. Readers familiar with the topic can look through the decompositions (\ref{eq:DecompDezeta})-(\ref{eq:DecompShear}) and proceed to the results in section~\ref{sec:Results}.

%***********************************************************************************************************
\subsection{Scalar}
Consider a scalar hydrodynamic field $X(x)$, whose linear response is obtained by the following operators, respectively from the local equilibrium and the dissipative part of the statistical operator:
\begin{equation}
\h{B}_X = b_X \, X(x)\, \h{\mathcal{B}}_X,\quad
\h{C}_X = c_X \, \int_\Omega \di\Omega\,X\, \h{\mathcal{C}}_X,
\end{equation}
where $b_X$ and $c_X$ are numbers. From the linear response theory we can factorize $X$ and define a function $G^{\mu}_{X}$ given by a thermal correlator:
\begin{subequations}\begin{align}
\Delta_{X,\,{\rm LTE}} \WFA^\mu =& X(x)\, b_X\, \left( \hWFA^\mu,\, \h{\mathcal{B}}_X\right)_{\rm LTE}
    \equiv X(x)\, G^{\mu}_{X,\,{\rm LTE}},\\
\Delta_{X,\,{\rm D}} \WFA^\mu =& X(x)\, c_X\, \left( \hWFA^\mu,\, \h{\mathcal{C}}_X\right)_{\rm D}
    \equiv X(x)\, G^{\mu}_{X,\,{\rm D}}.
\end{align}\end{subequations}
At first order in $X$, all the possible terms giving a vector are the following:
\begin{equation}
\begin{split}
\Delta_X \WFA^\mu = & \left[ a_{X u} u^\mu + a_{X k} \frac{k_\perp^\mu}{(k\cdot u)} \right] X
    \equiv G_X^\mu\,X = \left( G^{\mu}_{X,\,{\rm LTE}} + G^{\mu}_{X,\,{\rm D}} \right) \, X,
\end{split}
\end{equation}
and the coefficients are obtained with the following formulas
\begin{align}
a_{X u} = & u_\mu G^{\mu}_X, &
a_{X k} = & \frac{ (k\cdot u) k^\perp_\mu}{k_\perp^2} G^{\mu}_X,
\end{align}
where the definition of a perpendicular vector is given in eq.~(\ref{eq:ProjDef}).
When $\Delta_X \WFA\cdot k = 0$, we must have
\begin{equation}
a_{X k}  = - a_{X u} \frac{(k\cdot u)^2}{k_\perp^2}.
\end{equation}
%

%***********************************************************************************************************
\subsection{Vector}
The vector component of the irreducible representation under rotation of $u$ is a vector hydrodynamic field $V^\rho(x)$ that is orthogonal to $u$, i.e., $V\cdot u=0$.
The linear response to $V$ is obtained by the following operators, respectively from the local equilibrium and the dissipative part of the density operator:
\begin{equation}
\h{B}_V = b_V \, V_{\rho}(x)\, \h{\mathcal{B}}_V^{\rho},\quad
\h{C}_V = c_V \, \int_\Omega \di\Omega\,V_{\rho}\, \h{\mathcal{C}}_V^{\rho}\, .
\end{equation}
The linear response theory gives:
\begin{subequations}\begin{align}
\Delta_{V,\,{\rm LTE}} \WFA^\mu =& V_{\rho}(x)\, b_V\, \left( \hWFA^\mu,\, \h{\mathcal{B}}_V^{\rho}\right)_{\rm LTE}
    \equiv V_{\rho}(x)\, G^{\mu\rho}_{V,\,{\rm LTE}},\\
\Delta_{V,\,{\rm D}} \WFA^\mu =& V_{\rho}(x)\, c_V\, \left( \hWFA^\mu,\, \h{\mathcal{C}}_V^{\rho}\right)_{\rm D}
    \equiv V_{\rho}(x)\, G^{\mu\rho}_{V,\,{\rm D}}.
\end{align}\end{subequations}
At first order in $V$, all possible terms resulting in a vector are
\begin{equation}
\begin{split}
\Delta_V \WFA^\mu = & \left[ a_{V u} \frac{k_\perp^\rho u^\mu}{(k\cdot u)}
    + a_{V \Delta} \Delta^{\mu\rho}
    + a_{V k} Q^{\mu\rho}
    + a_{V \epsilon}\, \epsilon^{\mu\nu\rho\sigma} \frac{k^\perp_\nu u_\sigma}{(k\cdot u)}
    \right] V_{\rho}  \equiv G_V^{\mu\rho}\, V_\rho,
\end{split}
\end{equation}
where $\Delta$ is defined in eq.~(\ref{eq:ProjDef}) and $Q$ in eq.~(\ref{eq:QDef}).
It is straightforward to show that the coefficients are obtained with the following formulas
\begin{subequations}\begin{align}
a_{V u} = & \frac{(k\cdot u)}{k_\perp^2} u_\mu k^\perp_\rho G^{\mu\rho}_V, &
a_{V \Delta} = & \frac{1}{3} \Delta_{\mu\rho} G^{\mu\rho}_V,\\
a_{V k} = & \frac{3}{2} Q_{\mu\rho} G^{\mu\rho}_V, &
a_{V \epsilon} = & \frac{(k\cdot u)}{2 k_\perp^2} u^\lambda\epsilon_{\lambda\mu\tau\rho}k_\perp^\tau G^{\mu\rho}_V.
\end{align}\end{subequations}
When $\Delta_V \WFA\cdot k = 0$, we must have
\begin{equation}
a_{V u}  = \frac{2}{3} a_{V k} - a_{V \Delta}\, .
\end{equation}
%

%***********************************************************************************************************
\subsection{Symmetric tensor}
Consider a symmetric tensor hydrodynamic field $\Sigma^{\rho\sigma}(x)$ that is orthogonal to $u$ and traceless, such as those resulting from the irreducible components decomposition.
Its linear response is obtained by the following operators, respectively from the local equilibrium and the dissipative part of the density operator:
\begin{equation}
\h{B}_\Sigma = b_\Sigma \, \Sigma_{\rho\sigma}(x)\, \h{\mathcal{B}}_\Sigma^{\rho\sigma},\quad
\h{C}_\Sigma = c_\Sigma \, \int_\Omega \di\Omega\,\Sigma_{\rho\sigma}\, \h{\mathcal{C}}_\Sigma^{\rho\sigma}.
\end{equation}
We write the linear response to $\Sigma$ to a vector $\WFA^\mu$ as
\begin{subequations}\begin{align}
\Delta_{\Sigma,\,{\rm LTE}} \WFA^\mu =& \Sigma_{\rho\sigma}(x)\, b_\Sigma\,
    \left( \hWFA^\mu,\, \h{\mathcal{B}}_\Sigma^{\rho\sigma}\right)_{\rm LTE}
    \equiv \Sigma_{\rho\sigma}(x)\, G^{\mu\rho\sigma}_{\Sigma,\,{\rm LTE}},\\
\Delta_{\Sigma,\,{\rm D}} \WFA^\mu =& \Sigma_{\rho\sigma}(x)\, c_\Sigma\,
    \left( \hWFA^\mu,\, \h{\mathcal{C}}_\Sigma^{\rho\sigma}\right)_{\rm D}
    \equiv \Sigma_{\rho\sigma}(x)\, G^{\mu\rho\sigma}_{\Sigma,\,{\rm D}},
\end{align}\end{subequations}
The general form of a vector linear in $\Sigma$ is the following:
\begin{equation}
\begin{split}
\Delta_\Sigma \WFA^\mu = & \left[ a_{\Sigma u} \frac{k_\perp^\rho k_\perp^\sigma}{k_\perp^2} u^\mu
    + a_{\Sigma \Delta} \frac{\Delta^{\mu\rho} k_\perp^{\sigma}}{(k\cdot u)}
    + a_{\Sigma k} \frac{Q^{\mu\rho} k_\perp^{\sigma}}{(k\cdot u)}
    + a_{\Sigma \epsilon}\, \epsilon^{\mu\nu\alpha\rho} k_\perp^{\sigma} \frac{u_\nu k^\perp_\alpha}{(k\cdot u)^2}
    \right] \Sigma_{\rho\sigma}  \equiv G_\Sigma^{\mu\rho\sigma}\, \Sigma_{\rho\sigma},
\end{split}
\end{equation}
and the coefficients are obtained with the following formulas
\begin{subequations}\begin{align}
a_{\Sigma u} = & \frac{u_\mu k^\perp_\rho k^\perp_\sigma}{k_\perp^2} G^{\mu\rho\sigma}_\Sigma, &
a_{\Sigma \Delta} = & \frac{(k\cdot u)}{3 k_\perp^2} \Delta_{\mu\rho} k^\perp_{\sigma} G^{\mu\rho\sigma}_\Sigma,\\
a_{\Sigma k} = & \frac{3(k\cdot u)}{2 k_\perp^2} Q_{\mu\rho} k^\perp_{\sigma}G^{\mu\rho\sigma}_\Sigma, &
a_{\Sigma \epsilon} = & \frac{(k\cdot u)^2}{2 (k_\perp^2)^2} u^\lambda\epsilon_{\lambda\mu\tau\rho} k^\perp_{\sigma}k_\perp^\tau G^{\mu\rho\sigma}_\Sigma.
\end{align}\end{subequations}
When $\Delta_\Sigma \WFA\cdot k = 0$, we must have
\begin{equation}
a_{\Sigma u}  = \left( \frac{2}{3} a_{\Sigma k} - a_{\Sigma \Delta} \right) \frac{k^2_\perp}{(k\cdot u)^2}.
\end{equation}
%

%***********************************************************************************************************
\subsection{Antisymmetric tensor}
Now, consider an antisymmetric tensor hydrodynamic field $\Gamma^{\rho\sigma}(x)$ orthogonal to $u$.
In this case, it is more convenient to decompose it using two vectors orthogonal to $u$:
\begin{equation}
\Gamma^{\rho\sigma} = \alpha^\rho u^\sigma - \alpha^\sigma u^\rho + \phi^{\rho\sigma},
\end{equation}
where
\begin{equation}
\alpha^\mu = u_\rho \Gamma^{\mu\rho},\quad
\phi^{\rho\sigma} = \Delta^\rho_{\,\mu} \Delta^\sigma_{\,\nu} \Gamma^{\mu\nu}
\end{equation}
and $\phi$ can only be written as
\begin{equation}
\phi^{\rho\sigma} = \epsilon^{\rho\sigma\mu\nu}\, w_\mu\, u_\nu .
\end{equation}
Therefore, the linear response of an antisymmetric field can be reduced to the linear response of vectors. Indeed, with the above decomposition, we can rewrite the operators in the statistical operator as
\begin{subequations}\begin{align}
\h{B}_\Gamma =& b_\Gamma \, \Gamma_{\rho\sigma}(x)\, \h{\mathcal{B}}_\Gamma^{\rho\sigma}
    = b_\Gamma \left(\alpha_\rho u_\sigma - \alpha_\sigma u_\rho + \epsilon_{\rho\sigma\mu\nu}w^\mu u^\nu \right)  \h{\mathcal{B}}_\Gamma^{\rho\sigma}
    = - 2 b_\Gamma\, \alpha_\rho \h{\mathcal{B}}_\alpha^{\rho} - 2 b_\Gamma\, w_\rho \h{\mathcal{B}}_w^{\rho},\\
\h{C}_\Gamma =& c_\Gamma \, \int_\Omega \di\Omega\,\Gamma_{\rho\sigma}\, \h{\mathcal{C}}_\Gamma^{\rho\sigma}
    = -2 c_\Gamma\, \int_\Omega \di\Omega\, \alpha_\rho \h{\mathcal{C}}_\alpha^{\rho}
    - 2 c_\Gamma\, \int_\Omega \di\Omega\, w_\rho \h{\mathcal{C}}_w^{\rho},
\end{align}\end{subequations}
where
\begin{subequations}\begin{align}
\h{\mathcal{B}}_\alpha^{\rho} = & u_\lambda \h{\mathcal{B}}_\Gamma^{\lambda\rho},\quad
    \h{\mathcal{B}}_w^{\rho} = \frac{1}{2}\epsilon^{\alpha\beta\gamma\rho}u_\alpha\h{\mathcal{B}}_{\Gamma\,\beta\gamma},\\
\h{\mathcal{C}}_\alpha^{\rho} = & u_\lambda \h{\mathcal{C}}_\Gamma^{\lambda\rho},\quad
    \h{\mathcal{C}}_w^{\rho} = \frac{1}{2}\epsilon^{\alpha\beta\gamma\rho}u_\alpha\h{\mathcal{C}}_{\Gamma\,\beta\gamma}.
\end{align}\end{subequations}
Applying the linear response theory to this field, we obtain
\begin{subequations}\begin{align}
\Delta_{\Gamma,\,{\rm LTE}} \WFA^\mu =& \alpha_{\rho}(x)\, (-2 b_\Gamma)\, \left( \hWFA^\mu,\, \h{\mathcal{B}}_\alpha^{\rho}\right)_{\rm LTE} + w_{\rho}(x)\, (-2 b_\Gamma)\, \left( \hWFA^\mu,\, \h{\mathcal{B}}_w^{\rho}\right)_{\rm LTE}\\
    \equiv& \alpha_{\rho}\, G^{\mu\rho}_{\alpha,\,{\rm LTE}} + w_{\rho}\, G^{\mu\rho}_{w,\,{\rm LTE}},\\
\Delta_{\Gamma,\,{\rm D}} \WFA^\mu =& \alpha_{\rho}(x)\, (- 2 c_\Gamma)\, \left( \hWFA^\mu,\, \h{\mathcal{C}}_\alpha^{\rho}\right)_{\rm D} + w_{\rho}(x)\, (- 2 c_\Gamma)\, \left( \hWFA^\mu,\, \h{\mathcal{C}}_w^{\rho}\right)_{\rm D}\\
    \equiv & \alpha_{\rho}\, G^{\mu\rho}_{\alpha,\,{\rm D}} + w_{\rho}\, G^{\mu\rho}_{w,\,{\rm D}}.
\end{align}\end{subequations}
Using the results obtained above for the vector part, we obtain all the possible terms linear to $\Gamma$:
\begin{equation}
\Delta_\Gamma \WFA^\mu = \Delta_\alpha \WFA^\mu + \Delta_w \WFA^\mu
    = G^{\mu\rho}_\alpha \alpha_\rho + G^{\mu\rho}_w w_\rho,
\end{equation}
and
\begin{subequations}\begin{align}
\Delta_\alpha \WFA^\mu = & \left[ a_{\alpha \Delta} \Delta^{\mu\rho}
    - a_{\alpha u} \frac{k_\perp^\rho u^\mu}{(k\cdot u)}
    + a_{\alpha k} Q^{\mu\rho}
    + a_{\alpha \epsilon}\, \epsilon^{\mu\nu\rho\sigma} \frac{k^\perp_\nu u_\sigma}{(k\cdot u)}
    \right] \alpha_{\rho}  \equiv G_\alpha^{\mu\rho}\, \alpha_\rho, \\
\Delta_w \WFA^\mu = & \left[ a_{w \Delta} \Delta^{\mu\rho}
    - a_{w u} \frac{k_\perp^\rho u^\mu}{(k\cdot u)}
    + a_{w k} Q^{\mu\rho}
    + a_{w \epsilon}\, \epsilon^{\mu\nu\rho\sigma} \frac{k^\perp_\nu u_\sigma}{(k\cdot u)}
    \right] w_{\rho}  \equiv G_w^{\mu\rho}\, w_\rho,
\end{align}\end{subequations}
with the Kubo formulas
\begin{subequations}\begin{align}
a_{\alpha u} = & -\frac{(k\cdot u)}{k_\perp^2} u_\mu k^\perp_\rho G^{\mu\rho}_\alpha, &
a_{\alpha \Delta} = & \frac{1}{3} \Delta_{\mu\rho} G^{\mu\rho}_\alpha,\\
a_{\alpha k} = & \frac{3}{2} Q_{\mu\rho} G^{\mu\rho}_\alpha, &
a_{\alpha \epsilon} = & \frac{(k\cdot u)}{2 k_\perp^2} u^\lambda\epsilon_{\lambda\mu\tau\rho}k_\perp^\tau G^{\mu\rho}_\alpha,
\end{align}\end{subequations}
and
\begin{subequations}\begin{align}
a_{w u} = & -\frac{(k\cdot u)}{k_\perp^2} u_\mu k^\perp_\rho G^{\mu\rho}_w, &
a_{w \Delta} = & \frac{1}{3} \Delta_{\mu\rho} G^{\mu\rho}_w,\\
a_{w k} = & \frac{3}{2} Q_{\mu\rho} G^{\mu\rho}_w, &
a_{w \epsilon} = & \frac{(k\cdot u)}{2 k_\perp^2} u^\lambda\epsilon_{\lambda\mu\tau\rho}k_\perp^\tau G^{\mu\rho}_w.
\end{align}\end{subequations}
When $\Delta_\Gamma \WFA\cdot k = 0$, we must have
\begin{equation}
a_{\alpha k}  = \frac{3}{2} \left( a_{\alpha u} + a_{\alpha \Delta}\right),\quad
a_{w k}  = \frac{3}{2} \left( a_{w u} + a_{w \Delta}\right).
\end{equation}
%

%***********************************************************************************************************
\subsection{Rank-3 spin-like tensor}
\label{sec:rank3spin}
Lastly, we consider a hydrodynamic field $\Phi^{\lambda,\mu\nu}$ with $\Phi^{\lambda,\mu\nu}=-\Phi^{\lambda,\nu\mu}$, which has 24 degrees of freedom. We first need to decompose it in the irreducible representation under the $SO(3)$ group that leaves $u$ invariant.
To this scope, we project the tensor onto the vector $u$ and, taking the antisymmetry of $\mu\nu$ indices into account, we obtain
\begin{equation}
\Phi^{\lambda,\mu\nu} = u^\lambda \Gamma^{\mu\nu}
    + I^{\lambda\mu} u^\nu - I^{\lambda\nu} u^\mu
    + \Phi^{\mean{\lambda},\mean{\mu}\mean{\nu}},
\end{equation}
where
\begin{subequations}\begin{align}
\Gamma^{\mu\nu} =& u_\lambda \Phi^{\lambda\mu\nu},\\
I^{\lambda\mu} = & \Delta^\lambda_{\,\tau} \Delta^\mu_{\,\rho} u_\sigma \Phi^{\tau\rho\sigma},\\
\Phi^{\mean{\lambda},\mean{\mu}\mean{\nu}} = & \Delta^\lambda_{\,\tau} \Delta^\mu_{\,\rho} \Delta^\nu_{\,\sigma} \Phi^{\tau\rho\sigma}.
\end{align}\end{subequations}
The antisymmetric tensor $\Gamma^{\mu\nu}$ can be decomposed into:
\begin{equation}
\Gamma^{\mu\nu} = f^\mu u^\nu - f^\nu u^\mu + \epsilon^{\mu\nu\alpha\beta} u_\alpha \Upsilon_\beta,
\end{equation}
where
\begin{equation}
f^\mu = u_\lambda u_\nu \Phi^{\lambda\mu\nu},\quad
\Upsilon^\mu = \frac{1}{2} u^\tau \epsilon^{\mu\nu\rho\sigma} u_\nu \Phi_{\tau,\rho\sigma}.
\end{equation}
The rank-2 tensor $I^{\lambda\mu}$ orthogonal to $u$, can be decomposed in a scalar part (the trace), an antisymmetric tensor and a symmetric traceless tensor:
\begin{equation}
I^{\lambda\mu} = I \Delta^{\lambda\mu} + I_A^{\lambda\mu} + I_S^{\lambda\mu}
    = I \Delta^{\lambda\mu} + \epsilon^{\lambda\mu\alpha\beta} u_\alpha I_\beta + I_S^{\lambda\mu},
\end{equation}
where
\begin{subequations}\begin{align}
I =& \frac{1}{3} \Delta_{\lambda\mu} \Delta^\lambda_{\,\tau} \Delta^\mu_{\,\rho} u_\sigma \Phi^{\tau\rho\sigma}
    = \frac{1}{3} \Delta_{\tau\rho} u_\sigma \Phi^{\tau\rho\sigma},\\
I_A^{\lambda\mu} =& \frac{1}{2}\left(\Delta^\lambda_{\,\tau}\Delta^\mu_{\,\rho} - \Delta^\mu_{\,\tau}\Delta^\lambda_{\,\rho}\right) u_\sigma \Phi^{\tau\rho\sigma},\quad
I^\mu = -\frac{1}{2} u^\rho \epsilon^{\mu\nu\tau\sigma} u_\nu \Phi_{\tau,\rho\sigma},\\
I_S^{\lambda\mu} =& \frac{1}{2}\left(\Delta^\lambda_{\,\tau}\Delta^\mu_{\,\rho} + \Delta^\mu_{\,\tau}\Delta^\lambda_{\,\rho}-\frac{2}{3}\Delta_{\tau\rho}\Delta^{\lambda\mu}\right) u_\sigma \Phi^{\tau\rho\sigma}.
\end{align}\end{subequations}
Furthermore, $\Phi^{\mean{\lambda},\mean{\mu}\mean{\nu}}$ can be decomposed in a completely anti-symmetric part, a part symmetric in the exchange of first and second indices and a part symmetric in the exchange of first and third indices:
\begin{equation}
\Phi^{\mean{\lambda},\mean{\mu}\mean{\nu}} = \phi\, \epsilon^{\lambda\mu\nu\rho} u_\rho
    + \Phi_{S12}^{\lambda,\mu\nu}+ \Phi_{S13}^{\lambda,\mu\nu}
\end{equation}
obtained with
\begin{subequations}\begin{align}
\phi= & -\frac{u^\rho}{6} \epsilon_{\lambda\mu\nu\rho} \Phi^{\lambda,\mu\nu}, \\
\Phi_{S12}^{\lambda,\mu\nu} = & \frac{1}{3}\left[ \Delta^\lambda_\rho \Delta^\mu_\sigma \Delta^\nu_\alpha
    + \Delta^\mu_\rho \Delta^\lambda_\sigma \Delta^\nu_\alpha
    - \Delta^\nu_\rho \Delta^\mu_\sigma \Delta^\lambda_\alpha
    -\Delta^\mu_\rho \Delta^\nu_\sigma \Delta^\lambda_\alpha \right]  \Phi^{\rho,\sigma\alpha}, \\
\Phi_{S13}^{\lambda,\mu\nu} = & \frac{1}{3}\left[ \Delta^\lambda_\rho \Delta^\mu_\sigma \Delta^\nu_\alpha
    + \Delta^\nu_\rho \Delta^\mu_\sigma \Delta^\lambda_\alpha
    - \Delta^\mu_\rho \Delta^\lambda_\sigma \Delta^\nu_\alpha
    - \Delta^\mu_\rho \Delta^\nu_\sigma \Delta^\lambda_\alpha \right]  \Phi^{\rho,\sigma\alpha}.
\end{align}\end{subequations}
All together, we have
\begin{equation}
\begin{split}
\Phi^{\lambda,\mu\nu} =& u^\lambda \left( f^\mu u^\nu - f^\nu u^\mu \right)
    + u^\lambda \epsilon^{\mu\nu\alpha\beta} u_\alpha \Upsilon_\beta
    + \left( \Delta^{\lambda\mu} u^\nu - \Delta^{\lambda\nu} u^\mu\right) I\\
    &+ \left( \epsilon^{\lambda\mu\alpha\beta} u^\nu - \epsilon^{\lambda\nu\alpha\beta} u^\mu\right)u_\alpha I_\beta
    + \left( I_S^{\lambda\mu} u^\nu - I_S^{\lambda\nu} u^\mu\right)
    + \phi\, \epsilon^{\lambda\mu\nu\rho} u_\rho
    + \Phi_{S12}^{\lambda,\mu\nu} + \Phi_{S13}^{\lambda,\mu\nu}.
\end{split}
\end{equation}
One can check that the count of the degrees of freedom is, from left to right, $24 = 3 + 3 + 1 + 3 + 5 + 1 + 4 + 4$. The Schouten identity allow us to write the factor in front of $\Upsilon_\beta$ as
\begin{equation}
u^\lambda \epsilon^{\mu\nu\alpha\beta} = u^\mu \epsilon^{\lambda\alpha\beta\nu}
    + u^\nu \epsilon^{\mu\alpha\beta\lambda} - u^\alpha \epsilon^{\beta\lambda\mu\nu} 
    +u^\beta \epsilon^{\mu\lambda\nu\alpha},
\end{equation}
then, using $\Upsilon\cdot u=0$, we have
\begin{equation}
u^\lambda \epsilon^{\mu\nu\alpha\beta} u_\alpha \Upsilon_\beta = \epsilon^{\lambda\mu\nu\rho} \Upsilon_\rho
    -\left(u^\nu \epsilon^{\lambda\mu\alpha\beta}-u^\mu\epsilon^{\lambda\nu\alpha\beta}\right) u_\alpha \Upsilon_\beta,
\end{equation}
and we can write
\begin{equation}
\label{eq:rank3decomp}
\begin{split}
\Phi^{\lambda,\mu\nu} =& u^\lambda \left( f^\mu u^\nu - f^\nu u^\mu \right)
    + \epsilon^{\lambda\mu\nu\rho} \Upsilon_\rho
    + \left( \Delta^{\lambda\mu} u^\nu - \Delta^{\lambda\nu} u^\mu\right) I\\
    &+ \left( \epsilon^{\lambda\mu\alpha\beta} u^\nu - \epsilon^{\lambda\nu\alpha\beta} u^\mu\right)u_\alpha
            (I_\beta - \Upsilon_\beta)
    + \left( I_S^{\lambda\mu} u^\nu - I_S^{\lambda\nu} u^\mu\right)\\
    &+ \phi\, \epsilon^{\lambda\mu\nu\rho} u_\rho
    + \Phi_{S12}^{\lambda,\mu\nu} + \Phi_{S13}^{\lambda,\mu\nu}.
\end{split}
\end{equation}
This form is more convenient when $\Phi^{\lambda,\mu\nu}$ is completely antisymmetric, e.g. the canonical spin tensor. In this case $\Phi^{\lambda,\mu\nu}$ has only 4 degree of freedom and has $I^\mu=\Upsilon^\mu$, such that the decomposition simplify in
\begin{equation}
\Phi^{\{\lambda,\mu\nu\}} = \phi\, \epsilon^{\lambda\mu\nu\rho} u_\rho
    + \epsilon^{\lambda\mu\nu\rho} \Upsilon_\rho.
\end{equation}

Using the decomposition (\ref{eq:rank3decomp}), the local thermal equilibrium and dissipative operators in the statistical operator proportional to a hydrodynamic field $\Phi^{\lambda,\mu\nu}$ are also decomposed as
\begin{subequations}\begin{align}
\h{B}_\Phi =& b_\Phi \, \Phi_{\tau,\rho\sigma}(x)\, \h{\mathcal{B}}_\Phi^{\tau,\rho\sigma} \nonumber\\
= & 2 b_\Phi f_\rho \h{\mathcal{B}}_f^\rho - 2 b_\Phi \Upsilon_\rho \h{\mathcal{B}}_\Upsilon^\rho
    + 2 b_\Phi\, I\, \h{\mathcal{B}}_I - 2 b_\Phi (I_\rho -\Upsilon_\rho) \h{\mathcal{B}}_{I-\Upsilon}^\rho
    + 2 b_\Phi I_{S\,\rho\sigma} \h{\mathcal{B}}_{I_S}^{\rho\sigma}\\
  & + b_\Phi\,\phi\, \h{\mathcal{B}}_\phi
    + 3 b_\Phi \Phi^{\rm S12}_{\tau,\rho\sigma}\h{\mathcal{B}}_{\rm S12}^{\tau,\rho\sigma}
    + 3 b_\Phi \Phi^{\rm S13}_{\tau,\rho\sigma}\h{\mathcal{B}}_{\rm S13}^{\tau,\rho\sigma},\nonumber\\
\h{C}_\Phi =& c_\Phi \, \int_\Omega \di\Omega\,\Phi_{\tau,\rho\sigma}\, \h{\mathcal{C}}_\Phi^{\tau,\rho\sigma} \nonumber\\
= &  \int_\Omega \di\Omega\left[2 c_\Phi f_\rho \h{\mathcal{C}}_f^\rho - 2 c_\phi \Upsilon_\rho \h{\mathcal{C}}_\Upsilon^\rho
    + 2 c_\Phi\, I\, \h{\mathcal{C}}_I - 2 c_\Phi (I_\rho -\Upsilon_\rho) \h{\mathcal{C}}_{I-\Upsilon}^\rho
    + 2 c_\Phi I_{S\,\rho\sigma} \h{\mathcal{C}}_{I_S}^{\rho\sigma}\right.\\
  & \left.+ c_\Phi\,\phi\, \h{\mathcal{C}}_\phi
    + 3 c_\Phi \Phi^{\rm S12}_{\tau,\rho\sigma}\h{\mathcal{C}}_{\rm S12}^{\tau,\rho\sigma}
    + 3 c_\Phi \Phi^{\rm S13}_{\tau,\rho\sigma}\h{\mathcal{C}}_{\rm S13}^{\tau,\rho\sigma}\right],\nonumber
\end{align}\end{subequations}
where
\begin{subequations}\begin{align}
\h{\mathcal{B}}_f^\rho = & u^\lambda \Delta^{\mu\rho}u^\nu \h{\mathcal{B}}^\Phi_{\lambda,\mu\nu}, &
\h{\mathcal{B}}_\Upsilon^\rho = & \frac{1}{2} \epsilon^{\rho\lambda\mu\nu} \h{\mathcal{B}}^\Phi_{\lambda,\mu\nu}, &
\h{\mathcal{B}}_I = & \Delta^{\lambda\mu}u^\nu \h{\mathcal{B}}^\Phi_{\lambda,\mu\nu},\\
\h{\mathcal{C}}_f^\rho =& u^\lambda \Delta^{\mu\rho}u^\nu \h{\mathcal{C}}^\Phi_{\lambda,\mu\nu}, &
\h{\mathcal{C}}_\Upsilon^\rho =& \frac{1}{2} \epsilon^{\rho\lambda\mu\nu}\h{\mathcal{C}}^\Phi_{\lambda,\mu\nu},&
\h{\mathcal{C}}_I =& \Delta^{\lambda\mu}u^\nu \h{\mathcal{C}}^\Phi_{\lambda,\mu\nu},\\
\h{\mathcal{B}}_{I-\Upsilon}^\rho = & \epsilon^{\rho\lambda\mu\alpha}u_\alpha u^\nu\h{\mathcal{B}}^\Phi_{\lambda,\mu\nu}, &
\h{\mathcal{B}}_{I_S}^{\rho\sigma} = & \Delta^{\lambda\rho}\Delta^{\sigma\mu}u^\nu \h{\mathcal{B}}^\Phi_{\lambda,\mu\nu},&
\h{\mathcal{B}}_\phi = & \epsilon^{\lambda\mu\nu\rho} u_\rho \h{\mathcal{B}}^\Phi_{\lambda,\mu\nu},\\
\h{\mathcal{C}}_{I-\Upsilon}^\rho =& \epsilon^{\rho\lambda\mu\alpha}u_\alpha u^\nu\h{\mathcal{C}}^\Phi_{\lambda,\mu\nu},&
\h{\mathcal{C}}_{I_S}^{\rho\sigma} =& \Delta^{\lambda\rho}\Delta^{\sigma\mu}u^\nu \h{\mathcal{C}}^\Phi_{\lambda,\mu\nu}, &
\h{\mathcal{C}}_\phi =& \epsilon^{\lambda\mu\nu\rho} u_\rho \h{\mathcal{C}}^\Phi_{\lambda,\mu\nu},
\end{align}
and
\begin{align}
\h{\mathcal{B}}_{\rm S12}^{\lambda,\mu\nu} = & \frac{1}{3}\left[ \Delta^\lambda_\rho \Delta^\mu_\sigma \Delta^\nu_\alpha
    + \Delta^\mu_\rho \Delta^\lambda_\sigma \Delta^\nu_\alpha
    - \Delta^\nu_\rho \Delta^\mu_\sigma \Delta^\lambda_\alpha
    -\Delta^\mu_\rho \Delta^\nu_\sigma \Delta^\lambda_\alpha \right]  \h{\mathcal{B}}_\Phi^{\rho,\sigma\alpha}, \\
\h{\mathcal{C}}_{\rm S12}^{\lambda,\mu\nu} = & \frac{1}{3}\left[ \Delta^\lambda_\rho \Delta^\mu_\sigma \Delta^\nu_\alpha
    + \Delta^\mu_\rho \Delta^\lambda_\sigma \Delta^\nu_\alpha
    - \Delta^\nu_\rho \Delta^\mu_\sigma \Delta^\lambda_\alpha
    -\Delta^\mu_\rho \Delta^\nu_\sigma \Delta^\lambda_\alpha \right]  \h{\mathcal{C}}_\Phi^{\rho,\sigma\alpha},\\
\h{\mathcal{B}}_{\rm S13}^{\lambda,\mu\nu} = & \frac{1}{3}\left[ \Delta^\lambda_\rho \Delta^\mu_\sigma \Delta^\nu_\alpha
    + \Delta^\nu_\rho \Delta^\mu_\sigma \Delta^\lambda_\alpha
    - \Delta^\mu_\rho \Delta^\lambda_\sigma \Delta^\nu_\alpha
    - \Delta^\mu_\rho \Delta^\nu_\sigma \Delta^\lambda_\alpha \right]  \h{\mathcal{B}}_\Phi^{\rho,\sigma\alpha},\\
\h{\mathcal{C}}_{\rm S13}^{\lambda,\mu\nu} = & \frac{1}{3}\left[ \Delta^\lambda_\rho \Delta^\mu_\sigma \Delta^\nu_\alpha
    + \Delta^\nu_\rho \Delta^\mu_\sigma \Delta^\lambda_\alpha
    - \Delta^\mu_\rho \Delta^\lambda_\sigma \Delta^\nu_\alpha
    - \Delta^\mu_\rho \Delta^\nu_\sigma \Delta^\lambda_\alpha \right]  \h{\mathcal{C}}_\Phi^{\rho,\sigma\alpha}.
\end{align}\end{subequations}

The linear response theory of these operator result in
\begin{multline}
\Delta_{\Phi,\,{\rm LTE}} \WFA^\mu =
    f_{\rho}\, (2 b_\Phi)\, \corrALTE{\h{\mathcal{B}}_f^{\rho}}
    + \Upsilon_\rho (- 2 b_\Phi) \corrALTE{\h{\mathcal{B}}_\Upsilon^\rho}
    + I\, (2 b_\Phi)\, \corrALTE{\h{\mathcal{B}}_I} \\
   + (I_\rho -\Upsilon_\rho) (- 2 b_\Phi) \corrALTE{\h{\mathcal{B}}_{I-\Upsilon}^\rho}
    + I_{S\,\rho\sigma} (2 b_\Phi) \corrALTE{\h{\mathcal{B}}_{I_S}^{\rho\sigma}}\\
   + \phi\, (b_\Phi)\,\corrALTE{\h{\mathcal{B}}_\phi}
    + \Phi^{\rm S12}_{\tau,\rho\sigma}(3 b_\Phi) \corrALTE{\h{\mathcal{B}}_{\rm S12}^{\tau,\rho\sigma}}
    + \Phi^{\rm S13}_{\tau,\rho\sigma}(3 b_\Phi) \corrALTE{\h{\mathcal{B}}_{\rm S13}^{\tau,\rho\sigma}}\\
\equiv f_{\rho}\, G^{\mu\rho}_{f,\,{\rm LTE}} +
    + \Upsilon_\rho \, G^{\mu\rho}_{\Upsilon,\,{\rm LTE}}
    + I\, G^{\mu}_{I,\,{\rm LTE}}
    + (I_\rho -\Upsilon_\rho) \, G^{\mu\rho}_{I-\Upsilon,\,{\rm LTE}}
    + I_{S\,\rho\sigma}\, G^{\mu\rho\sigma}_{I_S,\,{\rm LTE}}\\
   + \phi\, G^{\mu}_{\phi,\,{\rm LTE}}
    + \Phi^{\rm S12}_{\tau,\rho\sigma}\, G^{\mu\tau\rho\sigma}_{\rm S12,\, LTE}
    + \Phi^{\rm S13}_{\tau,\rho\sigma}\, G^{\mu\tau\rho\sigma}_{\rm S13,\, LTE},
\end{multline}
for the local equilibrium part and in
\begin{equation}
\begin{split}    
\Delta_{\Phi,\,{\rm D}} \WFA^\mu =&
    f_{\rho}\, (2 c_\Phi)\, \corrAD{\h{\mathcal{C}}_f^{\rho}}
    + \Upsilon_\rho (- 2 c_\Phi) \corrAD{\h{\mathcal{C}}_\Upsilon^\rho}
    + I\, (2 c_\Phi)\, \corrAD{\h{\mathcal{C}}_I} \\
  & + (I_\rho -\Upsilon_\rho) (- 2 c_\Phi) \corrAD{\h{\mathcal{C}}_{I-\Upsilon}^\rho}
    + I_{S\,\rho\sigma} (2 c_\Phi) \corrAD{\h{\mathcal{C}}_{I_S}^{\rho\sigma}}\\
  & + \phi\, (c_\Phi)\,\corrAD{\h{\mathcal{C}}_\phi}
    + \Phi^{\rm S12}_{\tau,\rho\sigma}(3 c_\Phi) \corrAD{\h{\mathcal{C}}_{\rm S12}^{\tau,\rho\sigma}}
    + \Phi^{\rm S13}_{\tau,\rho\sigma}(3 c_\Phi) \corrAD{\h{\mathcal{C}}_{\rm S13}^{\tau,\rho\sigma}}\\
\equiv& f_{\rho}\, G^{\mu\rho}_{f,\,{\rm D}} +
    + \Upsilon_\rho \, G^{\mu\rho}_{\Upsilon,\,{\rm D}}
    + I\, G^{\mu}_{I,\,{\rm D}}
    + (I_\rho -\Upsilon_\rho) \, G^{\mu\rho}_{I-\Upsilon,\,{\rm D}}
    + I_{S\,\rho\sigma}\, G^{\mu\rho\sigma}_{I_S,\,{\rm D}}\\
  & + \phi\, G^{\mu}_{\phi,\,{\rm D}}
    + \Phi^{\rm S12}_{\tau,\rho\sigma}\, G^{\mu\tau\rho\sigma}_{\rm S12,\, D}
    + \Phi^{\rm S13}_{\tau,\rho\sigma}\, G^{\mu\tau\rho\sigma}_{\rm S13,\, D}
\end{split}
\end{equation}
for the dissipative part. We also decompose the contributions of the different components to the linear response of $\WFA^\mu$ as
\begin{equation}
\begin{split}
\Delta_\Phi \WFA^\mu =& \Delta_{f} \WFA^\mu + \Delta_{\Upsilon} \WFA^\mu
     + \Delta_{I} \WFA^\mu  + \Delta_{I-\Upsilon} \WFA^\mu
      + \Delta_{I_S} \WFA^\mu + \Delta_{\phi} \WFA^\mu
       + \Delta_{\rm S12} \WFA^\mu + \Delta_{\rm S13} \WFA^\mu.
\end{split}
\end{equation}
Using the results from the previous sections, all possible vectors that can be built proportional to these components of $\Phi$ are immediately obtained:
\begin{subequations}\begin{align}
\Delta_{f} \WFA^\mu = & \left[ a_{f u} \frac{k_\perp^\rho u^\mu}{(k\cdot u)}
    + a_{f \Delta} \Delta^{\mu\rho}
    + a_{f k} Q^{\mu\rho}
    + a_{f \epsilon}\, \epsilon^{\mu\nu\rho\sigma} \frac{k^\perp_\nu u_\sigma}{(k\cdot u)}
    \right] f_{\rho}  \equiv G_f^{\mu\rho}\, f_\rho,\\
\Delta_{\Upsilon} \WFA^\mu = & \left[ a_{\Upsilon u} \frac{k_\perp^\rho u^\mu}{(k\cdot u)}
    + a_{\Upsilon \Delta} \Delta^{\mu\rho}
    + a_{\Upsilon k} Q^{\mu\rho}
    + a_{\Upsilon \epsilon}\, \epsilon^{\mu\nu\rho\sigma} \frac{k^\perp_\nu u_\sigma}{(k\cdot u)}
    \right] \Upsilon_{\rho}  \equiv G_\Upsilon^{\mu\rho}\, \Upsilon_\rho,\\
\Delta_{I} \WFA^\mu = & \left[ a_{I u} u^\mu + a_{I k} \frac{k_\perp^\mu}{(k\cdot u)} \right] I
    \equiv G_I^\mu\,I,\\
\Delta_{I-\Upsilon} \WFA^\mu = & \left[ a_{I-\Upsilon u} \frac{k_\perp^\rho u^\mu}{(k\cdot u)}
    + a_{I-\Upsilon \Delta} \Delta^{\mu\rho}
    + a_{I-\Upsilon k} Q^{\mu\rho}
    + a_{I-\Upsilon \epsilon}\, \epsilon^{\mu\nu\rho\sigma} \frac{k^\perp_\nu u_\sigma}{(k\cdot u)}
    \right] (I_\rho-\Upsilon_{\rho}) \\
        \equiv& G_{I-\Upsilon}^{\mu\rho}\, (I_\rho-\Upsilon_\rho),\nonumber\\
\Delta_{I_S} \WFA^\mu = & \left[ a_{I_S u} \frac{k_\perp^\rho k_\perp^\sigma}{k_\perp^2} u^\mu
    + a_{I_S \Delta} \frac{\Delta^{\mu\rho} k_\perp^{\sigma}}{(k\cdot u)}
    + a_{I_S k} \frac{Q^{\mu\rho} k_\perp^{\sigma}}{(k\cdot u)}
    + a_{I_S \epsilon}\, \epsilon^{\mu\nu\alpha\rho} k_\perp^{\sigma} \frac{u_\nu k^\perp_\alpha}{(k\cdot u)^2} \right] I_{S\,\rho\sigma}\\
        \equiv & G_{I_S}^{\mu\rho\sigma}\, I_{S\,\rho\sigma},\nonumber\\
\Delta_\phi \WFA^\mu = & \left[ a_{\phi u} u^\mu + a_{\phi k} \frac{k_\perp^\mu}{(k\cdot u)} \right] \phi
    \equiv G_\phi^\mu\,\phi,
\end{align}\end{subequations}
where
\begin{subequations}\begin{align}
a_{f u} = & \frac{(k\cdot u)}{k_\perp^2} u_\mu k^\perp_\rho G^{\mu\rho}_f, &
    a_{f \Delta} =& \frac{1}{3} \Delta_{\mu\rho} G^{\mu\rho}_f, &
    a_{f k} = & \frac{3}{2} Q_{\mu\rho} G^{\mu\rho}_f,\\
a_{f \epsilon} =& \frac{(k\cdot u)}{2 k_\perp^2} u^\lambda\epsilon_{\lambda\mu\tau\rho}k_\perp^\tau G^{\mu\rho}_f,&
    a_{\Upsilon u} = & \frac{(k\cdot u)}{k_\perp^2} u_\mu k^\perp_\rho G^{\mu\rho}_\Upsilon, & 
    a_{\Upsilon \Delta} =& \frac{1}{3} \Delta_{\mu\rho} G^{\mu\rho}_\Upsilon, \\
a_{\Upsilon k} = & \frac{3}{2} Q_{\mu\rho} G^{\mu\rho}_\Upsilon, &
    a_{\Upsilon \epsilon} =& \frac{(k\cdot u)}{2 k_\perp^2} u^\lambda\epsilon_{\lambda\mu\tau\rho}k_\perp^\tau G^{\mu\rho}_\Upsilon, &
    a_{I u} = & u_\mu G^{\mu}_I, \\
a_{I k} =& \frac{ (k\cdot u) k^\perp_\mu}{k_\perp^2} G^{\mu}_I, &
    a_{I-\Upsilon u} = & \frac{(k\cdot u)}{k_\perp^2} u_\mu k^\perp_\rho G^{\mu\rho}_{I-\Upsilon}, &
    a_{I-\Upsilon \Delta} =& \frac{1}{3} \Delta_{\mu\rho} G^{\mu\rho}_{I-\Upsilon},\\
a_{I-\Upsilon k} = & \frac{3}{2} Q_{\mu\rho} G^{\mu\rho}_{I-\Upsilon}, &
    a_{I-\Upsilon \epsilon} = &\frac{(k\cdot u)}{2 k_\perp^2} u^\lambda\epsilon_{\lambda\mu\tau\rho}k_\perp^\tau G^{\mu\rho}_{I-\Upsilon}, &
    a_{I_S u} = & \frac{u_\mu k^\perp_\rho k^\perp_\sigma}{k_\perp^2} G^{\mu\rho\sigma}_{I_S},
\end{align}
and
\begin{align}
a_{I_S \Delta} =& \frac{(k\cdot u)}{3 k_\perp^2} \Delta_{\mu\rho} k^\perp_{\sigma} G^{\mu\rho\sigma}_{I_S}, &
    a_{I_S k} = & \frac{3(k\cdot u)}{2 k_\perp^2} Q_{\mu\rho} k^\perp_{\sigma}G^{\mu\rho\sigma}_{I_S},&
a_{\phi u} = & u_\mu G^{\mu}_\phi, \\
    a_{\phi k} =& \frac{ (k\cdot u) k^\perp_\mu}{k_\perp^2} G^{\mu}_\phi, &
a_{I_S \epsilon} =& \frac{(k\cdot u)^2}{2 (k_\perp^2)^2} u^\lambda\epsilon_{\lambda\mu\tau\rho} k^\perp_{\sigma}k_\perp^\tau G^{\mu\rho\sigma}_{I_S},
\end{align}\end{subequations}
and imposing $\Delta_\Phi \WFA\cdot k = 0$, we must have
\begin{align}
a_{f u}  =& \frac{2}{3} a_{f k} - a_{f \Delta}, &
a_{\Upsilon u}  =& \frac{2}{3} a_{\Upsilon k} - a_{\Upsilon \Delta},\\
a_{I k}  =& - a_{I u} \frac{(k\cdot u)^2}{k_\perp^2}, &
a_{I-\Upsilon u}  =& \frac{2}{3} a_{I-\Upsilon k} - a_{I-\Upsilon \Delta},\\
a_{I_S u}  =& \left( \frac{2}{3} a_{I_S k} - a_{I_S \Delta} \right) \frac{k^2_\perp}{(k\cdot u)^2}, &
a_{\phi k}  =& - a_{\phi u} \frac{(k\cdot u)^2}{k_\perp^2}.
\end{align}
The parts that do not reduce to previous cases are the linear response to $\Phi_{S12}$ and $\Phi_{S13}$.
In these cases the vectors that one can build are the followings:
\begin{subequations}\begin{align}
\Delta_{\rm S12} \WFA^\mu = &
\left[ a_{\rm{S12}\Delta} \Delta^{\tau\rho} \Delta^{\mu\sigma} + a_{\rm{S12}k} Q^{\tau\rho}\Delta^{\mu\sigma}
    + a_{\rm{S12}\Delta\epsilon} \Delta^{\mu\tau}\epsilon^{\lambda\nu\rho\sigma}\frac{u_\lambda k^\perp_\nu}{(k\cdot u)}\right.\\
    &\left.+ a_{\rm{S12}\epsilon} \epsilon^{\mu\nu\rho\sigma}\frac{k_\perp^{\tau}u_\nu}{(k\cdot u)} \right] \Phi^{S12}_{\tau,\rho\sigma}
    \equiv G_{\rm S12}^{\mu\tau\rho\sigma}\, \Phi^{S12}_{\tau\rho\sigma},\\
\Delta_{\rm S13} \WFA^\mu = &
\left[ a_{\rm{S13}\Delta} \Delta^{\tau\sigma} \Delta^{\mu\rho} + a_{\rm{S13}k} Q^{\tau\sigma}\Delta^{\mu\rho}
    + a_{\rm{S13}\Delta\epsilon} \Delta^{\mu\tau}\epsilon^{\lambda\nu\rho\sigma}\frac{u_\lambda k^\perp_\nu}{(k\cdot u)} \right. \\
    &\left.+ a_{\rm{S13}\epsilon} \epsilon^{\mu\nu\rho\sigma}\frac{k_\perp^{\tau}u_\nu}{(k\cdot u)} \right] \Phi^{S13}_{\tau,\rho\sigma}
    \equiv G_{\rm S13}^{\mu\tau\rho\sigma}\, \Phi^{S13}_{\tau\rho\sigma},
\end{align}\end{subequations}
where
\begin{subequations}\begin{align}
a_{\rm{S12}\Delta} = & \frac{1}{9}\Delta_{\mu\sigma}\Delta_{\tau\rho} G^{\mu\tau\rho\sigma}_{\rm S12}, &
a_{\rm{S12}k} = & \frac{9}{4} Q_{\mu\sigma} Q_{\tau\rho} G^{\mu\tau\rho\sigma}_{\rm S12},\\
a_{\rm{S12}\Delta\epsilon} = & \frac{(k\cdot u)}{2 k_\perp^2}\Delta_{\mu\rho}\epsilon_{\tau\sigma\alpha\beta}
    k_\perp^\alpha u^\beta G^{\mu\tau\rho\sigma}_{\rm S12}, &
a_{\rm{S12}\epsilon} = & \frac{(k\cdot u)k^\perp_\tau k_\perp^\alpha u^\beta k^\perp_\rho}{2(k_\perp)^4}
    \epsilon_{\mu\sigma\alpha\beta}G^{\mu\tau\rho\sigma}_{\rm S12},\\
a_{\rm{S13}\Delta} = & \frac{1}{9}\Delta_{\mu\rho}\Delta_{\tau\sigma} G^{\mu\tau\rho\sigma}_{\rm S13}, &
a_{\rm{S13}k} = & \frac{9}{4} Q_{\mu\rho} Q_{\tau\sigma} G^{\mu\tau\rho\sigma}_{\rm S13},\\
a_{\rm{S13}\Delta\epsilon} = & \frac{(k\cdot u)}{2 k_\perp^2}\Delta_{\mu\rho}\epsilon_{\tau\sigma\alpha\beta}
    k_\perp^\alpha u^\beta G^{\mu\tau\rho\sigma}_{\rm S13}, &
a_{\rm{S13}\epsilon} = & \frac{(k\cdot u)k^\perp_\tau k_\perp^\alpha u^\beta k^\perp_\rho}{2(k_\perp)^4}
    \epsilon_{\mu\sigma\alpha\beta}G^{\mu\tau\rho\sigma}_{\rm S13},
\end{align}\end{subequations}
and when $\Delta_\Phi \WFA\cdot k = 0$, we have
\begin{subequations}\begin{align}
a_{\rm{S12}\Delta} = & -\frac{1}{3} a_{\rm{S12}k},\quad a_{\rm{S12}\Delta\epsilon} = 0,\\
a_{\rm{S13}\Delta} = & -\frac{1}{3} a_{\rm{S13}k},\quad a_{\rm{S13}\Delta\epsilon} = 0.
\end{align}\end{subequations}
%

%***********************************************************************************************************
%***********************************************************************************************************
%***********************************************************************************************************
\section{Vector and axial Wigner functions in a dissipative fluid}
\label{sec:KuboA}
This section derives the first-order gradient expansion of the axial and vector parts of the Wigner function resulting from the statistical operator in eq.~(\ref{eq:StatOper}). For the spin polarization, only $W_+$, i.e., the particle part of the Wigner function, is needed; however, because the decomposition has the same form for both the complete and the particle parts, the subscript $+$ has been omitted. From the statistical operator~(\ref{eq:StatOper}), it follows that we have to consider the linear response theory of the following hydrodynamic fields:
\begin{equation*}
\de_\rho\zeta,\, \zeta_{A},\, \de_\rho\zeta_A,\,
\varpi_{\rho\sigma},\, \mf{S}_{\rho\sigma},\, \xi_{\rho\sigma},\,
 \de_\tau\mf{S}_{\rho\sigma}.
\end{equation*}
In order to apply the results obtained in the previous section, these hydrodynamic fields are further decomposed into their components of irreducible representation under rotations that keep $u$ invariant. The gradient of the vector chemical potential is decomposed as
\begin{equation}
\label{eq:DecompDezeta}
\de^\rho\zeta = u^\rho D\zeta + r^\rho,\quad r^\rho=\de^{\ortu\rho}\zeta,
\end{equation}
where $D=u^\mu \de_\mu$, $V^{\ortu\rho} = V^\rho_\perp = \Delta^\rho_\lambda V^\lambda$, and $\Delta^{\mu\nu} = \eta^{\mu\nu} - u^\mu u^\nu$ (see the notation after the Introduction); similarly for the gradient of the axial chemical potential
\begin{equation}
\label{eq:DecompDezetaA}
\de^\rho\zeta_A = u^\rho D\zeta_A + r_A^\rho,\quad r_A^\rho=\de^{\ortu\rho}\zeta_A,
\end{equation}
while $\zeta_A$ is a scalar and does not need to be decomposed further.
The thermal vorticity, being an antisymmetric tensor, can be decomposed with two vectors:
\begin{equation}
\label{eq:DecompVort}
\varpi_{\mu\nu} = -\frac{1}{2}\left( \de_\mu\beta_\nu - \de_\nu\beta_\mu\right)
    = \alpha_\mu u_\nu - \alpha_\nu u_\mu + \epsilon_{\mu\nu\rho\sigma}w^\rho u^\sigma,
\end{equation}
where the thermal acceleration $\alpha=|\beta| a$ and the thermal angular velocity $w=|\beta| \omega$ are obtained by
\begin{equation}
\alpha^\mu = \varpi^{\mu\nu} u_\nu= \frac{1}{2}\left(\beta a^\mu - \de_\perp^\mu\beta \right),\quad
w^\mu = \tilde{\varpi}^{\mu\nu} u_\nu = \frac{1}{2}\epsilon^{\mu\nu\rho\sigma} \varpi_{\rho\sigma} u_\nu,
\end{equation}
with $\beta=|\beta|=\sqrt{\beta^2}$, and $a^\mu=u\cdot\de u^\mu$ is the acceleration field, and the vectors are orthogonal to $u$: $u\cdot \alpha = u \cdot w=0$.
The spin potential is also an antisymmetric tensor, therefore it can be split into two three-vectors:
\begin{equation}
\label{eq:DecompSP}
\mf{S}_{\mu\nu} = \mf{a}_\mu u_\nu - \mf{a}_\nu u_\mu + \epsilon_{\mu\nu\rho\sigma}\mf{w}^\rho u^\sigma,
\end{equation}
where the spin acceleration $\mf{a}$ and spin angular velocity $\mf{w}$ are defined by
\begin{equation}
\mf{a}^\mu = \mf{S}^{\mu\nu} u_\nu,\quad \mf{w}^\mu = \tilde{\mf{S}}^{\mu\nu} u_\nu = \frac{1}{2}\epsilon^{\mu\nu\rho\sigma} \mf{S}_{\rho\sigma} u_\nu .
\end{equation}
The gradient of the spin potential $\de^\lambda\mf{S}^{\mu\nu}$ is a rank-3 spin like tensor. The decomposition of this type of tensor was derived in section~\ref{sec:rank3spin} and results in:
\begin{equation}
\label{eq:DecompdeSP}
\begin{split}
\de^\lambda\mf{S}^{\mu\nu} =& u^\lambda \left( f^\mu u^\nu - f^\nu u^\mu \right)
    + \epsilon^{\lambda\mu\nu\rho} \Upsilon_\rho
    + \left( \Delta^{\lambda\mu} u^\nu - \Delta^{\lambda\nu} u^\mu\right) I\\
    &+ \left( \epsilon^{\lambda\mu\alpha\beta} u^\nu - \epsilon^{\lambda\nu\alpha\beta} u^\mu\right)u_\alpha
            (I_\beta - \Upsilon_\beta)
    + \left( I_S^{\lambda\mu} u^\nu - I_S^{\lambda\nu} u^\mu\right)\\
    &+ \phi\, \epsilon^{\lambda\mu\nu\rho} u_\rho
    + \Phi_{S12}^{\lambda,\mu\nu} + \Phi_{S13}^{\lambda,\mu\nu}.
\end{split}
\end{equation}
Finally, the symmetric thermal shear
\begin{equation}
\xi_{\rho\sigma} = \frac{1}{2}\left( \de_\rho\beta_\sigma + \de_\sigma\beta_\rho\right)
\end{equation}
is decomposed as follows
\begin{equation}
\label{eq:DecompShear}
\xi_{\rho\sigma} = u_\rho u_\sigma D\beta + \frac{\Delta_{\rho\sigma}}{3}\beta\theta
    + \frac{1}{2} \left(u_\rho \Delta^\tau_{\,\sigma} + u_\sigma \Delta^\tau_{\,\rho} \right)
            \left(\beta Du_\tau + \de_\tau\beta  \right)
    +\Delta^{\lambda\tau}_{\rho\sigma} \beta \sigma_{\lambda\tau},
\end{equation}
where we remind that $\theta=\de_\mu u^\mu$, and $\sigma$ is the proper shear tensor, which is symmetric, traceless and orthogonal to $u$, that is
\begin{equation}
\sigma_{\rho\sigma} = \Delta_{\rho\sigma\mu\nu} \de^\mu u^\nu
= \left[\frac{1}{2}\left(\Delta_{\mu\rho}\Delta_{\nu\sigma}
	+\Delta_{\mu\sigma}\Delta_{\nu\rho}\right)
	-\frac{1}{3}\Delta_{\mu\nu}\Delta_{\rho\sigma}\right]\de^\mu u^\nu .
\end{equation}
Each of these components is considered a hydrodynamic field, denoted by $\mcU_{(\alpha)}$. The linear response $\Delta_\mcU \WFA^\mu$ and $\Delta_\mcU \WFV^\mu$, corresponding to their local equilibrium and dissipative operators
\begin{equation}
\h{B}_{\mcU} = b_{\mcU}\,\mcU_{(\alpha)}(x)\, \h{\mathcal{B}}_{\mcU}^{(\alpha)},\quad
\h{C}_{\mcU} = c_{\mcU}\int_\Omega \di\Omega\,\mcU_{(\alpha)}(x_2) \h{\mathcal{C}}_{\mcU}^{(\alpha)}(x_2),
\end{equation}
appearing in the statistical operator, are obtained in the following sections.

%***********************************************************************************************************
\subsection{Vector and axial chemical potentials}
The vector chemical potential $\zeta$ is included at the leading order and it is contained in the statistical operator (\ref{eq:MeanLO}) that we use to compute the thermal correlators~(\ref{eq:CorrDCoord}) and (\ref{eq:CorrLTECoord}).
Consider now the axial chemical potential $\zeta_A$ that is a scalar quantity. The corresponding operators, see eqs.~(\ref{eq:BLTEa}) and~(\ref{eq:CDissb}), in the statistical operator are
\begin{equation}
\label{eq:DecZetaA}
\begin{split}
\mcU_{(\alpha)}= X = \zeta_A,\quad
b_{\zeta_A} =& 1,\quad
\h{\mathcal{B}}_{\zeta_A} = \h{Q}_A = \int_\Sigma\di\Sigma_\lambda(y)\,\h{j}_A^\lambda(y),\\
c_{\zeta_A} =& -1,\quad
\h{\mathcal{C}}_{\zeta_A} = \de_\lambda \h{j}^\lambda_A.
\end{split}
\end{equation}
The gradient of the chemical potential is decomposed in eq.~(\ref{eq:DecompDezeta}) and consist in a scalar part and a vector part. The linear response is obtained from eqs.~(\ref{eq:BLTEc}) and~(\ref{eq:CDissa}); the scalar part is
\begin{equation}
\label{eq:DecDeZetaScalar}
\begin{split}
\mcU_{(\alpha)}= X= D\zeta,\quad
b_{D\zeta} =& 1,\quad
\h{\mathcal{B}}_{D\zeta} = \int_\Sigma\di\Sigma_\lambda(y)\,(y-x)^\rho u_\rho(x)\h{j}^\lambda(y),\\
c_{D\zeta} =& -1,\quad
\h{\mathcal{C}}_{D\zeta}(x_2) = u_\lambda(x)\h{j}^\lambda(x_2),
\end{split}
\end{equation}
while the vector is
\begin{equation}
\label{eq:DecDeZetaVector}
\begin{split}
\mcU_{(\alpha)}= V_\rho = r_\rho,\quad
b_{r} =& 1,\quad
\h{\mathcal{B}}_{r}^\rho = \int_\Sigma\di\Sigma_\lambda(y)\,(y-x)^{\ortu\rho} \h{j}^\lambda(y),\\
c_{r} =& -1,\quad
\h{\mathcal{C}}_{r}(x_2)^\rho = \h{j}^{\ortu\rho}(x_2).
\end{split}
\end{equation}
Similarly, for the gradient of the axial chemical potential we have the decomposition~(\ref{eq:DecompDezetaA}), and from eqs.~(\ref{eq:BLTEc}) and~(\ref{eq:CDissb}) we have for the scalar part
\begin{equation}
\label{eq:DecDeZetaAScalar}
\begin{split}
\mcU_{(\alpha)}= X = D\zeta_A,\quad
b_{D\zeta_A} =& 1,\quad
\h{\mathcal{B}}_{D\zeta_A} = \int_\Sigma\di\Sigma_\lambda(y)\,(y-x)^\rho u_\rho(x)\h{j}_A^\lambda(y),\\
c_{D\zeta_A} =& -1,\quad
\h{\mathcal{C}}_{D\zeta_A}(x_2) =  u_\rho(x)\left[\h{j}_A^\rho(x_2)
    + (x_2-x)^\rho \de_\lambda \h{j}_A^\lambda(x_2) \right],
\end{split}
\end{equation}
and for the vector part
\begin{equation}
\label{eq:DecDeZetaAVector}
\begin{split}
\mcU_{(\alpha)}= V_\rho =r_{A\,\rho},\quad
b_{r_A} =& 1,\quad
\h{\mathcal{B}}_{r_A}^\rho = \int_\Sigma\di\Sigma_\lambda(y)\,(y-x)^{\ortu\rho} \h{j}_A^\lambda(y),\\
c_{r_A} =& -1,\quad
\h{\mathcal{C}}_{r_A}(x_2)^\rho = \h{j}_A^{\ortu\rho}(x_2)
    + (x_2-x)^{\ortu\rho} \de_\lambda \h{j}_A^\lambda(x_2).
\end{split}
\end{equation}

Therefore, using the results of previous sections, the linear response theory of the axial and vector parts of the Wigner function can be written as
\begin{equation}
\begin{split}
\Delta_{\zeta_A,\de\zeta,\de\zeta_A} \WFA^\mu = & \left[ a_{\zeta_A u} u^\mu + a_{\zeta_A k} \frac{k_\perp^\mu}{(k\cdot u)} \right] \zeta_A
    +  \left[ \bar{a}_{D\zeta_A u} u^\mu + \bar{a}_{D\zeta_A k} \frac{k_\perp^\mu}{(k\cdot u)} \right] D\zeta_A\\
    &+  \left[ \bar{\mf{a}}^c_{D\zeta u} u^\mu + \bar{\mf{a}}^c_{D\zeta k} \frac{k_\perp^\mu}{(k\cdot u)} \right] D\zeta\\
    &+\left[ \bar{\mf{a}}^c_{r u} \frac{k_\perp^\rho u^\mu}{(k\cdot u)}
    + \bar{\mf{a}}^c_{r \Delta} \Delta^{\mu\rho}
    + \bar{\mf{a}}^c_{r k} Q^{\mu\rho}
    + a^c_{r \epsilon}\, \epsilon^{\mu\nu\rho\sigma} \frac{k^\perp_\nu u_\sigma}{(k\cdot u)}
    \right] \de_{\ortu\rho}\zeta\\
    &+\left[ \bar{a}_{r_A u} \frac{k_\perp^\rho u^\mu}{(k\cdot u)}
    + \bar{a}_{r_A \Delta} \Delta^{\mu\rho}
    + \bar{a}_{r_A k} Q^{\mu\rho}
    + \mf{a}_{r_A \epsilon}\, \epsilon^{\mu\nu\rho\sigma} \frac{k^\perp_\nu u_\sigma}{(k\cdot u)}
    \right] \de_{\ortu\rho}\zeta_A,
\end{split}
\end{equation}
and
\begin{equation}
\begin{split}
\Delta_{\zeta_A,\de\zeta,\de\zeta_A} \WFV^\mu = & \left[ \mf{v}^c_{\zeta_A u} u^\mu + \mf{v}^c_{\zeta_A k} \frac{k_\perp^\mu}{(k\cdot u)} \right] \zeta_A
    +  \left[ \bar{\mf{v}}^c_{D\zeta_A u} u^\mu + \bar{\mf{v}}^c_{D\zeta_A k} \frac{k_\perp^\mu}{(k\cdot u)} \right] D\zeta_A\\
    &+  \left[ \bar{v}_{D\zeta u} u^\mu + \bar{v}_{D\zeta k} \frac{k_\perp^\mu}{(k\cdot u)} \right] D\zeta\\
    & +\left[ \bar{v}_{r u} \frac{k_\perp^\rho u^\mu}{(k\cdot u)}
    + \bar{v}_{r \Delta} \Delta^{\mu\rho}
    + \bar{v}_{r k} Q^{\mu\rho}
    + \mf{v}_{r \epsilon}\, \epsilon^{\mu\nu\rho\sigma} \frac{k^\perp_\nu u_\sigma}{(k\cdot u)}
    \right] \de_{\ortu\rho}\zeta\\
    & +\left[ \bar{\mf{v}}^c_{r_A u} \frac{k_\perp^\rho u^\mu}{(k\cdot u)}
    + \bar{\mf{v}}_{r_A \Delta}^c \Delta^{\mu\rho}
    + \bar{\mf{v}}_{r_A k}^c Q^{\mu\rho}
    + v_{r_A \epsilon}^c\, \epsilon^{\mu\nu\rho\sigma} \frac{k^\perp_\nu u_\sigma}{(k\cdot u)}
    \right] \de_{\ortu\rho}\zeta_A.
\end{split}
\end{equation}
As explained in the notation section, the symbols used for the coefficients reflect their properties under discrete transformations $\group{P}$, $\group{T}$, and $\group{C}$, according to table~(\ref{tablecoeff}), which has been derived from eq.~(\ref{tab:PTCTransf}) and
\begin{equation}
\label{tab:PTCZeta}
\begin{array}{|l|ccc|cc|cc|cc|cc|cc|cc|}
\hline
\, & \, & \, & \, & \, & \, & \, & \, & \, & \, & \, & \, & \, & \, & \, & \, \\[-1em]
 & \zeta_A & D\zeta_A & \de_{\rho}\zeta_A & D\zeta & \de_{\rho}\zeta
	& \h{\mathcal{B}}_{\zeta_A} & \h{\mathcal{C}}_{\zeta_A} & \h{\mathcal{B}}_{D\zeta} & \h{\mathcal{C}}_{D\zeta} & \h{\mathcal{B}}_{D\zeta_A} & \h{\mathcal{C}}_{D\zeta_A} & \h{\mathcal{B}}_{r} & \h{\mathcal{C}}_{r} & \h{\mathcal{B}}_{r_A} & \h{\mathcal{C}}_{r_A} \\
\hline
\group{P} & - & - & (-,+) & + & (+,-) & - & - & + & + & - & - & - & - & + & + \\
\group{T} & + & - & (-,+) & - & (-,+) & + & - & - & + & - & + & + & - & + & - \\
\group{C} & + & + & (+,+) & - & (-,-) & + & + & - & - & + & + & - & - & + & + \\
\hline
\end{array}
\end{equation}
The corresponding Kubo formulas, as derived in previous sections, are given by
\begin{subequations}\begin{align}
a_{\zeta_A u} = & \corrLTE{u_\mu\hWFA^\mu}{\h{Q}_A}, &
   \mf{v}^c_{\zeta_A u} = & \corrLTE{u_\mu\hWFV^\mu}{\h{Q}_A}, \\
a_{\zeta_A k} = & \frac{ (k\cdot u)}{k_\perp^2}\corrLTE{ k^\perp_\mu\hWFA^\mu}{\h{Q}_A}, &
    \mf{v}^c_{\zeta_A k} = & \frac{ (k\cdot u)}{k_\perp^2}\corrLTE{ k^\perp_\mu\hWFV^\mu}{\h{Q}_A},\\
\bar{\mf{a}}^c_{D\zeta u} = & -\corrD{u_\mu\hWFA^\mu}{u_\lambda\h{j}^\lambda},
    & \bar{v}_{D\zeta u}= & -\corrD{u_\mu\hWFV^\mu}{u_\lambda\h{j}^\lambda},\\
\bar{\mf{a}}^c_{D\zeta k} = & -\frac{ (k\cdot u)}{k_\perp^2}\corrD{ k^\perp_\mu\hWFA^\mu}{u_\lambda\h{j}^\lambda},
    & \bar{v}_{D\zeta k}= & -\frac{ (k\cdot u)}{k_\perp^2}\corrD{ k^\perp_\mu\hWFV^\mu}{u_\lambda\h{j}^\lambda},
\end{align}
and
\begin{align}
\bar{a}_{D\zeta_A u} = & -\corrD{u_\mu\hWFA^\mu}{u_\rho(\h{j}_A^\rho+(x_2-x)^\rho\de_\lambda\h{j}_A^\lambda)},\\ \bar{\mf{v}}^c_{D\zeta_A u}= & -\corrD{u_\mu\hWFV^\mu}{u_\rho(\h{j}_A^\rho+(x_2-x)^\rho\de_\lambda\h{j}_A^\lambda)},\\
\bar{a}_{D\zeta_A k} = & -\frac{ (k\cdot u)}{k_\perp^2}\corrD{ k^\perp_\mu\hWFA^\mu}{u_\rho(\h{j}_A^\rho+(x_2-x)^\rho\de_\lambda\h{j}_A^\lambda)},\\
    \bar{\mf{v}}^c_{D\zeta_A k}= & -\frac{ (k\cdot u)}{k_\perp^2}\corrD{ k^\perp_\mu\hWFV^\mu}{u_\rho(\h{j}_A^\rho+(x_2-x)^\rho\de_\lambda\h{j}_A^\lambda)},
\end{align}\end{subequations}
and
\begin{subequations}\begin{align}
\bar{\mf{a}}^c_{r u} = & -\frac{(k\cdot u)}{k_\perp^2}\corrD{u_\mu\hWFA^\mu}{k^\perp_\rho \h{j}^\rho},
   & \bar{v}_{r u} = & -\frac{(k\cdot u)}{k_\perp^2}\corrD{u_\mu\hWFV^\mu}{k^\perp_\rho \h{j}^\rho}, \\
\bar{\mf{a}}^c_{r \Delta} = & -\frac{1}{3} \Delta_{\mu\rho}\corrD{\hWFA^\mu}{\h{j}^\rho},
    & \bar{v}_{r \Delta} = & -\frac{1}{3} \Delta_{\mu\rho}\corrD{\hWFV^\mu}{\h{j}^\rho},\\
\bar{\mf{a}}^c_{r k} = & -\frac{3}{2} Q_{\mu\rho}\corrD{\hWFA^\mu}{\h{j}^\rho},
    & \bar{v}_{r k} = & -\frac{3}{2} Q_{\mu\rho}\corrD{\hWFV^\mu}{\h{j}^\rho},\\
\label{eq:SHEKf}
a_{r \epsilon}^c = & \frac{(k\cdot u)}{2 k_\perp^2} u^\lambda\epsilon_{\lambda\mu\tau\rho}k_\perp^\tau\corrLTE{\hWFA^\mu}{\h{\mathcal{B}}_r^\rho},
    & \mf{v}_{r \epsilon} = & \frac{(k\cdot u)}{2 k_\perp^2} u^\lambda\epsilon_{\lambda\mu\tau\rho}k_\perp^\tau\corrLTE{\hWFV^\mu}{\h{\mathcal{B}}_r^\rho},\\
\label{eq:CSHEKf}
\mf{a}_{r_A \epsilon} =& \frac{(k\cdot u)}{2 k_\perp^2} u^\lambda\epsilon_{\lambda\mu\tau\rho}k_\perp^\tau\corrLTE{\hWFA^\mu}{\h{\mathcal{B}}_{r_A}^\rho},
    & v^c_{r_A \epsilon} = & \frac{(k\cdot u)}{2 k_\perp^2} u^\lambda\epsilon_{\lambda\mu\tau\rho}k_\perp^\tau\corrLTE{\hWFV^\mu}{\h{\mathcal{B}}_{r_A}^\rho},
\end{align}
and
\begin{align}
\bar{a}_{r_A u} = & -\frac{(k\cdot u)}{k_\perp^2}\corrD{u_\mu\hWFA^\mu}{k^\perp_\rho \h{j}_A^\rho+(x_2-x)^\rho\de_\lambda\h{j}_A^\lambda},\\
\bar{\mf{v}}^c_{r_A u} = & -\frac{(k\cdot u)}{k_\perp^2}\corrD{u_\mu\hWFV^\mu}{k^\perp_\rho \h{j}_A^\rho+(x_2-x)^\rho\de_\lambda\h{j}_A^\lambda},\\
\bar{a}_{r_A \Delta} = & -\frac{1}{3} \Delta_{\mu\rho}\corrD{\hWFA^\mu}{\h{j}_A^\rho+(x_2-x)^\rho\de_\lambda\h{j}_A^\lambda},\\
\bar{\mf{v}}_{r_A \Delta}^c = & -\frac{1}{3} \Delta_{\mu\rho}\corrD{\hWFV^\mu}{\h{j}_A^\rho+(x_2-x)^\rho\de_\lambda\h{j}_A^\lambda},\\
\bar{a}_{r_A k} = & -\frac{3}{2} Q_{\mu\rho}\corrD{\hWFA^\mu}{\h{j}_A^\rho+(x_2-x)^\rho\de_\lambda\h{j}_A^\lambda},\\
\bar{\mf{v}}^c_{r_A k} = & -\frac{3}{2} Q_{\mu\rho}\corrD{\hWFV^\mu}{\h{j}_A^\rho+(x_2-x)^\rho\de_\lambda\h{j}_A^\lambda}.
\end{align}\end{subequations}
It is important to stress that the coefficients that are odd under time reversal, that we remind it means that they connect $\WFA^\mu$ or $\WFV^\mu$ with a vector with opposite parity under time reversal \group{T}, can not be obtained from the local thermal equilibrium operators, as it can be explicitly checked from the transformations listed in~(\ref{tab:PTCTransf}) and~(\ref{tab:PTCZeta}). Indeed, the equilibrium statistical operator used to evaluate the thermal correlators~(\ref{eq:CorrDCoord}) and~(\ref{eq:CorrLTECoord}) has time reversal symmetry, from which follows that the coefficients can be non-vanishing only if the correlators is evaluated between quantities with the same parity under time reversal. It is always the case that only the operators from the dissipative part can give rise to odd time reversal coefficients and that is the reason we refer to them as dissipative. We also notice that all the dissipative corrections require a chiral coefficient or are sourced by the axial chemical potential.

The coefficient $a_{r\epsilon}^c$ give the spin-Hall effect (SHE) \cite{Liu:2020dxg} in spin polarization (also called the spin Nerst effect). As noted earlier, despite this coefficient is charged, it can be non-vanishing even at zero chemical potential $\zeta=0$. That is because it gives the particle part of the Wigner function and differs from charged coefficients giving the axial current $j^\mu_A$, which is instead obtained by integrating the particle and anti-particle $\WFA^\mu$ over $k$. Moreover, we can identify the terms proportional to  $\mf{a}_{r_A\epsilon}$, $\mf{v}_{r\epsilon}$ and $v^c_{r_A\epsilon}$ as analogous to the SHE and that we consequently call chiral spin-Hall (Nerst) effect, chiral electrical effect (CEE) and axial Hall effect (AHE), respectively. All of these effects are non-dissipative and half of them are chiral.

%***********************************************************************************************************
\subsection{Thermal vorticity}
We consider now the local equilibrium contributions from the thermal vorticity. The dissipative part will be included together with the spin potential. From eq.~(\ref{eq:BLTEa}), the operators associated with the vectors $w$ and $\alpha$ in the decomposition (\ref{eq:DecompVort}) are:
\begin{equation}
\label{eq:DecThermVort}
\begin{split}
\mcU_{(\alpha)}= \Gamma_{\rho\sigma}=\varpi_{\rho\sigma},\quad
b_{\varpi} =& \frac{1}{2},\quad
\h{\mathcal{B}}_{\varpi}^{\rho\sigma} = \h{J}^{\rho\sigma}_x
,\quad \h{\mathcal{B}}_\alpha^\rho=\h{K}_x^\rho,\quad \h{\mathcal{B}}_w^\rho=\h{J}_x^\rho,%\\
%c_{\varpi} =& 0,\quad
%\h{\mathcal{C}}_{\varpi}(x_2) =  0,
\end{split}
\end{equation}
where
\begin{equation}
\h{K}_x^{\rho} =  u_\lambda \h{J}_x^{\lambda\rho},\quad
    \h{J}_x^{\rho} = \frac{1}{2}\epsilon^{\alpha\beta\gamma\rho}u_\alpha\h{J}_{x\,\beta\gamma}
\end{equation}
are the boost and the angular momentum operators.
To make the notation more compact we define the following axial vectors
\begin{subequations}\begin{align}
A^\mu_{w\Delta} =& w^\mu =\Delta^{\mu\rho} w_\rho, &
A^\mu_{w u} =& \frac{(k\cdot w)}{(k\cdot u)} u^\mu =\frac{k^\rho_\perp u^\mu}{(k\cdot u)} w_\rho ,\\
A^\mu_{\alpha \epsilon} =& \epsilon^{\mu\nu\rho\sigma} \frac{k^\perp_\nu u_\sigma}{(k\cdot u)} \alpha_\rho, &
A^\mu_{w k} =& Q^{\mu\rho} w_\rho,
\end{align}\end{subequations}
and the following acceleration vectors:
\begin{subequations}\begin{align}
\mf{A}^\mu_{\alpha\Delta} =& \alpha^\mu =\Delta^{\mu\rho} \alpha_\rho,&
\mf{A}^\mu_{\alpha u} =& \frac{(k\cdot \alpha)}{(k\cdot u)} u^\mu =\frac{k^\rho_\perp u^\mu}{(k\cdot u)} \alpha_\rho ,\\
\mf{A}^\mu_{w\epsilon} =& \epsilon^{\mu\nu\rho\sigma} \frac{k^\perp_\nu u_\sigma}{(k\cdot u)} w_\rho,&
\mf{A}^\mu_{\alpha k} =& Q^{\mu\rho} \alpha_\rho \, .
\end{align}\end{subequations}
According to the transformation properties:
\begin{equation}
\begin{array}{|c|c|c|cc|}
\hline
\, & \, & \, & \, & \,\\[-1em]
 & w^\mu,\,A_X^\mu & a^\mu,\,\mf{A}_X^\mu & \h{K}^{\ortu\rho} & \h{J}^{\ortu\rho} \\
\hline
\group{P} & (-, +) & (+, -) & - & +\\
\group{T} & (+, -) & (-, +) & + & -\\
\group{C} & (+, +) & (+, +) & + & +\\
\hline
\end{array}
\end{equation}
the linear response to thermal vorticity for the axial and vector part of Wigner function results in
\begin{align}
\Delta_\varpi \WFA^\mu(x,k) = &
    a_{w \Delta} A^\mu_{w\Delta} - a_{w u} A^\mu_{w u}
    + a_{\alpha \epsilon} A^\mu_{\alpha \epsilon} + a_{w k} A^\mu_{w k}
    + \bar{\mf{a}}_{\alpha \Delta} \mf{A}^\mu_{\alpha\Delta}\\
    &- \bar{\mf{a}}_{\alpha u} \mf{A}^\mu_{\alpha u}
    +\bar{\mf{a}}_{w \epsilon} \mf{A}^\mu_{w \epsilon}
    + \bar{\mf{a}}_{\alpha k} \mf{A}^\mu_{\alpha k}, \nonumber\\
\Delta_\varpi \WFV^\mu(x,k) = &
    \mf{v}^c_{w \Delta} A^\mu_{w\Delta} - \mf{v}^c_{w u} A^\mu_{w u}
    + \mf{v}^c_{\alpha \epsilon} A^\mu_{\alpha \epsilon} + \mf{v}^c_{w k} A^\mu_{w k}\\
    &+ \bar{v}^c_{\alpha \Delta} \mf{A}^\mu_{\alpha\Delta}
    - \bar{v}^c_{\alpha u} \mf{A}^\mu_{\alpha u}
    + \bar{v}^c_{w \epsilon} \mf{A}^\mu_{w \epsilon}
    + \bar{v}^c_{\alpha k} \mf{A}^\mu_{\alpha k} . \nonumber
\end{align}
From previous results, their Kubo formulas are:
\begin{subequations}\begin{align}
a_{w u} = & \frac{(k\cdot u)}{k_\perp^2} \corrLTE{u_\mu\hWFA^\mu}{k^\perp_\rho \h{J}_x^\rho}, &
    \bar{\mf{a}}_{\alpha u} = & 0,\\
a_{w \Delta} = & -\frac{1}{3} \Delta_{\mu\rho} \corrLTE{\hWFA^\mu}{\h{J}_x^\rho}, &
    \bar{\mf{a}}_{\alpha \Delta} = & 0,\\
a_{w k} = & -\frac{3}{2} Q_{\mu\rho} \corrLTE{\hWFA^\mu}{\h{J}_x^\rho}, &
    \bar{\mf{a}}_{\alpha k} = & 0,\\
\bar{\mf{a}}_{w \epsilon} = & 0, &
    a_{\alpha \epsilon} = & -\frac{(k\cdot u)}{2 k_\perp^2} \corrLTE{\hWFA^\mu}{u^\lambda\epsilon_{\lambda\mu\tau\rho}k_\perp^\tau\h{K}_x^\rho},\\
%\end{align}
%
%and
%
%\begin{align}
\mf{v}^c_{w u} = & \frac{(k\cdot u)}{k_\perp^2} \corrLTE{u_\mu\hWFV^\mu}{k^\perp_\rho \h{J}_x^\rho}, &
    \bar{v}^c_{\alpha u} = & 0,\\
\mf{v}^c_{w \Delta} = & -\frac{1}{3} \Delta_{\mu\rho} \corrLTE{\hWFV^\mu}{\h{J}_x^\rho}, &
    \bar{v}^c_{\alpha \Delta} = & 0,\\
\mf{v}^c_{w k} = & -\frac{3}{2} Q_{\mu\rho} \corrLTE{\hWFV^\mu}{\h{J}_x^\rho}, &
    \bar{v}^c_{\alpha k} = & 0,\\
\bar{v}^c_{w \epsilon} = & 0, &
    \mf{v}^c_{\alpha \epsilon} = & -\frac{(k\cdot u)}{2 k_\perp^2} \corrLTE{\hWFV^\mu}{u^\lambda\epsilon_{\lambda\mu\tau\rho}k_\perp^\tau\h{K}_x^\rho}.
\end{align}\end{subequations}
The contributions from thermal vorticity are usually written in a different form. Taking advantage of the identity
\begin{equation}
\tilde\varpi^{\mu\nu} k_\nu = \frac{1}{2} \epsilon^{\mu\nu\alpha\beta}\varpi_{\alpha\beta}k_\nu
= -(k\cdot u)\left[ w^\mu - \frac{(k\cdot w)}{(k\cdot u)}u^\mu
    + \epsilon^{\mu\nu\rho\sigma} \frac{k^\perp_\nu u_\sigma}{(k\cdot u)} \alpha_\rho \right],
\end{equation}
we can re-write the Wigner functions as
\begin{align}
\label{eq:DeltaAVort}
\Delta_{\varpi}\WFA^\mu(x,k) =& -a_{\varpi} \frac{2\tilde\varpi^{\mu\nu}k_\nu}{(2\pi)^3}
    - (a_{w u} - a_{w \Delta}) A^\mu_{w u}
    + (a_{\alpha\epsilon} - a_{w \Delta}) A^\mu_{\alpha\epsilon} + a_{w k} A^\mu_{w k},\\
\Delta_{\varpi}\WFV^\mu(x,k) =& -\mf{v}^c_{\varpi} \frac{2\tilde\varpi^{\mu\nu}k_\nu}{(2\pi)^3}
    - (\mf{v}^c_{w u} - \mf{v}^c_{w \Delta}) A^\mu_{w u}
    + (\mf{v}^c_{\alpha\epsilon} - \mf{v}^c_{w \Delta}) A^\mu_{\alpha\epsilon} + \mf{v}^c_{w k} A^\mu_{w k},
\end{align}
where
\begin{subequations}\begin{align}
a_{\varpi} =& \frac{(2\pi)^3}{2(k\cdot u)} a_{w \Delta}
    = -\frac{(2\pi)^3}{6(k\cdot u)} \Delta_{\mu\rho} \corrLTE{\hWFA^\mu}{\h{J}_x^{\rho}}
    =  -\frac{(2\pi)^3}{6(k\cdot u)} \corrLTE{\hWFA^\mu}{\h{J}_{x\,\mu}},\\
\mf{v}^c_\varpi = & -\frac{(2\pi)^3}{6(k\cdot u)} \corrLTE{\hWFV^\mu}{\h{J}_{x\,\mu}} .
\end{align}\end{subequations}
The coefficients $a_\varpi$ and $\mf{v}^c$ are the one giving the axial vortical effect and the chiral vortical effect. The numerical factors defining $a_\varpi$ have been chosen such that for a free Dirac field the coefficient assumes the simple form:
\begin{equation}
a_{\varpi}^{\rm Free} = \delta(k^2-m^2)\theta(k\cdot u) n_F(\beta\cdot k)\left(1-n_F(\beta\cdot k)\right),
\end{equation}
with $n_F$ the Fermi-Dirac distribution function $n_F(z)=(\e^{z}+1)^{-1}$.
For a massive field, the spin polarization is directly related to the axial part of the Wigner function.
In that case, it must satisfy $k\cdot\Delta_\varpi\mathcal{A}(x,k)=0$, so:
\begin{equation}
\Delta_{\varpi}\WFA^\mu(x,k) = -a_{\varpi} \frac{2\tilde\varpi^{\mu\nu}k_\nu}{(2\pi)^3}
    - (a_{w u} - a_{w\Delta}) \left(A^\mu_{w u} + \frac{3}{2} A^\mu_{w k}\right)
    + (a_{\alpha\epsilon} - a_{w\delta}) A^\mu_{\alpha\epsilon} \, .
\end{equation}

As a check it will be shown that the equations above can reproduce the known conductivity of the axial vortical effect. Since the coefficients are Lorentz scalars, without loss of generality they can be evaluated in the local rest frame where $u=(1,\,\bm{0})$. In this frame, the coefficient $a_\varpi$ is given by
\begin{equation}
a_\varpi = \frac{(2\pi)^3}{6(k\cdot u)}
    \sum_i\left(\hWFA^i(0,k),\,\h{J}^i\right)_{{\rm LTE},T},
\end{equation}
where we used the translation invariance of the correlator. By invariance we just need to compute one component:
\begin{equation}
a_\varpi = \frac{(2\pi)^3}{2(k\cdot u)} \left(\hWFA^z(0,k),\,\h{J}^z\right)_{{\rm LTE},T}
    = \frac{(2\pi)^3}{2(k\cdot u)}\int_0^{|\beta(x)|}\frac{\di\tau}{|\beta(x)|}\mean{\h{J}^z_{[\tau/\beta]}\hWFA^z(0,k)}_{T,c}\, ,
\end{equation}
where we used the definition (\ref{eq:CorrLTECoord}). By integrating in $k$ we obtain the axial current, which result in the axial vortical effect, that is an axial current induced by rotation:
\begin{equation}
\Delta_{\varpi}j_A^\mu(x) = \int \di^4 k\,\Delta_{\varpi}\WFA^\mu(x,k)
    = \frac{1}{2} \varpi_{\rho\sigma}(x) \left(\h{J}^{\rho\sigma}(y),\,\h{j}_A^\mu(0)\right)_{\beta(x)} = W_A(x) w^\mu(x)+\dots
\end{equation}
From the equation above, the AVE conductivity $W_A$ is obtained with
\begin{equation}
    W_A = \frac{w_\mu}{w^2}\int \di^4 k\,\Delta_{w}\WFA^\mu(x,k).
\end{equation}
From eq.~(\ref{eq:DeltaAVort}), since $A^\mu_{\alpha\epsilon}$ is the response to acceleration and $w_\mu A^\mu_{wu} = w_\mu A^\mu_{wk} =0$, we are only left with
\begin{equation}
\begin{split}
W_A =&  - \frac{w_\mu\varpi_{\rho\sigma}}{w^2}\epsilon^{\mu\nu\rho\sigma}
    \int \frac{\di^4 k}{(2\pi)^3}\,k_\nu\,a_\varpi(x,k)\\
=& 2 \int \frac{\di^4 k}{(2\pi)^3}\,(k\cdot u)\,a_\varpi(x,k)
= \int \di^4 k \int_0^{\beta}\frac{\di\tau}{|\beta|}\mean{\h{J}^z_{[\tau/\beta]}\hWFA^z(0,k)}_{T,c}\, .
\end{split}
\end{equation}
For the free field, considering first the particle part, we obtain
\begin{equation}
\begin{split}
W^+_A =& 2 \int \frac{\di^4 k}{(2\pi)^3}\,(k\cdot u)\,a^{\rm Free}_\varpi(x,k)\\
 =& \int \frac{\di^4 k}{(2\pi)^3}\,2(k\cdot u)\,\delta(k^2-m^2)\theta(k\cdot u) n_F(\beta\cdot k)\left(1-n_F(\beta\cdot k)\right)\\
=& \int \frac{\di^3 k}{(2\pi)^3}\, n_F(|\beta| E_k)\left(1-n_F(|\beta| E_k)\right)
= -\frac{1}{|\beta|}\int \frac{\di^3 k}{(2\pi)^3}\, \frac{\de}{\de E_k}n_F(|\beta| E_k)\\
= & \frac{1}{2\pi|\beta|}\int\di k\,\frac{2 E_k^2-m^2}{E_k} n_F(|\beta| E_k).
\end{split}
\end{equation}
Then, adding the anti-particle part and including the chemical potential, we obtain the well known result~\cite{Buzzegoli:2017cqy}
\begin{equation}
W_A = \frac{1}{2\pi|\beta|}\int\di k\,\frac{2 E_k^2-m^2}{E_k} 
    \left[ n_F(|\beta| (E_k-\mu)) + n_F(|\beta| (E_k +\mu)) \right].
\end{equation}
%

%***********************************************************************************************************
\subsection{Spin potential}
The local thermal equilibrium and the dissipative contributions of the spin potential are actually proportional to the difference between the spin potential and the thermal vorticity. The operators~(\ref{eq:BLTEb}), (\ref{eq:CDissc}) and (\ref{eq:CDissd}) in terms of the vectors $w$, $\mf{w}$, $\alpha$ and $\mf{a}$ from the decompositions (\ref{eq:DecompVort}) and (\ref{eq:DecompSP}), for $\mcU_{(\alpha)}= \Gamma_{\rho\sigma} =\SP_{\rho\sigma}-\varpi_{\rho\sigma}$ are written as:
\begin{equation}
\label{eq:DecSpinPotential}
\begin{split}
b_{\SP-\varpi} =& \frac{1}{2},\quad
\h{\mathcal{B}}_{\SP-\varpi}^{\rho\sigma} = \h{S}^{\rho\sigma}_x,\quad
\h{\mathcal{B}}_{\mf{a}-\alpha}^\rho=u_\lambda \h{S}^{\lambda\rho},\quad
\h{\mathcal{B}}_{\mf{w}-w}^\rho=\frac{1}{2}\epsilon^{\alpha\beta\gamma\rho}u_\alpha \h{S}_{\beta\gamma},\\
c_{\SP-\varpi} =& 1,\quad
\h{\mathcal{C}}_{\SP-\varpi}(x_2) =  \h{T}_A^{\rho\sigma}(x_2),\quad
\h{\mathcal{C}}_{\mf{a}-\alpha}^\rho=u_\lambda \h{T}_A^{\lambda\rho},\quad
\h{\mathcal{C}}_{\mf{w}-w}^\rho=\frac{1}{2}\epsilon^{\alpha\beta\gamma\rho}u_\alpha \h{T}_{A\,\beta\gamma},
\end{split}
\end{equation}
where $\h{\mathcal{B}}_{\mf{a}-\alpha}$ and $\h{\mathcal{B}}_{\mf{w}-w}$ are the spin part of the total boost $\h{K}$ and angular moment $\h{J}$ operators. As done for thermal vorticity, we define the following axial vectors
\begin{subequations}\begin{align}
A^\mu_{\mf{w}-w\Delta} =& \Delta^{\mu\rho} (\mf{w}-w)_\rho,&
A^\mu_{\mf{w}-w u} =& \frac{k^\rho_\perp u^\mu}{(k\cdot u)} (\mf{w}-w)_\rho ,\\
A^\mu_{\mf{a}-\alpha \epsilon} =& \epsilon^{\mu\nu\rho\sigma} \frac{k^\perp_\nu u_\sigma}{(k\cdot u)} (\mf{a}-\alpha)_\rho,&
A^\mu_{\mf{w}-w k} =& Q^{\mu\rho} (\mf{w}-w)_\rho,
\end{align}\end{subequations}
and the following acceleration vectors:
\begin{subequations}\begin{align}
\mf{A}^\mu_{\mf{a}-\alpha\Delta} =& \Delta^{\mu\rho} (\mf{a}-\alpha)_\rho, &
\mf{A}^\mu_{\mf{a}-\alpha u} =& \frac{k^\rho_\perp u^\mu}{(k\cdot u)} (\mf{a}-\alpha)_\rho ,\\
\mf{A}^\mu_{\mf{w}-w\epsilon} =& \epsilon^{\mu\nu\rho\sigma} \frac{k^\perp_\nu u_\sigma}{(k\cdot u)} (\mf{w}-w)_\rho, &
\mf{A}^\mu_{\mf{a}-\alpha k} =& Q^{\mu\rho} (\mf{a}-\alpha)_\rho .
\end{align}\end{subequations}
According to the transformation properties:\begin{equation}
\begin{array}{|c|c|c|cc|cc|}
\hline
\, & \, & \, & \, & \, & \, & \,\\[-1em]
 & w^\mu,\,A_X^\mu & a^\mu,\,\mf{A}_X^\mu & \h{\mathcal{B}}_{\mf{a}-\alpha}^{\ortu\rho} & \h{\mathcal{B}}_{\mf{w}-w}^{\ortu\rho} & \h{\mathcal{C}}_{\mf{a}-\alpha}^{\ortu\rho} & \h{\mathcal{C}}_{\mf{w}-w}^{\ortu\rho} \\
\hline
\group{P} & (-, +) & (+, -) & - & + & - & +\\
\group{T} & (+, -) & (-, +) & + & - & - & +\\
\group{C} & (+, +) & (+, +) & + & + & + & +\\
\hline
\end{array}
\end{equation}
the linear response to the operators~(\ref{eq:DecSpinPotential}) for the axial and vector parts of the Wigner function is
\begin{align}
\Delta_{\SP-\varpi} \WFA^\mu(x,k) = &
    a_{\mf{w}-w \Delta} A^\mu_{\mf{w}-w\Delta} - a_{\mf{w}-w u} A^\mu_{\mf{w}-w u}
    + a_{\mf{a}-\alpha \epsilon} A^\mu_{\mf{a}-\alpha \epsilon} + a_{\mf{w}-w k} A^\mu_{\mf{w}-w k}\\
&+ \bar{\mf{a}}_{\mf{a}-\alpha \Delta} \mf{A}^\mu_{\mf{a}-\alpha\Delta}
    - \bar{\mf{a}}_{\mf{a}-\alpha u} \mf{A}^\mu_{\mf{a}-\alpha u}
    +\bar{\mf{a}}_{\mf{w}-w \epsilon} \mf{A}^\mu_{\mf{w}-w \epsilon}
    + \bar{\mf{a}}_{\mf{a}-\alpha k} \mf{A}^\mu_{\mf{a}-\alpha k}, \nonumber\\
\Delta_{\SP-\varpi} \WFV^\mu(x,k) = &
    \mf{v}^c_{\mf{w}-w \Delta} A^\mu_{\mf{w}-w\Delta} - \mf{v}^c_{\mf{w}-w u} A^\mu_{\mf{w}-w u}
    + \mf{v}^c_{\mf{a}-\alpha \epsilon} A^\mu_{\mf{a}-\alpha \epsilon} + \mf{v}^c_{\mf{w}-w k} A^\mu_{\mf{w}-w k}\\
&+ \bar{v}^c_{\mf{a}-\alpha \Delta} \mf{A}^\mu_{\mf{a}-\alpha\Delta}
    - \bar{v}^c_{\mf{a}-\alpha u} \mf{A}^\mu_{\mf{a}-\alpha u}
    + \bar{v}^c_{\mf{w}-w \epsilon} \mf{A}^\mu_{\mf{w}-w \epsilon}
    + \bar{v}^c_{\mf{a}-\alpha k} \mf{A}^\mu_{\mf{a}-\alpha k}. \nonumber
\end{align}
The Kubo formulas for the coefficients are:
\begin{subequations}\begin{align}
a_{\mf{w}-w u} = & \frac{(k\cdot u)}{k_\perp^2} \corrLTE{u_\mu\hWFA^\mu}{k^\perp_\rho \h{\mathcal{B}}_{\mf{w}-w}^\rho}, &
    \bar{\mf{a}}_{\mf{a}-\alpha u} = & 2\frac{(k\cdot u)}{k_\perp^2} \corrD{u_\mu\hWFA^\mu}{k^\perp_\rho \h{\mathcal{C}}_{\mf{a}-\alpha}^\rho},\\
a_{\mf{w}-w \Delta} = & -\frac{1}{3} \Delta_{\mu\rho} \corrLTE{\hWFA^\mu}{\h{\mathcal{B}}_{\mf{w}-w}x^\rho}, &
    \bar{\mf{a}}_{\mf{a}-\alpha \Delta} = & -\frac{2}{3} \Delta_{\mu\rho} \corrD{\hWFA^\mu}{\h{\mathcal{C}}_{\mf{a}-\alpha}x^\rho},\\
a_{\mf{w}-w k} = & -\frac{3}{2} Q_{\mu\rho} \corrLTE{\hWFA^\mu}{\h{\mathcal{B}}_{\mf{w}-w}^\rho}, &
    \bar{\mf{a}}_{\mf{a}-\alpha k} = & -3 Q_{\mu\rho} \corrD{\hWFA^\mu}{\h{\mathcal{C}}_{\mf{a}-\alpha}^\rho},
\end{align}
\begin{equation}
\bar{\mf{a}}_{\mf{w}-w \epsilon} = -\frac{(k\cdot u)}{k_\perp^2} \corrD{\hWFA^\mu}{u^\lambda\epsilon_{\lambda\mu\tau\rho}k_\perp^\tau\h{\mathcal{C}}_{\mf{w}-w}^\rho},\,
    a_{\mf{a}-\alpha \epsilon} = -\frac{(k\cdot u)}{2 k_\perp^2} \corrLTE{\hWFA^\mu}{u^\lambda\epsilon_{\lambda\mu\tau\rho}k_\perp^\tau\h{\mathcal{B}}_{\mf{a}-\alpha}^\rho},
\end{equation}
and
\begin{align}
\mf{v}^c_{\mf{w}-w u} = & \frac{(k\cdot u)}{k_\perp^2} \corrLTE{u_\mu\hWFV^\mu}{k^\perp_\rho \h{\mathcal{B}}_{\mf{w}-w}^\rho}, &
    \bar{v}^c_{\mf{a}-\alpha u} = & 2\frac{(k\cdot u)}{k_\perp^2} \corrD{u_\mu\hWFV^\mu}{k^\perp_\rho \h{\mathcal{C}}_{\mf{a}-\alpha}^\rho},\\
\mf{v}^c_{\mf{w}-w \Delta} = & -\frac{1}{3} \Delta_{\mu\rho} \corrLTE{\hWFV^\mu}{\h{\mathcal{B}}_{\mf{w}-w}^\rho}, &
    \bar{v}^c_{\mf{a}-\alpha \Delta} = & -\frac{2}{3} \Delta_{\mu\rho} \corrD{\hWFV^\mu}{\h{\mathcal{C}}_{\mf{a}-\alpha}^\rho},\\
\mf{v}^c_{\mf{w}-w k} = & -\frac{3}{2} Q_{\mu\rho} \corrLTE{\hWFV^\mu}{\h{\mathcal{B}}_{\mf{w}-w}^\rho}, &
    \bar{v}^c_{\mf{a}-\alpha k} = & -3 Q_{\mu\rho} \corrD{\hWFV^\mu}{\h{\mathcal{C}}_{\mf{a}-\alpha}^\rho},
\end{align}
\begin{equation}
\bar{v}^c_{\mf{w}-w \epsilon} = -\frac{(k\cdot u)}{k_\perp^2} \corrD{\hWFV^\mu}{u^\lambda\epsilon_{\lambda\mu\tau\rho}k_\perp^\tau\h{\mathcal{C}}_{\mf{w}-w}^\rho},\,
    \mf{v}^c_{\mf{a}-\alpha \epsilon} = -\frac{(k\cdot u)}{2 k_\perp^2} \corrLTE{\hWFV^\mu}{u^\lambda\epsilon_{\lambda\mu\tau\rho}k_\perp^\tau\h{\mathcal{B}}_{\mf{a}-\alpha}^\rho}.
\end{equation}\end{subequations}
Note that, in this case as well, all dissipative corrections to the axial Wigner function require chiral coefficients. As done before for the thermal vorticity, the Wigner functions can also be written as
\begin{align}
\Delta_{\SP-\varpi,{\rm LTE}}\WFA^\mu(x,k) =& -a_{\SP-\varpi} \frac{2(\tilde\SP^{\mu\nu}-\tilde\varpi^{\mu\nu})k_\nu}{(2\pi)^3}
    - (a_{\mf{w}-w u} - a_{\mf{w}-w \Delta}) A^\mu_{\mf{w}-w u}\\
   & + (a_{\mf{a}-\alpha\epsilon} - a_{\mf{w}-w \Delta}) A^\mu_{\mf{a}-\alpha\epsilon} + a_{\mf{w}-w k} A^\mu_{\mf{w}-w k},\nonumber\\
\Delta_{\SP-\varpi,{\rm LTE}}\WFV^\mu(x,k) =& -\mf{v}^c_{\SP-\varpi} \frac{2(\tilde\SP^{\mu\nu}-\tilde\varpi^{\mu\nu})k_\nu}{(2\pi)^3}
    - (\mf{v}^c_{\mf{w}-w u} - \mf{v}^c_{\mf{w}-w \Delta}) A^\mu_{\mf{w}-w u}\\
    &+ (\mf{v}^c_{\mf{a}-\alpha\epsilon} - \mf{v}^c_{\mf{w}-w \Delta}) A^\mu_{\mf{a}-\alpha\epsilon} + \mf{v}^c_{\mf{w}-w k} A^\mu_{\mf{w}-w k},\nonumber
\end{align}
where
\begin{subequations}\begin{align}
a_{\SP-\varpi} =& \frac{(2\pi)^3}{2(k\cdot u)} a_{\mf{w}-w \Delta}
    =  -\frac{(2\pi)^3}{6(k\cdot u)} \corrLTE{\hWFA_\mu}{\h{\mathcal{B}}_{\mf{w}-w}^{\mu}},\\
\mf{v}^c_{\SP-\varpi} = & -\frac{(2\pi)^3}{6(k\cdot u)} \corrLTE{\hWFV_\mu}{\h{\mathcal{B}}_{\mf{w}-w}^{\mu}} .
\end{align}\end{subequations}
For a massive field, the spin polarization is directly related to the axial part of the Wigner function.
In that case, it must satisfy $k\cdot\Delta_\varpi\mathcal{A}(x,k)=0$ and we can write:
\begin{equation}
\begin{split}
\Delta_{\SP-\varpi,{\rm LTE}}\WFA^\mu(x,k) =& -a_{\SP-\varpi} \frac{2(\tilde\SP^{\mu\nu}-\tilde\varpi^{\mu\nu})k_\nu}{(2\pi)^3}
    - (a_{\mf{w}-w u} - a_{\mf{w}-w\Delta}) \left(A^\mu_{\mf{w}-w u} + \frac{3}{2} A^\mu_{\mf{w}-w k}\right)\\
    &+ (a_{\mf{a}-\alpha\epsilon} - a_{\mf{w}-w\delta}) A^\mu_{\mf{a}-\alpha\epsilon} \, .
\end{split}
\end{equation}
%

%***********************************************************************************************************
\subsection{Thermal shear}
Using the decomposition of the thermal shear tensor~(\ref{eq:DecompShear}), the non-dissipative part (\ref{eq:BLTEa}) can be written as
\begin{equation}
\h{B}_\xi = -\frac{1}{2}\xi_{\rho\sigma}(x)\h{\Xi}_x^{\rho\sigma}=-\frac{1}{2}\left[
\h{\varepsilon}_{\Xi} D\beta-\h{p}_{\Xi}\beta\theta+\h{q}_\Xi^\sigma\left(\beta Du_\sigma + \partial_\sigma\beta \right)+\h{\pi}_\Xi^{\alpha\tau}\beta\sigma_{\alpha\tau}
\right],
\end{equation}
where
\begin{equation}
\h{\Xi}_x^{\rho\sigma}(\Sigma)=\int_\Sigma{\rm d}\Sigma_\lambda(y)
    \left[(y-x)^\rho\wT^{\lambda\sigma}(y) + (y-x)^\sigma\wT^{\lambda\rho}(y)\right]
\end{equation}
and we defined
\begin{equation}
\begin{split}
\h{\varepsilon}_\Xi =& u_\mu u_\nu \h{\Xi}^{\mu\nu},\quad
	\hat{p}_\Xi=-\frac{1}{3}\Delta_{\mu\nu} \h{\Xi}^{\mu\nu},\quad
\h{\pi}_\Xi^{\mu\nu} = \Delta_{\alpha\beta}^{\mu\nu} \h{\Xi}^{\alpha\beta},\quad
	\h{q}_\Xi^\mu  = u_\alpha\Delta_{\beta}^{\mu}\h{\Xi}^{\alpha\beta}.
\end{split}
\end{equation}
For the dissipative part~(\ref{eq:CDissa}) we have instead:
\begin{equation}
\h{\mathcal{C}}_{\xi}(x_2) = \h{\varepsilon} D\beta-\h{p}\beta\theta
    +\h{q}^\sigma\left(\beta Du_\sigma + \partial_\sigma\beta \right)
    +\h{\pi}^{\alpha\tau}\beta\sigma_{\alpha\tau},
\end{equation}
where
\begin{equation}
\begin{split}
\h{\varepsilon} =& u_\mu u_\nu \h{T}^{\mu\nu},\quad
	\hat{p}=-\frac{1}{3}\Delta_{\mu\nu} \h{T}^{\mu\nu},
\h{\pi}^{\mu\nu} = \Delta_{\alpha\beta}^{\mu\nu} \h{T}^{\alpha\beta},\quad
	\h{q}^\mu  = u_\alpha\Delta_{\beta}^{\mu}\h{T}_S^{\alpha\beta}.
\end{split}
\end{equation}
The contributions from thermal shear are then divided into two scalars, one vector and one spin-1 symmetric contributions. The scalar one proportional to gradients of temperature is
\begin{equation}
\label{eq:DecShearDBeta}
\begin{split}
\mcU_{(\alpha)}= X = D\beta,\quad
b_{D\beta} =& -\frac{1}{2},\quad
\h{\mathcal{B}}_{D\beta} = \h{\varepsilon}_\Xi,\\
c_{D\beta} =& 1,\quad
\h{\mathcal{C}}_{D\beta}(x_2) = \h{\varepsilon}(x_2),
\end{split}
\end{equation}
the one proportional to the expansion rate is
\begin{equation}
\label{eq:DecDeShearTheta}
\begin{split}
\mcU_{(\alpha)}= X =\beta\theta,\quad
b_{\theta} =& +\frac{1}{2},\quad
\h{\mathcal{B}}_{\theta} =\hat{p}_\Xi,\\
c_{\theta} =& -1,\quad
\h{\mathcal{C}}_{\theta}(x_2) = \hat{p}(x_2),
\end{split}
\end{equation}
the vector one is
\begin{equation}
\label{eq:DecShearVector}
\begin{split}
\mcU_{(\alpha)}= V_\rho = \beta Du_\rho + \partial_\rho\beta ,\quad
b_{q} =& -\frac{1}{2},\quad
\h{\mathcal{B}}_{q} = \h{q}_\Xi^\rho,\\
c_{q} =& 1,\quad
\h{\mathcal{C}}_{q}(x_2) = \h{q}^\rho(x_2),
\end{split}
\end{equation}
and finally the one proportional to the shear tensor is
\begin{equation}
\label{eq:DecSpinShearSSpin1}
\begin{split}
\mcU_{(\alpha)}= \Sigma_{\rho\sigma}=\beta\sigma_{\rho\sigma},\quad
b_{\sigma} =& -\frac{1}{2},\quad
\h{\mathcal{B}}_{\sigma}^{\rho\sigma} = \h{\pi}^{\rho\sigma}_\Xi,\\
c_{\sigma} =& 1,\quad
\h{\mathcal{C}}_{\sigma}(x_2) =  \h{\pi}^{\rho\sigma}(x_2).
\end{split}
\end{equation}
After introducing the vectors
\begin{subequations}\begin{align}
A^\mu_{\sigma\epsilon} =& \epsilon^{\mu\nu\alpha\rho}k_\perp^{\sigma}\frac{u_\nu k_\alpha}{(k\cdot u)} \beta\sigma_{\rho\sigma},&
\mf{A}^\mu_{\sigma u} =& (k\cdot u) \frac{k_\perp^\rho k_\perp^\sigma}{k_\perp^2}u^\mu\,\beta\sigma_{\rho\sigma},\\
\mf{A}^\mu_{\sigma \Delta} =& \Delta^{\mu\rho} k_\perp^\sigma \beta\sigma_{\rho\sigma},&
\mf{A}^\mu_{\sigma k} =& Q^{\mu\rho} k_\perp^{\sigma}\,\beta\sigma_{\rho\sigma},
\end{align}\end{subequations}
that transform as follows
\begin{equation}
\begin{array}{|l|ccc|c|c|c|c|c|c|c|c|c|}
\hline
\, & \, & \, & \, & \, & \, & \, & \, & \, & \, & \, & \, & \, \\[-1em]
 & \beta & D\beta & \theta & Du^{\ortu\mu},\,\de^{\ortu\mu}\beta & \WFA^\mu,\,A^\mu_{\sigma\epsilon} & a^\mu,\,\mf{A}^\mu_{\sigma X}
 & \h{\varepsilon}_\Xi,\,\h{p}_\Xi & \h{\varepsilon},\,\h{p} & \h{q}^{\ortu\mu}_\Xi & \h{q}^{\ortu\mu} & \h{\pi}_\Xi^{\rho\sigma} & \h{\pi}^{\rho\sigma}\\
\hline
\group{P} & + & + & + & - & (-,+) & (+,-) & + & + & - & - & + & + \\
\group{T} & + & - & - & + & (+,-) & (-,+) & - & + & + & - & - & + \\
\group{C} & + & + & + & + & (+,+) & (+,+)  & + & + & + & + & + & + \\
\hline
\end{array}
\end{equation}
the linear response of the thermal shear is
\begin{subequations}\begin{align}
\Delta_{D\beta}\WFA^\mu = & \left[ \bar{\mf{a}}_{D\beta u} u^\mu + \bar{\mf{a}}_{D\beta k} \frac{k_\perp^\mu}{(k\cdot u)} \right] D\beta, &
\Delta_{D\beta}\WFV^\mu = & \left[ \bar{v}^c_{D\beta u} u^\mu + \bar{v}^c_{D\beta k} \frac{k_\perp^\mu}{(k\cdot u)} \right] D\beta,\\
\Delta_{\theta}\WFA^\mu = & \left[ \bar{\mf{a}}_{\theta u} u^\mu + \bar{\mf{a}}_{\theta k} \frac{k_\perp^\mu}{(k\cdot u)} \right] \beta\theta, &
\Delta_{\theta}\WFV^\mu = & \left[ \bar{v}^c_{\theta u} u^\mu + \bar{v}^c_{\theta k} \frac{k_\perp^\mu}{(k\cdot u)} \right] \beta\theta,
\end{align}
\begin{equation}
\Delta_{\sigma}\WFA^\mu = \bar{\mf{a}}_{\sigma u}\mf{A}^\mu_{\sigma u}
    + \bar{\mf{a}}_{\sigma \Delta}\mf{A}^\mu_{\sigma\Delta}
    + \bar{\mf{a}}_{\sigma k} \mf{A}^\mu_{\sigma k} + a_{\sigma\epsilon} A^\mu_{\sigma\epsilon},\,
\Delta_{\sigma}\WFV^\mu = \bar{v}^c_{\sigma u}\mf{A}^\mu_{\sigma u}
    + \bar{v}^c_{\sigma \delta}\mf{A}^\mu_{\sigma\Delta}
    + \bar{v}^c_{\sigma k} \mf{A}^\mu_{\sigma k} + \mf{v}^c_{\sigma\epsilon} A^\mu_{\sigma\epsilon},
\end{equation}
and
\begin{align}
\Delta_{q}\WFA^\mu = & \left[ \bar{\mf{a}}_{q u} \frac{k_\perp^\rho u^\mu}{(k\cdot u)}
    + \bar{\mf{a}}_{q \Delta} \Delta^{\mu\rho}
    + \bar{\mf{a}}_{q k} Q^{\mu\rho}
    + a_{q \epsilon}\, \epsilon^{\mu\nu\rho\sigma} \frac{k^\perp_\nu u_\sigma}{(k\cdot u)}
    \right] (\beta Du_{\rho}+\de_\rho\beta),\\    
\Delta_{q}\WFV^\mu = & \left[ \bar{v}^c_{q u} \frac{k_\perp^\rho u^\mu}{(k\cdot u)}
    + \bar{v}^c_{q \Delta} \Delta^{\mu\rho}
    + \bar{v}^c_{q k} Q^{\mu\rho}
    + \mf{v}^c_{q \epsilon}\, \epsilon^{\mu\nu\rho\sigma} \frac{k^\perp_\nu u_\sigma}{(k\cdot u)}
    \right] (\beta Du_{\rho}+\de_\rho\beta).
\end{align}\end{subequations}
The coefficients are obtained with the Kubo formulas:
\begin{subequations}\begin{align}
\bar{\mf{a}}_{D\beta u} = & \corrD{u_\mu\hWFA^\mu}{\h{\varepsilon}}, &
\bar{\mf{a}}_{D\beta k} = & \frac{ (k\cdot u)}{k_\perp^2} \corrD{k^\perp_\mu\hWFA^\mu}{\h{\varepsilon}},\\
\bar{\mf{a}}_{\theta u} = & -\corrD{u_\mu\hWFA^\mu}{\h{p}}, &
\bar{\mf{a}}_{\theta k} = & -\frac{ (k\cdot u)}{k_\perp^2} \corrD{k^\perp_\mu\hWFA^\mu}{\h{p}},\\
\bar{\mf{a}}_{q u} = & \frac{(k\cdot u)}{k_\perp^2} \corrD{u_\mu\hWFA^\mu}{k^\perp_\rho\h{q}^\rho}, &
\bar{\mf{a}}_{q \Delta} = & \frac{1}{3} \Delta_{\mu\rho} \corrD{\hWFA^\mu}{\h{q}^\rho},\\
\bar{\mf{a}}_{q k} = & \frac{3}{2} Q_{\mu\rho} \corrD{\hWFA^\mu}{\h{q}^\rho}, &
a_{q \epsilon} = & -\frac{(k\cdot u)}{4 k_\perp^2}  \corrLTE{\hWFA^\mu}{u^\lambda\epsilon_{\lambda\mu\tau\rho}k_\perp^\tau\h{q}_\Xi^\rho},\\
%\end{align}
%
%
%\begin{align}
\bar{\mf{a}}_{\sigma u} = & \frac{1}{k_\perp^2} \corrD{u_\mu\hWFA^\mu}{k^\perp_\rho k^\perp_\sigma\h{\pi}^{\rho\sigma}}, &
\bar{\mf{a}}_{\sigma \Delta} = & \frac{(k\cdot u)}{3 k_\perp^2} \Delta_{\mu\rho}  \corrD{\hWFA^\mu}{k^\perp_{\sigma}\h{\pi}^{\rho\sigma}},\\
\bar{\mf{a}}_{\sigma k} = & \frac{3(k\cdot u)}{2 k_\perp^2} Q_{\mu\rho} \corrD{\hWFA^\mu}{k^\perp_{\sigma}\h{\pi}^{\rho\sigma}}, &
a_{\sigma \epsilon} = & -\frac{(k\cdot u)^2}{4 (k_\perp^2)^2}  \corrLTE{\hWFA^\mu}{u^\lambda\epsilon_{\lambda\mu\tau\rho} k^\perp_{\sigma}k_\perp^\tau\h{\pi}_\Xi^{\rho\sigma}},
\end{align}\end{subequations}
and
\begin{subequations}\begin{align}
\bar{v}^c_{D\beta u} = & \corrD{u_\mu\hWFV^\mu}{\h{\varepsilon}}, &
\bar{v}^c_{D\beta k} = & \frac{ (k\cdot u)}{k_\perp^2} \corrD{k^\perp_\mu\hWFV^\mu}{\h{\varepsilon}},\\
\bar{v}^c_{\theta u} = & -\corrD{u_\mu\hWFV^\mu}{\h{p}}, &
\bar{v}^c_{\theta k} = & -\frac{ (k\cdot u)}{k_\perp^2} \corrD{k^\perp_\mu\hWFV^\mu}{\h{p}},\\
\bar{\mf{a}}_{q u} = & \frac{(k\cdot u)}{k_\perp^2} \corrD{u_\mu\hWFV^\mu}{k^\perp_\rho\h{q}^\rho}, &
\bar{v}^c_{q \Delta} = & \frac{1}{3} \Delta_{\mu\rho} \corrD{\hWFV^\mu}{\h{q}^\rho},\\
\bar{v}^c_{q k} = & \frac{3}{2} Q_{\mu\rho} \corrD{\hWFV^\mu}{\h{q}^\rho}, &
\mf{v}^c_{q \epsilon} = & -\frac{(k\cdot u)}{4 k_\perp^2}  \corrLTE{\hWFV^\mu}{u^\lambda\epsilon_{\lambda\mu\tau\rho}k_\perp^\tau\h{q}_\Xi^\rho},\\
\bar{v}^c_{\sigma u} = & \frac{1}{k_\perp^2} \corrD{u_\mu\hWFA^\mu}{k^\perp_\rho k^\perp_\sigma\h{\pi}^{\rho\sigma}}, &
\bar{v}^c_{\sigma \Delta} = & \frac{(k\cdot u)}{3 k_\perp^2} \Delta_{\mu\rho}  \corrD{\hWFA^\mu}{k^\perp_{\sigma}\h{\pi}^{\rho\sigma}},\\
\bar{v}^c_{\sigma k} = & \frac{3(k\cdot u)}{2 k_\perp^2} Q_{\mu\rho} \corrD{\hWFA^\mu}{k^\perp_{\sigma}\h{\pi}^{\rho\sigma}}, &
\mf{v}^c_{\sigma \epsilon} = & -\frac{(k\cdot u)^2}{4 (k_\perp^2)^2}  \corrLTE{\hWFA^\mu}{u^\lambda\epsilon_{\lambda\mu\tau\rho} k^\perp_{\sigma}k_\perp^\tau\h{\pi}_\Xi^{\rho\sigma}}.
\end{align}\end{subequations}
We again stress that all the dissipative corrections are given in terms of chiral coefficients.
\newpage
%***********************************************************************************************************
\subsection{Gradients of spin potential}
The operators related to the linear response of the gradient of the spin potential are given by~(\ref{eq:BLTEb}) and (\ref{eq:CDissb}), i.e.,
\begin{equation}
\label{eq:DecDeSpinPotential}
\begin{split}
\mcU_{(\alpha)}= \Phi_{\tau,\rho\sigma} =\de_\tau\SP_{\rho\sigma},\quad
b_{\de\SP} =& 1,\quad
\h{\mathcal{B}}_{\de\SP}^{\tau,\rho\sigma} = \int_\Sigma\di\Sigma_\lambda(y)\,(y-x)^\tau \h{S}^{\lambda,\rho\sigma}(y),\\
c_{\de\SP} =& -\frac{1}{2},\quad
\h{\mathcal{C}}_{\de\SP}(x_2) =  \h{S}^{\tau,\rho\sigma}(x_2)-2(x_2-x)^\tau\h{T}^{\rho\sigma}_A(x_2).
\end{split}
\end{equation}
In eq.~(\ref{eq:DecompdeSP}) we decomposed  $\de_\tau \SP_{\rho\sigma}$ in terms of $f,\,\Upsilon,\,I,\,I-\Upsilon,\,I_S,\,\phi$, $\Phi_{\rm S12}$ and $\Phi_{\rm S13}$. Correspondingly, the eq.~(\ref{eq:DecDeSpinPotential}) defines the operators
\begin{equation}
\h{\mathcal{B}}_f^\rho ,\,
\h{\mathcal{B}}_\Upsilon^\rho ,\,
\h{\mathcal{B}}_I ,\,
\h{\mathcal{B}}_{I-\Upsilon}^\rho ,\,
\h{\mathcal{B}}_{I_S}^{\rho\sigma} ,\,
\h{\mathcal{B}}_\phi ,\,
\h{\mathcal{B}}_{\rm S12}^{\tau,\rho\sigma} ,\,
\h{\mathcal{B}}_{\rm S13}^{\tau,\rho\sigma} ,
\end{equation}
and
\begin{equation}
\h{\mathcal{C}}_f^\rho ,\,
\h{\mathcal{C}}_\Upsilon^\rho ,\,
\h{\mathcal{C}}_I ,\,
\h{\mathcal{C}}_{I-\Upsilon}^\rho ,\,
\h{\mathcal{C}}_{I_S}^{\rho\sigma} ,\,
\h{\mathcal{C}}_\phi ,\,
\h{\mathcal{C}}_{\rm S12}^{\tau,\rho\sigma} ,\,
\h{\mathcal{C}}_{\rm S13}^{\tau,\rho\sigma}.
\end{equation}
These hydrodynamic fields and the corresponding operators transform as follows:
\begin{equation}
\begin{array}{|l|ccc|ccc|ccc|ccc|}
\hline
\, & \, & \, & \, & \, & \, & \, & \, & \, & \, & \, & \, & \, \\[-1em]
 & f^\rho & \h{\mathcal{B}}_f^\rho & \h{\mathcal{C}}_f^\rho
 & \Upsilon^\rho & \h{\mathcal{B}}_\Upsilon^\rho & \h{\mathcal{C}}_\Upsilon^\rho
 & I & \h{\mathcal{B}}_I & \h{\mathcal{C}}_I
 & I^\rho-\Upsilon^\rho & \h{\mathcal{B}}_{I-\Upsilon}^\rho & \h{\mathcal{C}}_{I-\Upsilon}^\rho\\
\hline
\group{P} & - & - & - & + & + & + & + & + & + & + & + & + \\
\group{T} & - & - & + & + & + & - & + & + & - & + & + & - \\
\group{C} & + & + & + & + & + & + & + & + & + & + & + & + \\
\hline
\end{array}
\end{equation}
and
\begin{equation}
\begin{array}{|l|ccc|ccc|ccc|ccc|}
\hline
\, & \, & \, & \, & \, & \, & \, & \, & \, & \, & \, & \, & \, \\[-1em]
 & I_S^{\rho\sigma} & \h{\mathcal{B}}_{I_S}^{\rho\sigma} & \h{\mathcal{C}}_{I_S}^{\rho\sigma}
 & \phi & \h{\mathcal{B}}_\phi & \h{\mathcal{C}}_\phi
 & \Phi^{\tau,\rho\sigma}_{\rm S12} & \h{\mathcal{B}}_{\rm S12}^{\tau,\rho\sigma} & \h{\mathcal{C}}_{\rm S12}^{\tau,\rho\sigma}
 & \Phi^{\tau,\rho\sigma}_{\rm S13} & \h{\mathcal{B}}_{\rm S13}^{\tau,\rho\sigma} & \h{\mathcal{C}}_{\rm S13}^{\tau,\rho\sigma}\\
\hline
\group{P} & + & + & + & - & - & - & - & - & - & - & - & - \\
\group{T} & + & + & - & - & - & + & - & - & + & - & - & + \\
\group{C} & + & + & + & + & + & + & + & + & + & + & + & + \\
\hline
\end{array}
\end{equation}
Accordingly, with the results of the previous sections, we can write the linear response to the gradient of spin potential to the axial of the Wigner function and their coefficients:
\begin{subequations}\begin{align}
\Delta_{f} \WFA^\mu = & \left[ \mf{a}_{f u} \frac{k_\perp^\rho u^\mu}{(k\cdot u)}
    + \mf{a}_{f \Delta} \Delta^{\mu\rho}
    + \mf{a}_{f k} Q^{\mu\rho}
    + \bar{a}_{f \epsilon}\, \epsilon^{\mu\nu\rho\sigma} \frac{k^\perp_\nu u_\sigma}{(k\cdot u)}
    \right] f_{\rho},\\
\Delta_{\Upsilon} \WFA^\mu = & \left[ \bar{a}_{\Upsilon u} \frac{k_\perp^\rho u^\mu}{(k\cdot u)}
    + \bar{a}_{\Upsilon \Delta} \Delta^{\mu\rho}
    + \bar{a}_{\Upsilon k} Q^{\mu\rho}
    + \mf{a}_{\Upsilon \epsilon}\, \epsilon^{\mu\nu\rho\sigma} \frac{k^\perp_\nu u_\sigma}{(k\cdot u)}
    \right] \Upsilon_{\rho},\\
\Delta_{I} \WFA^\mu = & \left[ \mf{a}_{I u} u^\mu + \mf{a}_{I k} \frac{k_\perp^\mu}{(k\cdot u)} \right] I,\\
\Delta_{I-\Upsilon} \WFA^\mu = & \left[ \bar{a}_{I-\Upsilon u} \frac{k_\perp^\rho u^\mu}{(k\cdot u)}
    + \bar{a}_{I-\Upsilon \Delta} \Delta^{\mu\rho}
    + \bar{a}_{I-\Upsilon k} Q^{\mu\rho}
    + \mf{a}_{I-\Upsilon \epsilon}\, \epsilon^{\mu\nu\rho\sigma} \frac{k^\perp_\nu u_\sigma}{(k\cdot u)}
    \right] (I_\rho-\Upsilon_{\rho}),\\
\Delta_\phi \WFA^\mu = & \left[ \bar{a}_{\phi u} u^\mu + \bar{a}_{\phi k} \frac{k_\perp^\mu}{(k\cdot u)} \right] \phi,\\
\Delta_{I_S} \WFA^\mu = & \left[ \mf{a}_{I_S u} \frac{k_\perp^\rho k_\perp^\sigma}{k_\perp^2} u^\mu
    + \mf{a}_{I_S \Delta} \frac{\Delta^{\mu\rho} k_\perp^{\sigma}}{(k\cdot u)}
    + \mf{a}_{I_S k} \frac{Q^{\mu\rho} k_\perp^{\sigma}}{(k\cdot u)}
    + \bar{a}_{I_S \epsilon}\, \epsilon^{\mu\nu\alpha\rho} k_\perp^{\sigma} \frac{u_\nu k^\perp_\alpha}{(k\cdot u)^2}
    \right] I_{S\,\rho\sigma},
\end{align}
\begin{align}
\Delta_{\rm S12} \WFA^\mu = &
\left[ \left(\mf{a}_{\rm{S12}\Delta} \Delta^{\tau\rho} + \mf{a}_{\rm{S12}k} Q^{\tau\rho}\right)\Delta^{\mu\sigma}
    + \bar{a}_{\rm{S12}\Delta\epsilon} \Delta^{\mu\tau}\epsilon^{\lambda\nu\rho\sigma}\frac{u_\lambda k^\perp_\nu}{(k\cdot u)}\right. \\ & \left.
    + \bar{a}_{\rm{S12}\epsilon} \epsilon^{\mu\nu\rho\sigma}\frac{k_\perp^{\tau}u_\nu}{(k\cdot u)} \right] \Phi^{S12}_{\tau,\rho\sigma},\nonumber
\end{align}
\begin{align}
\Delta_{\rm S13} \WFA^\mu = &
\left[ \left(\mf{a}_{\rm{S13}\Delta} \Delta^{\tau\sigma} + \mf{a}_{\rm{S13}k} Q^{\tau\sigma}\right)\Delta^{\mu\rho}
    + \bar{a}_{\rm{S13}\Delta\epsilon} \Delta^{\mu\tau}\epsilon^{\lambda\nu\rho\sigma}\frac{u_\lambda k^\perp_\nu}{(k\cdot u)}\right. \\ & \left.
    + \bar{a}_{\rm{S13}\epsilon} \epsilon^{\mu\nu\rho\sigma}\frac{k_\perp^{\tau}u_\nu}{(k\cdot u)} \right] \Phi^{S13}_{\tau,\rho\sigma},\nonumber
\end{align}\end{subequations}
and similarly for the vector part of the Wigner function
\begin{subequations}\begin{align}
\Delta_{f} \WFV^\mu = & \left[ v^c_{f u} \frac{k_\perp^\rho u^\mu}{(k\cdot u)}
    + v^c_{f \Delta} \Delta^{\mu\rho}
    + v^c_{f k} Q^{\mu\rho}
    + \bar{\mf{v}}^c_{f \epsilon}\, \epsilon^{\mu\nu\rho\sigma} \frac{k^\perp_\nu u_\sigma}{(k\cdot u)}
    \right] f_{\rho},\\
\Delta_{\Upsilon} \WFV^\mu = & \left[ \bar{\mf{v}}^c_{\Upsilon u} \frac{k_\perp^\rho u^\mu}{(k\cdot u)}
    + \bar{\mf{v}}^c_{\Upsilon \Delta} \Delta^{\mu\rho}
    + \bar{\mf{v}}^c_{\Upsilon k} Q^{\mu\rho}
    + v^c_{\Upsilon \epsilon}\, \epsilon^{\mu\nu\rho\sigma} \frac{k^\perp_\nu u_\sigma}{(k\cdot u)}
    \right] \Upsilon_{\rho},\\
\Delta_{I} \WFV^\mu = & \left[ v^c_{I u} u^\mu + v^c_{I k} \frac{k_\perp^\mu}{(k\cdot u)} \right] I,\\
\Delta_{I-\Upsilon} \WFV^\mu = & \left[ \bar{\mf{v}}^c_{I-\Upsilon u} \frac{k_\perp^\rho u^\mu}{(k\cdot u)}
    + \bar{\mf{v}}^c_{I-\Upsilon \Delta} \Delta^{\mu\rho}
    + \bar{\mf{v}}^c_{I-\Upsilon k} Q^{\mu\rho}
    + v^c_{I-\Upsilon \epsilon}\, \epsilon^{\mu\nu\rho\sigma} \frac{k^\perp_\nu u_\sigma}{(k\cdot u)}
    \right] (I_\rho-\Upsilon_{\rho}),\\
\Delta_\phi \WFV^\mu = & \left[ \bar{\mf{v}}^c_{\phi u} u^\mu + \bar{\mf{v}}^c_{\phi k} \frac{k_\perp^\mu}{(k\cdot u)} \right] \phi,\\
\Delta_{I_S} \WFV^\mu = & \left[ v^c_{I_S u} \frac{k_\perp^\rho k_\perp^\sigma}{k_\perp^2} u^\mu
    + v^c_{I_S \Delta} \frac{\Delta^{\mu\rho} k_\perp^{\sigma}}{(k\cdot u)}
    + v^c_{I_S k} \frac{Q^{\mu\rho} k_\perp^{\sigma}}{(k\cdot u)}
    + \bar{\mf{v}}^c_{I_S \epsilon}\, \epsilon^{\mu\nu\alpha\rho} k_\perp^{\sigma} \frac{u_\nu k^\perp_\alpha}{(k\cdot u)^2}
    \right] I_{S\,\rho\sigma},\\
\Delta_{\rm S12} \WFV^\mu = &
\left[ v^c_{\rm{S12}\Delta} \Delta^{\tau\rho} \Delta^{\mu\sigma} + v^c_{\rm{S12}k} Q^{\tau\rho}\Delta^{\mu\sigma}
    + \bar{\mf{v}}^c_{\rm{S12}\Delta\epsilon} \Delta^{\mu\tau}\epsilon^{\lambda\nu\rho\sigma}\frac{u_\lambda k^\perp_\nu}{(k\cdot u)}
    \right. \\ & \left.+ \bar{\mf{v}}^c_{\rm{S12}\epsilon} \epsilon^{\mu\nu\rho\sigma}\frac{k_\perp^{\tau}u_\nu}{(k\cdot u)} \right] \Phi^{S12}_{\tau,\rho\sigma}\nonumber\\
\Delta_{\rm S13} \WFV^\mu = &
\left[ v^c_{\rm{S13}\Delta} \Delta^{\tau\sigma} \Delta^{\mu\rho} + v^c_{\rm{S13}k} Q^{\tau\sigma}\Delta^{\mu\rho}
    + \bar{\mf{v}}^c_{\rm{S13}\Delta\epsilon} \Delta^{\mu\tau}\epsilon^{\lambda\nu\rho\sigma}\frac{u_\lambda k^\perp_\nu}{(k\cdot u)}
    \right. \\ & \left.+ \bar{\mf{v}}^c_{\rm{S13}\epsilon} \epsilon^{\mu\nu\rho\sigma}\frac{k_\perp^{\tau}u_\nu}{(k\cdot u)} \right] \Phi^{S13}_{\tau,\rho\sigma},\nonumber
\end{align}\end{subequations}
Based on what we obtained in the previous sections, the Kubo formulas for the coefficients related to the gradient of spin potential are: for the scalars, vectors and rank-2 tensors
\begin{subequations}\begin{align}
\mf{a}_{f u} = & 2\frac{(k\cdot u)}{k_\perp^2} \corrLTE{u_\mu \hWFA^\mu}{k^\perp_\rho \h{\mathcal{B}}^\rho_f}, &
    \mf{a}_{f \Delta} =& \frac{2}{3} \Delta_{\mu\rho} \corrLTE{\hWFA^\mu}{\h{\mathcal{B}}^\rho_f},\\
\mf{a}_{f k} = & 3 Q_{\mu\rho} \corrLTE{\hWFA^\mu}{\h{\mathcal{B}}^\rho_f}, & 
    \bar{a}_{f \epsilon} =& -\frac{(k\cdot u)}{2 k_\perp^2} u^\lambda\epsilon_{\lambda\mu\tau\rho}k_\perp^\tau \corrD{\hWFA^\mu}{\h{\mathcal{C}}^\rho_f},\\
\bar{a}_{\Upsilon u} = & \frac{(k\cdot u)}{k_\perp^2} \corrD{u_\mu \hWFA^\mu}{k^\perp_\rho \h{\mathcal{C}}^\rho_\Upsilon}, & 
    \bar{a}_{\Upsilon \Delta} =& \frac{1}{3} \Delta_{\mu\rho} \corrD{\hWFA^\mu}{\h{\mathcal{C}}^\rho_\Upsilon}, \\
\bar{a}_{\Upsilon k} = & \frac{3}{2} Q_{\mu\rho} \corrD{\hWFA^\mu}{\h{\mathcal{C}}^\rho_\Upsilon}, &
    \mf{a}_{\Upsilon \epsilon} =& -\frac{(k\cdot u)}{k_\perp^2} u^\lambda\epsilon_{\lambda\mu\tau\rho}k_\perp^\tau \corrLTE{\hWFA^\mu}{\h{\mathcal{B}}^\rho_\Upsilon}, \\
\mf{a}_{I u} = & 2 \corrLTE{u_\mu \hWFA^\mu}{\h{\mathcal{B}}_I}, &
    \mf{a}_{I k} =& 2\frac{(k\cdot u)}{k_\perp^2} \corrLTE{k^\perp_\mu \hWFA^\mu}{\h{\mathcal{B}}_I},\\
\bar{a}_{I-\Upsilon u} = & \frac{(k\cdot u)}{k_\perp^2} \corrD{u_\mu \hWFA^\mu}{k^\perp_\rho \h{\mathcal{C}}^\rho_{I-\Upsilon}}, &
    \bar{a}_{I-\Upsilon \Delta} =& \frac{1}{3} \Delta_{\mu\rho} \corrD{\hWFA^\mu}{\h{\mathcal{C}}^\rho_{I-\Upsilon}},\\
\bar{a}_{I-\Upsilon k} = & \frac{3}{2} Q_{\mu\rho} \corrD{\hWFA^\mu}{\h{\mathcal{C}}^\rho_{I-\Upsilon}}, &
    \mf{a}_{I-\Upsilon \epsilon} = &-\frac{(k\cdot u)}{k_\perp^2} u^\lambda\epsilon_{\lambda\mu\tau\rho}k_\perp^\tau \corrLTE{\hWFA^\mu}{\h{\mathcal{B}}^\rho_{I-\Upsilon}}, \\
\mf{a}_{I_S u} = & 2\corrLTE{u_\mu \hWFA^\mu}{\frac{k^\perp_\rho k^\perp_\sigma}{k_\perp^2}\h{\mathcal{B}}^{\rho\sigma}_{I_S}}, &
    \mf{a}_{I_S \Delta} =& \frac{2(k\cdot u)}{3 k_\perp^2} \Delta_{\mu\rho} \corrLTE{\hWFA^\mu}{\h{\mathcal{B}}^{\rho\sigma}_{I_S} k^\perp_{\sigma}}, \\
\mf{a}_{I_S k} = & \frac{3(k\cdot u)}{k_\perp^2} Q_{\mu\rho} \corrLTE{\hWFA^\mu}{\h{\mathcal{B}}^{\rho\sigma}_{I_S} k^\perp_{\sigma}}, &
    \bar{a}_{I_S \epsilon} =& -\frac{(k\cdot u)^2}{2 (k_\perp^2)^2} u^\lambda\epsilon_{\lambda\mu\tau\rho}k_\perp^\tau \corrD{\hWFA^\mu}{\h{\mathcal{C}}^{\rho\sigma}_{I_S} k^\perp_{\sigma}}, \\
\bar{a}_{\phi u} = & -\frac{1}{2} \corrD{u_\mu \hWFA^\mu}{\h{\mathcal{C}}_{\phi}}, &
    \bar{a}_{\phi k} =& -\frac{ (k\cdot u)}{2k_\perp^2} \corrD{k^\perp_\mu \hWFA^\mu}{\h{\mathcal{C}}_{\phi}},
\end{align}\end{subequations}
and for the rank-3 irreducible tensors
\begin{subequations}\begin{align}
\mf{a}_{\rm{S12}\Delta} = & \frac{1}{3}\Delta_{\mu\sigma}\Delta_{\tau\rho}\corrLTE{\hWFA^\mu}{\h{\mathcal{B}}^{\tau,\rho\sigma}_{\rm S12}}, &
\mf{a}_{\rm{S12}k} = & \frac{27}{4} Q_{\mu\sigma} Q_{\tau\rho} \corrLTE{\hWFA^\mu}{\h{\mathcal{B}}^{\tau,\rho\sigma}_{\rm S12}},\\
\bar{a}_{\rm{S12}\Delta\epsilon} = & -\frac{3(k\cdot u)}{4 k_\perp^2}\Delta_{\mu\rho}\epsilon_{\tau\sigma\alpha\beta}
    k_\perp^\alpha u^\beta \corrD{\hWFA^\mu}{\h{\mathcal{C}}^{\tau,\rho\sigma}_{\rm S12}}, \\
\bar{a}_{\rm{S12}\epsilon} = & -\frac{3(k\cdot u)k^\perp_\tau k_\perp^\alpha u^\beta k^\perp_\rho}{4(k_\perp)^4}
    \epsilon_{\mu\sigma\alpha\beta} \corrD{\hWFA^\mu}{\h{\mathcal{C}}^{\tau,\rho\sigma}_{\rm S12}},\\
\mf{a}_{\rm{S13}\Delta} = & \frac{1}{3}\Delta_{\mu\sigma}\Delta_{\tau\rho}\corrLTE{\hWFA^\mu}{\h{\mathcal{B}}^{\tau,\rho\sigma}_{\rm S13}}, &
\mf{a}_{\rm{S13}k} = & \frac{27}{4} Q_{\mu\sigma} Q_{\tau\rho} \corrLTE{\hWFA^\mu}{\h{\mathcal{B}}^{\tau,\rho\sigma}_{\rm S13}},\\
\bar{a}_{\rm{S13}\Delta\epsilon} = & -\frac{3(k\cdot u)}{4 k_\perp^2}\Delta_{\mu\rho}\epsilon_{\tau\sigma\alpha\beta}
    k_\perp^\alpha u^\beta \corrD{\hWFA^\mu}{\h{\mathcal{C}}^{\tau,\rho\sigma}_{\rm S13}}, \\
\bar{a}_{\rm{S13}\epsilon} = & -\frac{3(k\cdot u)k^\perp_\tau k_\perp^\alpha u^\beta k^\perp_\rho}{4(k_\perp)^4}
    \epsilon_{\mu\sigma\alpha\beta} \corrD{\hWFA^\mu}{\h{\mathcal{C}}^{\tau,\rho\sigma}_{\rm S13}},
\end{align}\end{subequations}
and similarly for the vector part of the Wigner function
\begin{subequations}\begin{align}
v^c_{f u} = & 2\frac{(k\cdot u)}{k_\perp^2} \corrLTE{u_\mu \hWFV^\mu}{k^\perp_\rho \h{\mathcal{B}}^\rho_f}, &
    v^c_{f \Delta} =& \frac{2}{3} \Delta_{\mu\rho} \corrLTE{\hWFV^\mu}{\h{\mathcal{B}}^\rho_f},\\
v^c_{f k} = & 3 Q_{\mu\rho} \corrLTE{\hWFV^\mu}{\h{\mathcal{B}}^\rho_f}, & 
    \bar{\mf{v}}^c_{f \epsilon} =& -\frac{(k\cdot u)}{2 k_\perp^2} u^\lambda\epsilon_{\lambda\mu\tau\rho}k_\perp^\tau \corrD{\hWFV^\mu}{\h{\mathcal{C}}^\rho_f},\\
\bar{\mf{v}}^c_{\Upsilon u} = & \frac{(k\cdot u)}{k_\perp^2} \corrD{u_\mu \hWFV^\mu}{k^\perp_\rho \h{\mathcal{C}}^\rho_\Upsilon}, & 
    \bar{\mf{v}}^c_{\Upsilon \Delta} =& \frac{1}{3} \Delta_{\mu\rho} \corrD{\hWFV^\mu}{\h{\mathcal{C}}^\rho_\Upsilon}, \\
\bar{\mf{v}}^c_{\Upsilon k} = & \frac{3}{2} Q_{\mu\rho} \corrD{\hWFV^\mu}{\h{\mathcal{C}}^\rho_\Upsilon}, &
    v^c_{\Upsilon \epsilon} =& -\frac{(k\cdot u)}{k_\perp^2} u^\lambda\epsilon_{\lambda\mu\tau\rho}k_\perp^\tau \corrLTE{\hWFV^\mu}{\h{\mathcal{B}}^\rho_\Upsilon}, \\
v^c_{I u} = & 2 \corrLTE{u_\mu \hWFV^\mu}{\h{\mathcal{B}}_I}, &
    v^c_{I k} =& 2\frac{(k\cdot u)}{k_\perp^2} \corrLTE{k^\perp_\mu \hWFV^\mu}{\h{\mathcal{B}}_I},\\
\bar{\mf{v}}^c_{I-\Upsilon u} = & \frac{(k\cdot u)}{k_\perp^2} \corrD{u_\mu \hWFV^\mu}{k^\perp_\rho \h{\mathcal{C}}^\rho_{I-\Upsilon}}, &
    \bar{\mf{v}}^c_{I-\Upsilon \Delta} =& \frac{1}{3} \Delta_{\mu\rho} \corrD{\hWFV^\mu}{\h{\mathcal{C}}^\rho_{I-\Upsilon}},\\
\bar{\mf{v}}^c_{I-\Upsilon k} = & \frac{3}{2} Q_{\mu\rho} \corrD{\hWFV^\mu}{\h{\mathcal{C}}^\rho_{I-\Upsilon}}, &
    v^c_{I-\Upsilon \epsilon} = &-\frac{(k\cdot u)}{k_\perp^2} u^\lambda\epsilon_{\lambda\mu\tau\rho}k_\perp^\tau \corrLTE{\hWFV^\mu}{\h{\mathcal{B}}^\rho_{I-\Upsilon}}, \\
v^c_{I_S u} = & 2\corrLTE{u_\mu \hWFV^\mu}{\frac{k^\perp_\rho k^\perp_\sigma}{k_\perp^2}\h{\mathcal{B}}^{\rho\sigma}_{I_S}}, &
    v^c_{I_S \Delta} =& \frac{2(k\cdot u)}{3 k_\perp^2} \Delta_{\mu\rho} \corrLTE{\hWFV^\mu}{\h{\mathcal{B}}^{\rho\sigma}_{I_S} k^\perp_{\sigma}}, \\
v^c_{I_S k} = & \frac{3(k\cdot u)}{k_\perp^2} Q_{\mu\rho} \corrLTE{\hWFV^\mu}{\h{\mathcal{B}}^{\rho\sigma}_{I_S} k^\perp_{\sigma}}, &
    \bar{\mf{v}}^c_{I_S \epsilon} =& -\frac{(k\cdot u)^2}{2 (k_\perp^2)^2} u^\lambda\epsilon_{\lambda\mu\tau\rho}k_\perp^\tau \corrD{\hWFV^\mu}{\h{\mathcal{C}}^{\rho\sigma}_{I_S} k^\perp_{\sigma}}, \\
\bar{\mf{v}}^c_{\phi u} = & -\frac{1}{2} \corrD{u_\mu \hWFV^\mu}{\h{\mathcal{C}}_{\phi}}, &
    \bar{\mf{v}}^c_{\phi k} =& -\frac{ (k\cdot u)}{2k_\perp^2} \corrD{k^\perp_\mu \hWFV^\mu}{\h{\mathcal{C}}_{\phi}},
\end{align}\end{subequations}
and also
\begin{subequations}\begin{align}
v^c_{\rm{S12}\Delta} = & \frac{1}{3}\Delta_{\mu\sigma}\Delta_{\tau\rho}\corrLTE{\hWFV^\mu}{\h{\mathcal{B}}^{\tau,\rho\sigma}_{\rm S12}}, &
v^c_{\rm{S12}k} = & \frac{27}{4} Q_{\mu\sigma} Q_{\tau\rho} \corrLTE{\hWFV^\mu}{\h{\mathcal{B}}^{\tau,\rho\sigma}_{\rm S12}},\\
\bar{\mf{v}}^c_{\rm{S12}\Delta\epsilon} = & -\frac{3(k\cdot u)}{4 k_\perp^2}\Delta_{\mu\rho}\epsilon_{\tau\sigma\alpha\beta}
    k_\perp^\alpha u^\beta \corrD{\hWFV^\mu}{\h{\mathcal{C}}^{\tau,\rho\sigma}_{\rm S12}}, \\
\bar{\mf{v}}^c_{\rm{S12}\epsilon} = & -\frac{3(k\cdot u)k^\perp_\tau k_\perp^\alpha u^\beta k^\perp_\rho}{4(k_\perp)^4}
    \epsilon_{\mu\sigma\alpha\beta} \corrD{\hWFV^\mu}{\h{\mathcal{C}}^{\tau,\rho\sigma}_{\rm S12}},\\
v^c_{\rm{S13}\Delta} = & \frac{1}{3}\Delta_{\mu\sigma}\Delta_{\tau\rho}\corrLTE{\hWFV^\mu}{\h{\mathcal{B}}^{\tau,\rho\sigma}_{\rm S13}}, &
v^c_{\rm{S13}k} = & \frac{27}{4} Q_{\mu\sigma} Q_{\tau\rho} \corrLTE{\hWFV^\mu}{\h{\mathcal{B}}^{\tau,\rho\sigma}_{\rm S13}},\\
\bar{\mf{v}}^c_{\rm{S13}\Delta\epsilon} = & -\frac{3(k\cdot u)}{4 k_\perp^2}\Delta_{\mu\rho}\epsilon_{\tau\sigma\alpha\beta}
    k_\perp^\alpha u^\beta \corrD{\hWFV^\mu}{\h{\mathcal{C}}^{\tau,\rho\sigma}_{\rm S13}}, \\
\bar{\mf{v}}^c_{\rm{S13}\epsilon} = & -\frac{3(k\cdot u)k^\perp_\tau k_\perp^\alpha u^\beta k^\perp_\rho}{4(k_\perp)^4}
    \epsilon_{\mu\sigma\alpha\beta} \corrD{\hWFV^\mu}{\h{\mathcal{C}}^{\tau,\rho\sigma}_{\rm S13}}.
\end{align}\end{subequations}
As we might have expected, the gradients of spin potential are the only hydrodynamic fields capable of generating dissipative effects on the axial part of the Wigner function that do not require a breaking of parity.

%***********************************************************************************************************
%***********************************************************************************************************
%***********************************************************************************************************
\section{Results and discussion}
\label{sec:Results}
This section presents the results and discusses them.

%***********************************************************************************************************
\subsection{The axial part of Wigner function up to first order in hydrodynamic gradients}
The previous sections obtained all the linear contributions of the particle axial part of the Wigner function, which is connected to the spin polarization through eq.~(\ref{eq:SMass}). The decomposition and the definitions of the hydrodynamic fields that can induce polarization are given in eqs.~(\ref{eq:DecompDezeta})-(\ref{eq:DecompShear}) and the associated Kubo formula are explicitly written through section~\ref{sec:KuboA}. To present the results, all possible contributions are divided into four categories based on two properties: whether they are dissipative contributions and whether they require a breaking of the parity symmetry. In general, there are 13 non-dissipative and non-chiral possible contributions to the axial Wigner function:
\begin{subequations}\label{eq:DWFALTE}\begin{align}
\label{eq:DWFALTEzeta}
\Delta_{\rm LTE} \WFA_+^\mu = &
	\left[ a_{\zeta_A u} u^\mu + a_{\zeta_A k} \frac{k_\perp^\mu}{(k\cdot u)} \right] \zeta_A
    + a^c_{r \epsilon}\, \epsilon^{\mu\nu\rho\sigma} \frac{k^\perp_\nu u_\sigma}{(k\cdot u)} \de_{\ortu\rho}\zeta\\
\label{eq:DWFALTEvort}
   &-a_{\varpi} \frac{2\tilde\varpi^{\mu\nu}k_\nu}{(2\pi)^3}
    - (a_{w u} - a_{w \Delta}) \frac{(k\cdot w)}{(k\cdot u)} u^\mu\\
\label{eq:DWFALTEvort2}
    &+ (a_{\alpha\epsilon} - a_{w \Delta}) \epsilon^{\mu\nu\rho\sigma} \frac{k^\perp_\nu u_\sigma}{(k\cdot u)}\alpha_\rho + a_{w k} Q^{\mu\rho} w_\rho \\
\label{eq:DWFALTESP1}
   &-a_{\SP-\varpi} \frac{2(\tilde\SP^{\mu\nu}-\tilde\varpi^{\mu\nu})k_\nu}{(2\pi)^3}
    - (a_{\mf{w}-w u} - a_{\mf{w}-w \Delta}) \frac{k^\rho_\perp u^\mu}{(k\cdot u)} (\mf{w}-w)_\rho\\
\label{eq:DWFALTESP2}
   &+ (a_{\mf{a}-\alpha\epsilon} - a_{\mf{w}-w \Delta}) \epsilon^{\mu\nu\rho\sigma} \frac{k^\perp_\nu u_\sigma}{(k\cdot u)} (\mf{a}-\alpha)_\rho
   + a_{\mf{w}-w k}  Q^{\mu\rho} (\mf{w}-w)_\rho \\
\label{eq:DWFALTEshear}
   & + a_{q \epsilon}\, \epsilon^{\mu\nu\rho\sigma} \frac{k^\perp_\nu u_\sigma}{(k\cdot u)} (\beta Du_{\rho}+\de_\rho\beta) + a_{\sigma\epsilon} \epsilon^{\mu\nu\alpha\rho}k_\perp^{\sigma}\frac{u_\nu k_\alpha}{(k\cdot u)} \beta\sigma_{\rho\sigma},
\end{align}\end{subequations}
and 15 non-dissipative and chiral coefficients
\begin{subequations}\label{eq:DWFALTEChi}\begin{align}
\Delta_{{\rm LTE},\chi} \WFA_+^\mu = &
    \mf{a}_{r_A \epsilon}\, \epsilon^{\mu\nu\rho\sigma} \frac{k^\perp_\nu u_\sigma}{(k\cdot u)} \de_{\ortu\rho}\zeta_A\\
   &+ \left[ \mf{a}_{f u} \frac{k_\perp^\rho u^\mu}{(k\cdot u)}
    + \mf{a}_{f \Delta} \Delta^{\mu\rho}
    + \mf{a}_{f k} Q^{\mu\rho} \right] f_{\rho}
    + \mf{a}_{\Upsilon \epsilon}\, \epsilon^{\mu\nu\rho\sigma} \frac{k^\perp_\nu u_\sigma}{(k\cdot u)}
    \Upsilon_{\rho} \\
   &+ \left[ \mf{a}_{I u} u^\mu + \mf{a}_{I k} \frac{k_\perp^\mu}{(k\cdot u)} \right] I
    + \mf{a}_{I-\Upsilon \epsilon}\, \epsilon^{\mu\nu\rho\sigma} \frac{k^\perp_\nu u_\sigma}{(k\cdot u)} (I_\rho-\Upsilon_{\rho})\\
   &+ \left[ \mf{a}_{I_S u} \frac{k_\perp^\rho k_\perp^\sigma}{k_\perp^2} u^\mu
    + \mf{a}_{I_S \Delta} \frac{\Delta^{\mu\rho} k_\perp^{\sigma}}{(k\cdot u)} 
    + \mf{a}_{I_S k} \frac{Q^{\mu\rho} k_\perp^{\sigma}}{(k\cdot u)} \right] I_{S\,\rho\sigma} \\
   &+ \left[ \mf{a}_{\rm{S12}\Delta} \Delta^{\tau\rho} \Delta^{\mu\sigma}
    + \mf{a}_{\rm{S12}k} Q^{\tau\rho}\Delta^{\mu\sigma} \right] \Phi^{S12}_{\tau,\rho\sigma}\\
   &+ \left[ \mf{a}_{\rm{S13}\Delta} \Delta^{\tau\sigma} \Delta^{\mu\rho}
    + \mf{a}_{\rm{S13}k} Q^{\tau\sigma}\Delta^{\mu\rho} \right] \Phi^{S13}_{\tau,\rho\sigma}.
\end{align}\end{subequations}
Regarding dissipative contributions, 19 non-chiral possible contributions were found
\begin{subequations}\label{eq:DWFADiss}\begin{align}
\Delta_{\rm D} \WFA_+^\mu = &
    \left[ \bar{a}_{D\zeta_A u} u^\mu + \bar{a}_{D\zeta_A k} \frac{k_\perp^\mu}{(k\cdot u)} \right] D\zeta_A 
    + \left[ \bar{a}_{r_A u} \frac{k_\perp^\rho u^\mu}{(k\cdot u)}
    	+ \bar{a}_{r_A \Delta} \Delta^{\mu\rho}
    	+ \bar{a}_{r_A k} Q^{\mu\rho} \right] \de_{\ortu\rho}\zeta_A\\
    &+ \bar{a}_{f \epsilon}\, \epsilon^{\mu\nu\rho\sigma} \frac{k^\perp_\nu u_\sigma}{(k\cdot u)} f_\rho
     + \left[ \bar{a}_{\Upsilon u} \frac{k_\perp^\rho u^\mu}{(k\cdot u)}
              + \bar{a}_{\Upsilon \Delta} \Delta^{\mu\rho}
              + \bar{a}_{\Upsilon k} Q^{\mu\rho} \right] \Upsilon_{\rho} \\
   &+ \left[ \bar{a}_{I-\Upsilon u} \frac{k_\perp^\rho u^\mu}{(k\cdot u)}
    + \bar{a}_{I-\Upsilon \Delta} \Delta^{\mu\rho}
    + \bar{a}_{I-\Upsilon k} Q^{\mu\rho} \right] (I_\rho-\Upsilon_{\rho}) \\
   &+ \left[ \bar{a}_{\phi u} u^\mu + \bar{a}_{\phi k} \frac{k_\perp^\mu}{(k\cdot u)} \right] \phi
    + \bar{a}_{I_S \epsilon}\, \epsilon^{\mu\nu\alpha\rho} k_\perp^{\sigma} \frac{u_\nu k^\perp_\alpha}{(k\cdot u)^2} I_{S\,\rho\sigma}\\
   &+ \left[ \bar{a}_{\rm{S12}\Delta\epsilon} \Delta^{\mu\tau}\epsilon^{\lambda\nu\rho\sigma}\frac{u_\lambda k^\perp_\nu}{(k\cdot u)}
    + \bar{a}_{\rm{S12}\epsilon} \epsilon^{\mu\nu\rho\sigma}\frac{k_\perp^{\tau}u_\nu}{(k\cdot u)} \right] \Phi^{S12}_{\tau,\rho\sigma}\\
   &+ \left[ \bar{a}_{\rm{S13}\Delta\epsilon} \Delta^{\mu\tau}\epsilon^{\lambda\nu\rho\sigma}\frac{u_\lambda k^\perp_\nu}{(k\cdot u)}
    + \bar{a}_{\rm{S13}\epsilon} \epsilon^{\mu\nu\rho\sigma}\frac{k_\perp^{\tau}u_\nu}{(k\cdot u)} \right] \Phi^{S13}_{\tau,\rho\sigma},
\end{align}\end{subequations}
and 19 possible chiral contributions
\begin{subequations}\label{eq:DWFADissChi}\begin{align}
\Delta_{{\rm D},\chi} \WFA_+^\mu = &
     \left[ \bar{\mf{a}}^c_{D\zeta u} u^\mu + \bar{\mf{a}}^c_{D\zeta k} \frac{k_\perp^\mu}{(k\cdot u)} \right] D\zeta
    + \left[ \bar{\mf{a}}^c_{r u} \frac{k_\perp^\rho u^\mu}{(k\cdot u)}
    	+ \bar{\mf{a}}^c_{r \Delta} \Delta^{\mu\rho}
    	+ \bar{\mf{a}}^c_{r k} Q^{\mu\rho} \right] \de_{\ortu\rho}\zeta\\
   &+ \bar{\mf{a}}_{\mf{a}-\alpha \Delta} \Delta^{\mu\rho} (\mf{a}-\alpha)_\rho
    - \bar{\mf{a}}_{\mf{a}-\alpha u} \frac{k^\rho_\perp u^\mu}{(k\cdot u)} (\mf{a}-\alpha)_\rho
    +\bar{\mf{a}}_{\mf{w}-w \epsilon} \epsilon^{\mu\nu\rho\sigma} \frac{k^\perp_\nu u_\sigma}{(k\cdot u)} (\mf{w}-w)_\rho \\
   &+ \bar{\mf{a}}_{\mf{a}-\alpha k} Q^{\mu\rho} (\mf{a}-\alpha)_\rho
    + \left[ \bar{\mf{a}}_{D\beta u} u^\mu + \bar{\mf{a}}_{D\beta k} \frac{k_\perp^\mu}{(k\cdot u)} \right] D\beta
    + \left[ \bar{\mf{a}}_{\theta u} u^\mu + \bar{\mf{a}}_{\theta k} \frac{k_\perp^\mu}{(k\cdot u)} \right] \beta\theta \\
   &+ \left[ \bar{\mf{a}}_{q u} \frac{k_\perp^\rho u^\mu}{(k\cdot u)}
    + \bar{\mf{a}}_{q \Delta} \Delta^{\mu\rho}
    + \bar{\mf{a}}_{q k} Q^{\mu\rho} \right] (\beta Du_{\rho}+\de_\rho\beta)
    + \bar{\mf{a}}_{\sigma u}(k\cdot u) \frac{k_\perp^\rho k_\perp^\sigma}{k_\perp^2}u^\mu\,\beta\sigma_{\rho\sigma}\\
   &+ \bar{\mf{a}}_{\sigma \Delta}\Delta^{\mu\rho} k_\perp^\sigma \beta\sigma_{\rho\sigma}
    + \bar{\mf{a}}_{\sigma k} Q^{\mu\rho} k_\perp^{\sigma}\,\beta\sigma_{\rho\sigma}.
\end{align}\end{subequations}

The non-dissipative, non-chiral contributions given in eq.~(\ref{eq:DWFALTE}) have already been obtained in previous works. The first terms in eq.~(\ref{eq:DWFALTEzeta}) describe a spin polarization induced by a chiral imbalance and, for a free field, is given by~\cite{Becattini:2020xbh}
\begin{equation}
a_{\zeta_A k}^{\rm Free} = \frac{g_A}{2}\frac{2(k\cdot u)}{(2\pi)^3}\delta(k^2-m^2)\theta(k\cdot u) n_F(\beta\cdot k)\left(1-n_F(\beta\cdot k)\right)=-\frac{(k\cdot u)^2}{k_\perp^2}a_{\zeta_A u}^{\rm Free},
\end{equation}
where $g_A$ is the axial charge of the baryon species. Taking advantage of this effect, it is possible to probe the presence of a chiral imbalance in the QGP by measuring the correlation between helicities of particles emitted in the same event~\cite{Becattini:2020xbh,Du:2008zzb}. The last terms in eq.~(\ref{eq:DWFALTEzeta}) is the spin Hall effect~\cite{Liu:2020dxg,Buzzegoli:2022qrr} and for a non-interacting Dirac field is given by:
\begin{equation}
\label{eq:SHEcond}
a_{r\epsilon}^{c{\rm Free}} = -\frac{2\delta(k^2-m^2)\theta(k\cdot u)}{(2\pi)^3} n_F(\beta\cdot k)\left(1-n_F(\beta\cdot k)\right),
\end{equation}
and we will discuss this effect further in the next section. The contributions induced by thermal vorticity are given in the lines~(\ref{eq:DWFALTEvort}) and (\ref{eq:DWFALTEvort2}) and are the only one surviving at global equilibrium. For a free field they were first obtained in~\cite{Becattini:2013fla}, and they result in
\begin{subequations}
\begin{align}
a^{\rm Free}_{\varpi} =& \delta(k^2-m^2)\theta(k\cdot u) n_F(\beta\cdot k)\left(1-n_F(\beta\cdot k)\right),\\
a^{\rm Free}_{w u} =& a^{\rm Free}_{w \Delta},\quad
a^{\rm Free}_{\alpha\epsilon} = a^{\rm Free}_{w \Delta},\quad
a^{\rm Free}_{w k} =0.
\end{align}\end{subequations}
It is interesting to note that among all the four possible terms in eqs.~(\ref{eq:DWFALTEvort}) and (\ref{eq:DWFALTEvort2}), only the first one is non-vanishing for non-interacting particles. Interactions can break the degeneracy of the coefficients and give rise to the additional contributions. These additional effects and their Kubo formulas were first obtained in~\cite{Liu:2021uhn}.
The contributions coming from the spin potential are given in eqs.~(\ref{eq:DWFALTESP1}) and~(\ref{eq:DWFALTESP2}). In contrast to the previous coefficients, those related to the spin potential depend on the pseudo-gauge chosen to express the energy-momentum tensor and the spin tensor in the statistical operator. For a non-interacting Dirac field, choosing the Canonical spin tensor, one obtains~\cite{Buzzegoli:2021wlg,Liu:2021nyg}
\begin{equation}\begin{split}
\Delta_{\SP-\varpi,{\rm LTE}}\WFA^\mu_+ =& \epsilon^{\lambda\rho\sigma\tau} u_\tau \frac{\Delta^\mu_\tau k^2-k^\mu k_\tau}{(k\cdot u)}(\SP_{\rho\sigma}-\varpi_{\rho\sigma})\\
&\times\frac{\delta(k^2-m^2)\theta(k\cdot u)}{(2\pi)^3} n_F(\beta\cdot k)\left(1-n_F(\beta\cdot k)\right),
\end{split}
\end{equation}
from which one can obtain the different coefficients. Lastly, the thermal shear induced polarization is given in eq.~(\ref{eq:DWFALTEshear}). These coefficients are also pseudo-gauge dependent~\cite{Buzzegoli:2021wlg}, and they were first obtained in~\cite{Liu:2021uhn,Becattini:2021suc}. For a free Dirac field the result is the same in the Canonical and the Belinfante pseudo-gauge:
\begin{equation}
a_{\sigma\epsilon}^{\rm Free} = \frac{2\delta(k^2-m^2)\theta(k\cdot u)}{(2\pi)^3} n_F(\beta\cdot k)\left(1-n_F(\beta\cdot k)\right) = \frac{2 a_{q\epsilon}^{\rm Free}}{(k\cdot u)} ,
\end{equation}
instead in the de Groot–van Leeuwen–van Weert or the Hilgevoord-Wouthuysen pseudo-gauge the coefficients are vanishing~\cite{Buzzegoli:2021wlg}. We stress again that despite these results were already known, we showed that at first order in gradients these are all the possible contributions and the coefficients are given by Kubo formulas valid for any quantum field theory. We leave for future studies the possible phenomenological consequences of considering the spin polarization in an interacting fluid.

To the best of our knowledge, the non-dissipative chiral contributions to spin polarization are given in eq.~(\ref{eq:DWFALTEChi}) for the first time. The first term is the chiral analogue of the spin Hall effect and the other are induced by the gradient of the spin potential. Moreover, this is also the first time that the dissipative contributions of spin polarization, (\ref{eq:DWFADiss}) and (\ref{eq:DWFADissChi}), have been identified. It is important to stress that dissipative effects induced by thermal shear, which include the expansion rate, the shear tensor, gradients of temperature, and of fluid velocity, are only possible in the presence of parity symmetry breaking, either through interactions or by the presence of chiral imbalance. For this reason one expect these contributions to be small and therefore not of phenomenological relevance. The only effects possible without parity breaking or a gradient of an axial chemical potential are those induced by the gradient of spin potential in eq.~(\ref{eq:DWFADiss}). Among the non-dissipative coefficients, only the one related to the gradients of the vector and axial chemical potential are independent of the pseudo-gauge chosen.

%***********************************************************************************************************
\subsection{Chiral spin Hall effect}
\label{sec:CSHE}
\begin{figure}[tbph]
\centering
\includegraphics[width=0.9\textwidth]{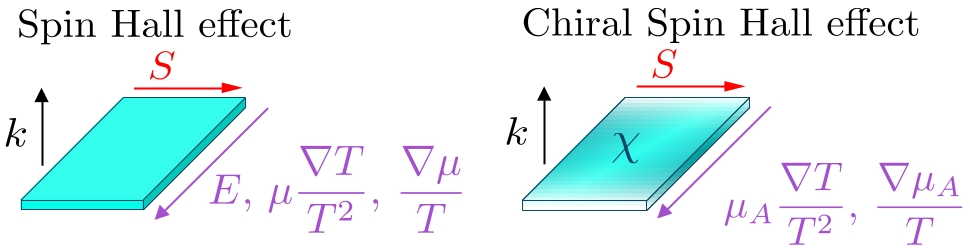}\\
\includegraphics[width=0.9\textwidth]{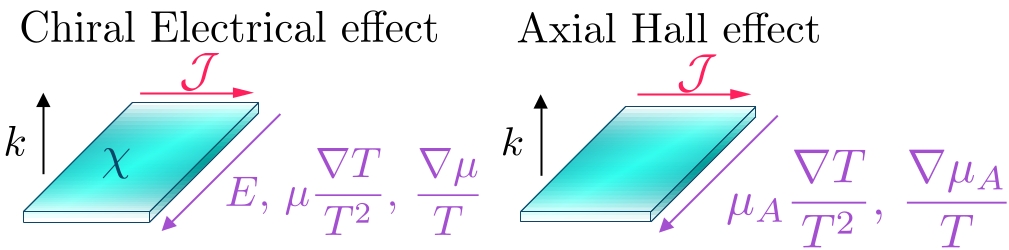}
\caption{The figure illustrates the effects similar to the spin Hall effect (SHE) for the spin polarization of a particle with momentum $k$ in the $z$ direction. In purple along the $y$ direction, the hydrodynamic fields inducing the effects. The effects on the top part induce a contribution to the axial part of the Wigner function, which is related to spin polarization $S$. The effects on the bottom generate terms in the vector part of the Wigner function $\mathcal{J}$, which is related to spin polarization of massless fields. The symbol $\chi$ in the material indicates that the coefficient related to the effect require a breaking of parity. Replacing $k$ with a magnetic field $B$, the hydrodynamic fields with an electric field $E$, and $S$ and $\mathcal{J}$ with an electric potential, we obtain the classical Hall effect.}
\label{fig:CSHE}
\end{figure}
Within the non-dissipative effects we discuss in more depth the analogues of the spin Hall effect (SHE). The SHE describe a spin polarization induced by the gradient of $\zeta$. Since in a system embedded in an external electromagnetic field the thermal global equilibrium is reached if the comoving electric field equals the gradient of $\zeta$, that is $\de^\mu\zeta(x)=F^{\nu\mu}\beta_\nu$, a spin polarization is induced in the same way by an electric field. In heavy-ion collisions the presence of the SHE can be detected by looking at the azimuthal angle behavior of the difference between the local spin polarization of $\Lambda$s and $\bar{\Lambda}$s~\cite{Liu:2020dxg}. Indeed, the SHE conductivity $a^c_{r\epsilon}$ is a charged coefficients and give opposite effect for the $\Lambda$ and the $\bar{\Lambda}$. The analysis carried out in this work revealed that the gradients of $\zeta$ and $\zeta_A$ can induce similar effects in both the axial and vector part of the Wigner function. For the axial part of Wigner function we found that they induce the following terms
\begin{equation*}
\Delta_{\rm SHE} \mathcal{A}_+^\mu(x,k)=\epsilon^{\mu\nu\rho\sigma}\frac{k^\perp_\nu u_\sigma}{(k\cdot u)}\left[ a^c_{r\epsilon}(k)\de_\rho\zeta + \mathfrak{a}_{r_A\epsilon}(k)\de_\rho\zeta_A \right].
\end{equation*}
These are relevant for the spin polarization of a massive field, see eq.~(\ref{eq:SMass}). We call the axial Wigner function induced by $\de_\rho\zeta_A$ the chiral spin Hall effect (CSHE), its corresponding coefficients is chiral and is sourced by gradient of the axial chemical potential. Similarly to the spin polarization induced by the axial chemical potential in eq.~(\ref{eq:DWFALTEzeta}), these effect could be used to detect anisotropies of the axial chemical potential, deriving from the topological states of the  QCD vacuum, by looking at the local spin polarization or helicity. For a massless field, also the vector part of the Wigner function play a role in spin polarization, see eq.~(\ref{eq:SMassless}). The gradients of the chemical potentials induce the following terms:
\begin{equation*}
\Delta_{\rm SHE} \mathcal{V}_+^\mu(x,k)=\epsilon^{\mu\nu\rho\sigma}\frac{k^\perp_\nu u_\sigma}{(k\cdot u)}\left[ \mathfrak{v}_{r\epsilon}(k)\de_\rho\zeta + v^c_{r_A\epsilon}(k)\de_\rho\zeta_A \right].
\end{equation*}
Similarly to the chiral magnetic effect, we call the vector part of the Wigner function induced by the gradient of $\zeta$, or by an electric field, the chiral electrical effect (CEE). The corresponding coefficient $\mathfrak{v}_{r\epsilon}$ is chiral. We call the term induced by $\de_\rho\zeta_A$ the axial Hall effect (AHE), whose coefficient $v^c_{r_A\epsilon}$ is not chiral, but it requires a non-homogeneous chiral imbalance to generate the effect. The SHE is the only effect that does not require a breaking of parity. Instead, all of them are non-dissipative and require the fluid to be out-of-equilibrium. A summary of the effects is given in figure~\ref{fig:CSHE}.
To obtain a spin current the corrections to the axial Wigner function must be plugged in the expression for the spin vector~(\ref{eq:SMass}) obtaining in the rest frame of the system:
\begin{equation}
\vec{S}(k) = \vec{k}\times \left\langle \left\langle  a_{r \epsilon}^c
\left(\frac{\vec{\nabla}\mu}{T} + \mu\vec{\nabla}\frac{1}{T} \right) \right\rangle\right\rangle + \vec{k}\times \left\langle \left\langle  \mathfrak{a}_{r_A\epsilon}
\left(\frac{\vec{\nabla}\mu_A}{T} + \mu_A\vec{\nabla}\frac{1}{T} \right) \right\rangle\right\rangle .
\end{equation}
In the context of spintronics~\cite{Maekawa2017-um}, the CSHE refer to the generation of a spin current induced by the inherent chirality of the material—stemming from structural properties~\cite{AdvancementInChiralSpintronics} (like the chiral induced Spin Selectivity Effect (CISS)~\cite{ChiralInducedSpinSelectivity} and the Quantum Spin Hall Effect (QSHE)~\cite{QSHE}), magnetic configurations~\cite{Zhang_2018} (like the Topological Spin Hall Effect (TSHE)~\cite{THE}, or atomic vibrations (chiral phonons)~\cite{Kim2023}. The CSHE has been observed in diverse materials including HgTe quantum wells~\cite{Konig2007} , van der Waals materials like WTe2~\cite{Wu2018} , chiral organic molecules such as DNA~\cite{ChiralInducedSpinSelectivity}, ferromagnetic metals with static skyrmion lattices~\cite{Kolincio2021} , and 2D hybrid organic-inorganic perovskites~\cite{Kim2023}.

The values of these coefficients can be calculated using the Kubo formulas in eqs. (\ref{eq:SHEKf}) and (\ref{eq:CSHEKf}). In a fluid with a chiral imbalance $\zeta_A(x)$ and a chemical potential $\zeta(x)$, these coefficients have been computed in appendix~\ref{App:ChiralSHE} for a massless non-interacting Dirac field or equivalently for the high temperature limit $T\gg m$. The results are, for the SHE and the CSHE:
\begin{align}
a^c_{r\epsilon}=&-\frac{\delta(k^2)\theta(k\cdot u)}{(2\pi)^3}\left[n^R_F(x,k)(1-n^R_F(x,k)) + n^L_F(x,k)(1-n^L_F(x,k))\right],\\
\mathfrak{a}_{r_A\epsilon}=&-\frac{\delta(k^2)\theta(k\cdot u)}{(2\pi)^3}\left[n^R_F(x,k)(1-n^R_F(x,k)) - n^L_F(x,k)(1-n^L_F(x,k))\right],
\end{align}
and for the CEE and the AHE:
\begin{align}
\mathfrak{v}_{r\epsilon}=&-\frac{\delta(k^2)\theta(k\cdot u)}{(2\pi)^3}\left[n^R_F(x,k)(1-n^R_F(x,k)) - n^L_F(x,k)(1-n^L_F(x,k))\right],\\
v^c_{r_A\epsilon}=&-\frac{\delta(k^2)\theta(k\cdot u)}{(2\pi)^3}\left[n^R_F(x,k)(1-n^R_F(x,k)) + n^L_F(x,k)(1-n^L_F(x,k))\right],
\end{align}
where
\begin{equation}
\label{eq:nfchiral}
n^{\chi}_F(x,k) = \frac{1}{\e^{\beta(x)\cdot k -\zeta(x)-\chi\zeta_A(x)}+1},\quad
\chi=\begin{cases}
+1 & R\\
-1 & L\\
\end{cases}\, .
\end{equation}
We note that in the absence of chiral imbalance the SHE conductivity reduces to the know value in eq.~(\ref{eq:SHEcond}), and that the chiral coefficients $\mathfrak{a}_{r_A\epsilon}$ and $\mathfrak{v}_{r\epsilon}$ and the non-chiral coefficients $a^c_{r\epsilon}$ and $v^c_{r_A\epsilon}$ have the same values. For the chiral coefficients we can expect that their values are connected to the chiral anomaly. An other question is whether these effects survives in the axial and vector currents obtained by integration the Wigner function over the momentum $k$. In the local rest frame, by noticing that the coefficients depend only on the modulus $|\bm{k}|$ and that the effects are odd in $k_\perp$, we conclude that these effect do not generate a net axial and vector current, that is
\begin{equation}
\Delta_{\rm SHE} \wj^\mu_A(x) =\int \di^4 k \, \Delta_{\rm SHE} \mathcal{A}^\mu(x,k) = 0,\quad
\Delta_{\rm SHE} \wj^\mu(x) =\int \di^4 k \, \Delta_{\rm SHE} \mathcal{V}^\mu(x,k) = 0\, .
\end{equation}
Similarly, these effects can only be detected in the local spin polarization, that is as a function of momentum, and do not appear in the global spin polarization, the one averaged over the momenta of all particles.

%********************************************************************************************************
\subsection{Spin polarization of a free massless fermion}
Now, consider the spin polarization of a massless fermion given by eq.~(\ref{eq:SMassless}).
At first order in gradient expansion the spin polarization becomes eq.~(\ref{eq:SMasslessExp}) or in compact notation
\begin{equation}
\label{eq:MasslessSpinPolExpand}
S^\mu(|\bm{k}|,\bm{k}) = \frac{k^\mu}{2}\left[\frac{A_0}{V_0}\left(1 -\frac{\Delta V}{V_0}\right)
    +\frac{\Delta A}{V_0}+\cdots \right],
\end{equation}
where $A_0$ and $\Delta A$ are the leading order and the first corrections of the gradient expansion:
\begin{equation}
A(q) = \int_\Sigma \di\Sigma\cdot k\, q_\alpha\WFA_+^\alpha(x,k) = A_0 + \Delta A,
\end{equation}
and similarly for $V$. We notice that, exactly as for the massive case, the first order dissipative corrections to the axial part of the Wigner function require a breaking of parity. Therefore, the dissipative contribution to the massless spin polarization coming from $\Delta A$ also requires a breaking of parity. Instead, $\Delta V$ can have non chiral dissipative corrections, see eq.~(\ref{eq:DWFVDiss}). However, at first order, the dissipative corrections to the massless spin polarization coming from $\Delta V$ are proportional to the ratio of $A_0$ and $V_0$, and in order to have a non-vanishing $A_0$, a violation of parity symmetry is needed. In conclusion, also for the massless case, first order dissipative corrections are allowed only by breaking parity.

As the spin polarization of a massless fermion with the correct formulas~(\ref{eq:SMassless}) and (\ref{eq:SMasslessExp}) has only been studied at global equilibrium~\cite{Palermo:2023cup}, we compute some contributions for a free Dirac field in presence of chiral imbalance~\cite{Buzzegoli:2020ycf,Buzzegoli:2018wpy}. The leading order Wigner function evaluated with the statistical operator~(\ref{eq:SOeqmuA}) is
\begin{equation}
\mean{\h{W}_+(x,k)}_\beta = \left(\frac{1+\gamma^5}{2} n_F^R(x,k) + \frac{1-\gamma^5}{2} n_F^L(x,k)\right)
    \gamma^\nu k_\nu \frac{\delta(k^2)\theta(k\cdot u)}{(2\pi)^3},
\end{equation}
with the distribution functions given in eq. (\ref{eq:nfchiral}),
from which it follows that the leading order axial and vector parts of the Wigner function are
\begin{subequations}\label{eq:V0andA0}\begin{align}
%\mean{\h{\mathcal{R}}_+^\mu}_\beta =&\frac{4\delta(k^2)\theta(k\cdot u)}{(2\pi)^3}\frac{n_F^R(x,k)}{2}k^\mu ,&
\mean{\h{\mathcal{V}}_+^\mu}_\beta =&\frac{4\delta(k^2)\theta(k\cdot u)}{(2\pi)^3}
    \frac{n_F^R(x,k)+n_F^L(x,k)}{2}k^\mu ,\\
%\mean{\h{\mathcal{L}}_+^\mu}_\beta =&\frac{4\delta(k^2)\theta(k\cdot u)}{(2\pi)^3}\frac{n_F^L(x,k)}{2}k^\mu ,&
\mean{\h{\mathcal{A}}_+^\mu}_\beta =&\frac{4\delta(k^2)\theta(k\cdot u)}{(2\pi)^3}
    \frac{n_F^R(x,k)-n_F^L(x,k)}{2}k^\mu.
\end{align}\end{subequations}
To write down the expression we need to choose a vector $q$, such that $q^2=0$ and $q\cdot k\neq 0$. Denoting with $\mf{p}=(\kappa,0,0,\kappa)$ the standard form and with $[k]$ the standard Lorentz transformation of $k=(|\bm{k}|,\bm{k})$, such that $k^\mu=[k]^\mu_{\,\nu}\mf{p}^\nu$, a convenient choice for $q$ is $q^\mu=\bar{k}^\mu=[k]^\mu_{\,\nu}\bar{\mf{p}}^\nu$ where
$\mf{p}=(\kappa,0,0,-\kappa)$. In this way we have the decomposition $k^\mu = (k\cdot u)u^\mu + k_\perp^\mu$ and $\bar{k}^\mu = (k\cdot u)u^\mu - k_\perp^\mu$. With this choice and from eqs.~(\ref{eq:V0andA0}) and~(\ref{eq:MasslessSpinPolExpand}), we readily obtain
\begin{equation}
S^\mu_0(|\bm{k}|,\bm{k}) =  \frac{k^\mu}{2} \frac{A_0}{V_0}
=  \frac{k^\mu}{2} \frac{\int_\Sigma \di\Sigma\cdot k\, \left[n_F(\beta\cdot k - \zeta-\zeta_A)- n_F(\beta\cdot k - \zeta+\zeta_A) \right]}{\int_\Sigma \di\Sigma\cdot k\, \left[n_F(\beta\cdot k - \zeta-\zeta_A) + n_F(\beta\cdot k - \zeta+\zeta_A) \right]},
\end{equation}
which vanish in absence of parity breaking, i.e. $\zeta_A=0$.
We now consider the first order contribution induced by thermal vorticity; for the axial and vector part of Wigner function of a free field, we found:
\begin{equation}
\Delta_\varpi \WFA_+^\mu = -a^{\rm Free}_{\varpi} \frac{2\tilde\varpi^{\mu\nu}k_\nu}{(2\pi)^3},\quad
\Delta_\varpi \WFV_+^\mu = -\mf{v}^{c{\rm Free}}_{\varpi} \frac{2\tilde\varpi^{\mu\nu}k_\nu}{(2\pi)^3},
\end{equation}
where
\begin{align}
a^{\rm Free}_{\varpi} =& \delta(k^2)\theta(k\cdot u) 
    \frac{n_F^R(\beta\cdot k)\left(1-n_F^R(\beta\cdot k)\right)
        +n_F^L(\beta\cdot k)\left(1-n_F^L(\beta\cdot k)\right)}{2},\\
\mf{v}^{c{\rm Free}}_{\varpi} =& \delta(k^2)\theta(k\cdot u) 
    \frac{n_F^R(\beta\cdot k)\left(1-n_F^R(\beta\cdot k)\right)
        -n_F^L(\beta\cdot k)\left(1-n_F^L(\beta\cdot k)\right)}{2}.
\end{align}
Choosing $q=\bar{k}$ and defining
\begin{equation}
H = \frac{\epsilon^{\mu\nu\rho\sigma}\bar{k}_\mu k_\nu \varpi_{\rho\sigma}}{2(k\cdot\bar{k})},
\end{equation}
we obtain
\begin{equation}
\begin{split}
\Delta_\varpi S^\mu(|\bm{k}|,\,\bm{k}) =& \frac{k^\mu}{2}\frac{\int_\Sigma \di\Sigma\cdot k\,\frac{H}{2}\left(n_F^R\left(1-n_F^R\right)+n_F^L\left(1-n_F^L\right)\right)}{\int_\Sigma \di\Sigma\cdot k\,\left(n_F^R+n_F^L\right)}\\
& +\frac{k^\mu}{2}\frac{A_0}{V_0}\frac{\int_\Sigma \di\Sigma\cdot k\,\frac{H}{2}\left(n_F^R\left(1-n_F^R\right)-n_F^L\left(1-n_F^L\right)\right)}{\int_\Sigma \di\Sigma\cdot k\, \left(n_F^R+n_F^L\right)}.
\end{split}
\end{equation}
This expression at global equilibrium with $\zeta_A=0$ reduces to
\begin{equation}
\Delta_\varpi S^\mu(|\bm{k}|,\,\bm{k}) = -\frac{k^\mu}{2}\frac{H}{2}\left(1-n_F\right)
+\mathcal{O}\left(\varpi^2\right)
\end{equation}
in accordance with~\cite{Palermo:2023cup}.

Other contributions can be obtained in a similar fashion. However, one must be careful on how to choose $q$ as it might be orthogonal to terms contributing to the axial and vector parts of the Wigner function. For instance the spin Hall effect term contracted with $q=\bar{k}$ is vanishing: $\epsilon^{\mu\nu\rho\sigma} \bar{k}_\mu k^\perp_\nu u_\sigma=\epsilon^{\mu\nu\rho\sigma}\bar{k}^\perp_\mu k^\perp_\nu u_\sigma = -\epsilon^{\mu\nu\rho\sigma}k^\perp_\mu k^\perp_\nu u_\sigma=0$.
In these cases, it is more convenient to choose $q^\mu=\tilde{k}^\mu=[k]^\mu_{\,\nu}\tilde{\mf{p}}^\nu$ with $\tilde{\mf{p}}=(\kappa,\kappa,0,0)$, which has the decomposition $\tilde{k}^\mu = (k\cdot u)u^\mu + \tilde{k}_\perp^\mu$, with $\tilde{k}_\perp\cdot k_\perp = 0$. With this choice and with the results of previous section we obtain:
\begin{equation}
\begin{split}
\Delta_{\rm SHE} S^\mu&(|\bm{k}|,\,\bm{k}) =
-\frac{k^\mu}{4}\epsilon^{\alpha\nu\rho\sigma}\tilde{k}_\alpha k_\nu\\ &\times
\frac{\int_\Sigma \frac{\di\Sigma\cdot k}{(k\cdot u)}\,u_\sigma \left[n_F^R\left(1-n_F^R\right)\left(1-\frac{A_0}{V_0}\right)+n_F^L\left(1-n_F^L\right)\left(1+\frac{A_0}{V_0}\right)\right]\de_\rho\zeta}{\int_\Sigma \di\Sigma\cdot k\, (\tilde{k}\cdot k)  \left(n_F^R+n_F^L\right)}\\
& -\frac{k^\mu}{4}\epsilon^{\alpha\nu\rho\sigma}\tilde{k}_\alpha k_\nu
\frac{\int_\Sigma \frac{\di\Sigma\cdot k}{(k\cdot u)}\,u_\sigma \left[n_F^R\left(1-n_F^R\right)\left(1-\frac{A_0}{V_0}\right)-n_F^L\left(1-n_F^L\right)\left(1+\frac{A_0}{V_0}\right)\right]\de_\rho\zeta_A }{\int_\Sigma \di\Sigma\cdot k\, (\tilde{k}\cdot k)  \left(n_F^R+n_F^L\right)}.
\end{split}
\end{equation}
%

%********************************************************************************************************
\subsection{The vector part of Wigner function up to first order in hydrodynamic gradients}
The results for the vector part of the Wigner function are also reported.
We found that in general there are 15 non-dissipative and non-chiral possible contributions to the vector Wigner function:
\begin{subequations}\label{eq:DWFVLTE}\begin{align}
\Delta_{\rm LTE} \WFV_+^\mu = &
	 v_{r_A \epsilon}^c\, \epsilon^{\mu\nu\rho\sigma} \frac{k^\perp_\nu u_\sigma}{(k\cdot u)} \de_{\ortu\rho}\zeta_A
+\left[ v^c_{f u} \frac{k_\perp^\rho u^\mu}{(k\cdot u)} + v^c_{f \Delta} \Delta^{\mu\rho} + v^c_{f k} Q^{\mu\rho} \right] f_{\rho} \\
&+ v^c_{\Upsilon \epsilon}\, \epsilon^{\mu\nu\rho\sigma} \frac{k^\perp_\nu u_\sigma}{(k\cdot u)} \Upsilon_{\rho}
+\left[ v^c_{I u} u^\mu + v^c_{I k} \frac{k_\perp^\mu}{(k\cdot u)} \right] I
+ v^c_{I-\Upsilon \epsilon}\, \epsilon^{\mu\nu\rho\sigma} \frac{k^\perp_\nu u_\sigma}{(k\cdot u)}  (I_\rho-\Upsilon_{\rho})\\
&+\left[ v^c_{I_S u} \frac{k_\perp^\rho k_\perp^\sigma}{k_\perp^2} u^\mu + v^c_{I_S \Delta} \frac{\Delta^{\mu\rho} k_\perp^{\sigma}}{(k\cdot u)} + v^c_{I_S k} \frac{Q^{\mu\rho} k_\perp^{\sigma}}{(k\cdot u)} \right] I_{S\,\rho\sigma}\\
& +\left[ v^c_{\rm{S12}\Delta} \Delta^{\tau\rho} \Delta^{\mu\sigma} + v^c_{\rm{S12}k} Q^{\tau\rho}\Delta^{\mu\sigma}\right] \Phi^{S12}_{\tau,\rho\sigma}\\
&+\left[ v^c_{\rm{S13}\Delta} \Delta^{\tau\sigma} \Delta^{\mu\rho} + v^c_{\rm{S13}k} Q^{\tau\sigma}\Delta^{\mu\rho} \right] \Phi^{S13}_{\tau,\rho\sigma},
\end{align}\end{subequations}
and 13 non-dissipative and chiral coefficients
\begin{subequations}\label{eq:DWFVLTEChi}\begin{align}
\Delta_{{\rm LTE},\chi} \WFV_+^\mu = &
    \left[ \mf{v}^c_{\zeta_A u} u^\mu + \mf{v}^c_{\zeta_A k} \frac{k_\perp^\mu}{(k\cdot u)} \right] \zeta_A
+ \mf{v}_{r \epsilon}\, \epsilon^{\mu\nu\rho\sigma} \frac{k^\perp_\nu u_\sigma}{(k\cdot u)}\de_{\ortu\rho}\zeta\\
& -\mf{v}^c_{\varpi} \frac{2\tilde\varpi^{\mu\nu}k_\nu}{(2\pi)^3}
- (\mf{v}^c_{w u} - \mf{v}^c_{w \Delta})  \frac{(k\cdot w)}{(k\cdot u)} u^\mu\\
    &+ (\mf{v}^c_{\alpha\epsilon} - \mf{v}^c_{w \Delta}) \epsilon^{\mu\nu\rho\sigma} \frac{k^\perp_\nu u_\sigma}{(k\cdot u)} \alpha_\rho
	+ \mf{v}^c_{w k} Q^{\mu\rho} w_\rho\\    
&-\mf{v}^c_{\SP-\varpi} \frac{2(\tilde\SP^{\mu\nu}-\tilde\varpi^{\mu\nu})k_\nu}{(2\pi)^3}
- (\mf{v}^c_{\mf{w}-w u} - \mf{v}^c_{\mf{w}-w \Delta}) \frac{k^\rho_\perp u^\mu}{(k\cdot u)} (\mf{w}-w)_\rho\\
&+ (\mf{v}^c_{\mf{a}-\alpha\epsilon} - \mf{v}^c_{\mf{w}-w \Delta}) \epsilon^{\mu\nu\rho\sigma} \frac{k^\perp_\nu u_\sigma}{(k\cdot u)} (\mf{a}-\alpha)_\rho
 + \mf{v}^c_{\mf{w}-w k}  Q^{\mu\rho} (\mf{w}-w)_\rho\\
&+ \mf{v}^c_{q \epsilon}\, \epsilon^{\mu\nu\rho\sigma} \frac{k^\perp_\nu u_\sigma}{(k\cdot u)} (\beta Du_{\rho}+\de_\rho\beta)
+ \mf{v}^c_{\sigma\epsilon} \epsilon^{\mu\nu\alpha\rho}k_\perp^{\sigma}\frac{u_\nu k_\alpha}{(k\cdot u)} \beta\sigma_{\rho\sigma} .
\end{align}\end{subequations}
Regarding dissipative contributions, we found 19 non-chiral possible contributions
\begin{subequations}\label{eq:DWFVDiss}\begin{align}
\Delta_{\rm D} \WFV_+^\mu = &
    \left[ \bar{v}_{D\zeta u} u^\mu + \bar{v}_{D\zeta k} \frac{k_\perp^\mu}{(k\cdot u)} \right] D\zeta
+\left[ \bar{v}_{r u} \frac{k_\perp^\rho u^\mu}{(k\cdot u)}  + \bar{v}_{r \Delta} \Delta^{\mu\rho} + \bar{v}_{r k} Q^{\mu\rho}\right] \de_{\ortu\rho}\zeta \\
&+ \bar{v}^c_{\mf{a}-\alpha \Delta} \Delta^{\mu\rho} (\mf{a}-\alpha)_\rho
- \bar{v}^c_{\mf{a}-\alpha u} \frac{k^\rho_\perp u^\mu}{(k\cdot u)} (\mf{a}-\alpha)_\rho
+ \bar{v}^c_{\mf{w}-w \epsilon} \epsilon^{\mu\nu\rho\sigma} \frac{k^\perp_\nu u_\sigma}{(k\cdot u)} (\mf{w}-w)_\rho\\
&+ \bar{v}^c_{\mf{a}-\alpha k} Q^{\mu\rho} (\mf{a}-\alpha)_\rho
+\left[ \bar{v}^c_{D\beta u} u^\mu + \bar{v}^c_{D\beta k} \frac{k_\perp^\mu}{(k\cdot u)} \right] D\beta  +\left[ \bar{v}^c_{\theta u} u^\mu + \bar{v}^c_{\theta k} \frac{k_\perp^\mu}{(k\cdot u)} \right] \beta\theta\\
&+ \bar{v}^c_{\sigma u} (k\cdot u) \frac{k_\perp^\rho k_\perp^\sigma}{k_\perp^2}u^\mu\,\beta\sigma_{\rho\sigma}
+ \bar{v}^c_{\sigma \delta} \Delta^{\mu\rho} k_\perp^\sigma \beta\sigma_{\rho\sigma}
+ \bar{v}^c_{\sigma k} Q^{\mu\rho} k_\perp^{\sigma}\,\beta\sigma_{\rho\sigma}\\
&+ \left[ \bar{v}^c_{q u} \frac{k_\perp^\rho u^\mu}{(k\cdot u)} + \bar{v}^c_{q \Delta} \Delta^{\mu\rho} + \bar{v}^c_{q k} Q^{\mu\rho}\right] (\beta Du_{\rho}+\de_\rho\beta)
,
\end{align}\end{subequations}
and 19 possible chiral contributions
\begin{subequations}\label{eq:DWFVDissChi}\begin{align}
\Delta_{{\rm D},\chi} \WFV_+^\mu = &
    \left[ \bar{\mf{v}}^c_{D\zeta_A u} u^\mu + \bar{\mf{v}}^c_{D\zeta_A k} \frac{k_\perp^\mu}{(k\cdot u)} \right] D\zeta_A
+\left[ \bar{\mf{v}}^c_{r_A u} \frac{k_\perp^\rho u^\mu}{(k\cdot u)} + \bar{\mf{v}}_{r_A \Delta}^c \Delta^{\mu\rho} + \bar{\mf{v}}_{r_A k}^c Q^{\mu\rho}\right] \de_{\ortu\rho}\zeta_A\\
&+ \bar{\mf{v}}^c_{f \epsilon}\, \epsilon^{\mu\nu\rho\sigma} \frac{k^\perp_\nu u_\sigma}{(k\cdot u)} f_{\rho}
+ \left[ \bar{\mf{v}}^c_{\Upsilon u} \frac{k_\perp^\rho u^\mu}{(k\cdot u)} + \bar{\mf{v}}^c_{\Upsilon \Delta} \Delta^{\mu\rho} + \bar{\mf{v}}^c_{\Upsilon k} Q^{\mu\rho} \right] \Upsilon_{\rho}\\
&+\left[ \bar{\mf{v}}^c_{I-\Upsilon u} \frac{k_\perp^\rho u^\mu}{(k\cdot u)} + \bar{\mf{v}}^c_{I-\Upsilon \Delta} \Delta^{\mu\rho} + \bar{\mf{v}}^c_{I-\Upsilon k} Q^{\mu\rho}\right] (I_\rho-\Upsilon_{\rho})
 + \left[ \bar{\mf{v}}^c_{\phi u} u^\mu + \bar{\mf{v}}^c_{\phi k} \frac{k_\perp^\mu}{(k\cdot u)} \right] \phi\\
&+ \bar{\mf{v}}^c_{I_S \epsilon}\, \epsilon^{\mu\nu\alpha\rho} k_\perp^{\sigma} \frac{u_\nu k^\perp_\alpha}{(k\cdot u)^2} I_{S\,\rho\sigma}
+ \left[ \bar{\mf{v}}^c_{\rm{S12}\Delta\epsilon} \Delta^{\mu\tau}\epsilon^{\lambda\nu\rho\sigma}\frac{u_\lambda k^\perp_\nu}{(k\cdot u)}
    + \bar{\mf{v}}^c_{\rm{S12}\epsilon} \epsilon^{\mu\nu\rho\sigma}\frac{k_\perp^{\tau}u_\nu}{(k\cdot u)} \right] \Phi^{S12}_{\tau,\rho\sigma}\\
&+\left[ \bar{\mf{v}}^c_{\rm{S13}\Delta\epsilon} \Delta^{\mu\tau}\epsilon^{\lambda\nu\rho\sigma}\frac{u_\lambda k^\perp_\nu}{(k\cdot u)}
    + \bar{\mf{v}}^c_{\rm{S13}\epsilon} \epsilon^{\mu\nu\rho\sigma}\frac{k_\perp^{\tau}u_\nu}{(k\cdot u)} \right] \Phi^{S13}_{\tau,\rho\sigma}.
\end{align}\end{subequations}
The same quantities leading to non-dissipative (dissipative) contributions to the axial part also give non-dissipative (dissipative) contributions to the axial part but with opposite chirality. For this reason, the gradients of the spin potential can give a local thermal equilibrium, hence non-dissipative, contribution to the vector part.

%********************************************************************************************************
%********************************************************************************************************
%********************************************************************************************************
\section{Summary and outlook}
\label{sec:Conclusions}
In summary, the first-order gradient expansion of $\WFA^\mu$ and $\WFV^\mu$, that is the axial and vector parts of the Wigner function of a fermion, has been obtained, including the dissipative effects. These quantities are related to the spin polarization of a massive or massless fermion in a fluid trough eq.~(\ref{eq:SMass}) and eq.~(\ref{eq:SMassless}) respectively. The main results are given in eqs.~(\ref{eq:DWFALTE})-(\ref{eq:DWFADissChi}) for the axial part and in eqs.~(\ref{eq:DWFVLTE})-(\ref{eq:DWFVDissChi}) for the vector part. The definitions and decompositions of the hydrodynamic fields used to write down all the contributions are given in eqs.~(\ref{eq:DecompDezeta})-(\ref{eq:DecompShear}). All the contributions are proportional to coefficients that are functions of the momentum $k$ and of the point $x$ where the Wigner function is evaluated. The values of these coefficients depend on the microscopic theory considered and can be evaluated with the momentum dependent Kubo formulas listed in section~\ref{sec:KuboA}. All of these effects and their coefficients have been classified as either non-dissipative or dissipative, and either chiral or non-chiral. Dissipative coefficients are those induced by the dissipative part of the statistical operator (i.e., the part responsible for the increase of the entropy of the system), while the non-dissipative comes from the local thermal equilibrium part, which does not increase entropy. Chiral coefficients are those that are not vanishing only if the equilibrium statistical operator used to evaluate them is not symmetric under parity transformation, for instance if it has a chiral imbalance or it contains interactions that break parity.

We found that up to first order, all the dissipative corrections to the massive and massless spin polarization are chiral contributions, with the only exception of the spin polarization induced by the gradient of the spin potential. From this observation we can expect that the chiral dissipative effects can be neglected for the spin polarization in heavy-ion collisions. Instead, if the spin degrees of freedom are not equilibrated, the gradient of spin potential can have a seizable impact on the data. We leave the calculation of these dissipative coefficients and their phenomenological relevance to future works.

For the non-dissipative coefficients, we found that in addition to the usual spin polarization induced by vorticity, there could be other independent contributions to spin polarization induced by vorticity if interactions break the degeneracy of some coefficients that cancel out for free fields. This situation is similar to the spin alignment induced by the thermal shear. It is allowed by symmetry, but it turns out to be vanishing for a free field~\cite{Zhang:2024mhs}. However, it could be non-vanishing because of interactions~\cite{Li:2022vmb,Wagner:2022gza,Dong:2023cng,Grossi:2024pyh} or of finite volume effects. Moreover, we found that in addition to the spin Hall effect, where spin polarization is induce by the gradient of $\mu/T$, there are other chiral effects that generate polarization and a vector part of the Wigner function with the gradients of $\mu/T$ and $\mu_A/T$.

The insights gained from this work have direct phenomenological implications for understanding spin phenomena in various collision systems.
One critical application lies in the study of proton-lead (p-Pb) collisions. Unlike nucleus-nucleus collisions, p-Pb collisions are characterized by a medium that is inherently more out-of-equilibrium. The dissipative effects meticulously derived in this paper are then expected to be more prominent and relevant in p-Pb collisions. The Kubo formulas provided offer crucial theoretical predictions for measurable spin phenomena in these systems, potentially explaining discrepancies observed in current experimental data~\cite{CMS:2025nqr,Yi:2024kwu}.

Furthermore, the findings presented in this paper are highly relevant to understanding spin phenomena at lower collision energies. It has been observed from ideal spin hydrodynamics that at lower energies the relaxation time for spin is larger~\cite{Wagner:2024fhf}. This indicates that the spin degrees of freedom may not fully equilibrate, necessitating the inclusion of the spin potential and its gradients in theoretical descriptions. Our formulas provide a framework for more accurate spin polarization predictions beyond the local thermal equilibrium descriptions at these energies.

Finally, the chiral spin Hall effect could be used to probe the topological configurations of Quantum Chromodynamics (QCD). This effect, where local spin polarization is induced by gradients of temperature and axial chemical potential in a direction orthogonal to both the particle's momentum and the gradients, offers a novel observable. Its phenomenological application could involve utilizing this effect to study the anisotropies of the topological configurations of QCD. In the presence of a chiral imbalance, as might arise from sphaleron transitions in the early stages of heavy-ion collisions, the chiral spin Hall effect provides a direct link between macroscopic spin observables and the underlying topological charge fluctuations. Detecting and characterizing this effect experimentally could provide invaluable information about the vacuum structure of QCD, and elucidate the dynamics of chiral anomaly in hot and dense nuclear matter. This opens up exciting new avenues for exploring the interplay between hydrodynamics, spin, and the topological sectors of QCD.
%%%%%%%%%%%%%%%%%%%%%%%%%%%%%%%%
%\bigskip

\acknowledgments
I am grateful to V. Ambrus, F. Becattini, M. Chernodub, D. Roselli and E. Grossi for many fruitful discussions.
This work was supported by the European Union - NextGenerationEU through grant No. 760079/23.05.2023, funded by the Romanian Ministry of Research, Innovation and Digitization through Romania's National Recovery and Resilience Plan, call no. PNRR-III-C9-2022-I8.

%%%%%%%%%%%%%%%%%%%%%%%%%%%%%%%%%%%%%%%%%%%%%%%%%%%%%%%%%%%%%%%%%%%%%%%%%%%%%%%%%%%%%%%%%%%%%%%%
%%%%%%%%%%%%%%%%%%%%%%%%%%%%%%%%%%%%%%%%%%%%%%%%%%%%%%%%%%%%%%%%%%%%%%%%%%%%%%%%%%%%%%%%%%%%%%%%
%%%%%%%%%%%%%%%%%%%%%%%%%%%%%%%%%%%%%%%%%%%%%%%%%%%%%%%%%%%%%%%%%%%%%%%%%%%%%%%%%%%%%%%%%%%%%%%%

\newpage
\appendix
%********************************************************************************************
%********************************************************************************************
%********************************************************************************************
\section{Kubo formulas in momentum space}
\label{app:Kubo}
This appendix expresses the correlation functions (Kubo formulas) obtained with linear response theory in coordinate space as retarded Green's functions in momentum space.
For the dissipative part we obtained the correlator in eq. (\ref{eq:CorrDCoord}) that is
\begin{equation}
\left( \h{X},\, \h{Y}\right)_{\rm D} =  \frac{\I}{\beta}
    \int_{-\infty}^t \di^4 x_2 \int_{-\infty}^{t_2} \di s \,
    \left\langle\left[\h{X}(x),\, \h{Y}(s,\, \bm{x}_2) \right] \right\rangle_{\beta(x)} .
\end{equation}
Since all the transport coefficient obtained are scalar quantities under Lorentz transformations, we compute this correlator in the local rest frame where $u=(1,\bm{0})$.
It is convenient to introduce a positive frequency $\omega'$ inside the time integrals and obtain the correlator in the limit $\omega'\to 0$. We then consider the functional
\begin{equation}
C[\h{X},\,\h{Y}](x;\,\omega') = \frac{\I}{\beta}
    \int_{-\infty}^t \di^4 x_2 \int_{-\infty}^{t_2} \di s \, \e^{\I\omega'(t-t_2)}
    \left\langle\left[\h{X}(x),\, \h{Y}(s,\, \bm{x}_2) \right] \right\rangle_{T(x)}.
\end{equation}
Noticing that $s<t_2<t$, we can express the thermal correlator as the retarded Green's function
\begin{equation}
\label{eqapp:GRCoord}
G^R_{\h{X}\h{Y}}(t-s,\,\bm{x}-\bm{x}_2) = -\I\theta(t-s)
    \left\langle\left[\h{X}(x),\, \h{Y}(s,\bm{x}_2) \right] \right\rangle_{T(x)},
\end{equation}
obtaining
\begin{equation}
C[\h{X},\,\h{Y}](x;\,\omega') = -\frac{1}{\beta}
    \int_{-\infty}^t \di^4 x_2 \int_{-\infty}^{t_2} \di s \, \e^{\I\omega'(t-t_2)}
    G^R_{\h{X}\h{Y}}(t-s,\,\bm{x}-\bm{x}_2) .
\end{equation}
With a shift of integration variables: $\bm{x}_2 \to \bm{x}_2 - \bm{x}$, $t_2 \to t_2 - t$ and $s\to s - t_2$, we obtain 
\begin{equation}
C[\h{X},\,\h{Y}](x;\,\omega') = -\frac{1}{\beta}
    \int_{-\infty}^0 \di t_2 \, \e^{-\I\omega' t_2} \int_{-\infty}^{t_2} \di s
    \int \di^3\bm{x}\, G^R_{\h{X}\h{Y}}(-s,\,-\bm{x}_2),
\end{equation}
which can be expressed as the Fourier transform of the retarded Green's function
\begin{equation}
G^R_{\h{X}\h{Y}}(t,\bm{x}) = \int \frac{\di\omega}{2\pi} \int \frac{\di^3 p}{(2\pi)^3}
    \e^{-\I(\omega t - \bm{p}\cdot \bm{x})} G^R_{\h{X}\h{Y}}(\omega,\,\bm{p}).
\end{equation}
In the large volume limit we have
\begin{equation}
\int \di^3\bm{x}\, G^R_{\h{X}\h{Y}}(-s,\,-\bm{x}_2) =
    \lim_{\bm{p}\to 0} \int \frac{\di\omega}{2\pi}\, \e^{\I\omega s} G^R_{\h{X}\h{Y}}(\omega,\,\bm{p})
\end{equation}
from which follows that
\begin{equation}
C[\h{X},\,\h{Y}](x;\,\omega') = -\frac{1}{\beta}\lim_{\bm{p}\to 0} \int \frac{\di\omega}{2\pi}
    \int_{-\infty}^0 \di t_2 \, \e^{-\I\omega' t_2}  G^R_{\h{X}\h{Y}}(\omega,\,\bm{p})
   \int_{-\infty}^{t_2} \di s\, \e^{\I\omega s}.
\end{equation}
The integral in $s$ can be computed with a shift $\omega\to\omega-\I\delta$, with a positive $\delta$ and taking the limit $\delta\to 0^+$; we obtain
\begin{equation}
\int_{-\infty}^{t_2} \di s\, \e^{\I\omega s} =
    \lim_{\delta\to 0^+} \int_{-\infty}^{t_2} \di s\, \e^{(\I\omega + \delta) s}
    = \lim_{\delta\to 0^+} \frac{\e^{(\I\omega+\delta)t_2}}{\I\omega + \delta}
    =-\frac{\I\e^{\I\omega t_2}}{\omega-\I 0^+},
\end{equation}
and thus
\begin{equation}
\begin{split}
C[\h{X},\,\h{Y}](x;\,\omega') =& \frac{\I}{\beta}\lim_{\bm{p}\to 0} \int \frac{\di\omega}{2\pi}
    G^R_{\h{X}\h{Y}}(\omega,\,\bm{p}) \int_{-\infty}^0 \di t_2 \,
   \frac{\I\e^{\I(\omega-\omega') t_2}}{\omega-\I 0^+} \\
   = & \frac{1}{\beta}\lim_{\bm{p}\to 0} \int \frac{\di\omega}{2\pi}
   \frac{1}{\omega-\I 0^+}\frac{1}{\omega-\omega'-\I 0^+} G^R_{\h{X}\h{Y}}(\omega,\,\bm{p})\\
   = & \frac{1}{\beta}\lim_{\bm{p}\to 0} \oint \frac{\di\omega}{2\pi}
   \frac{1}{\omega-\I 0^+}\frac{1}{\omega-\omega'-\I 0^+} G^R_{\h{X}\h{Y}}(\omega,\,\bm{p}),
\end{split}
\end{equation}
where the path in the complex integral is closed in the upper half-plane where the retarded Green's function is analytic. We also assume that the retarded Green's function goes to zero faster than $\omega^{-1}$ such that the upper half-circle vanishes at infinity. The complex integral can then be computed with the residue theorem and our integrand has only two poles. We obtain
\begin{equation}
\begin{split}
C[\h{X},\,\h{Y}](x;\,\omega') =& \frac{2\pi\I}{\beta}\lim_{\bm{p}\to 0}\left[
    \text{Res}\left( \frac{1}{\omega-\I 0^+}\frac{1}{\omega-\omega'-\I 0^+} \frac{G^R_{\h{X}\h{Y}}(\omega,\,\bm{p})}{2\pi} \right)_{\omega=\omega'+\I0^+}\right.\\ & + \left.
    \text{Res}\left( \frac{1}{\omega-\I 0^+}\frac{1}{\omega-\omega'-\I 0^+} \frac{G^R_{\h{X}\h{Y}}(\omega,\,\bm{p})}{2\pi} \right)_{\omega=+\I0^+}\right]\\
    = & \frac{\I}{\beta}\lim_{\bm{p}\to 0} \frac{G^R_{\h{X}\h{Y}}(\omega',\,\bm{p})-G^R_{\h{X}\h{Y}}(0,\,\bm{p})}{\omega'} .
\end{split}
\end{equation}
Now, in the limit of vanishing $\omega'$, we finally obtain
\begin{equation}
\left( \h{X},\, \h{Y}\right)_{\rm D} =  \lim_{\omega'\to 0} \frac{\I}{\beta}\lim_{\bm{p}\to 0} \frac{G^R_{\h{X}\h{Y}}(\omega',\,\bm{p})-G^R_{\h{X}\h{Y}}(0,\,\bm{p})}{\omega'}
=\frac{\I}{\beta}\left.\frac{\di}{\di\omega}\right|_{\omega=0}\lim_{\bm{p}\to 0}G^R_{\h{X}\h{Y}}(\omega,\,\bm{p}).
\end{equation}
Since the operators $\h{X}$ and $\h{Y}$ are hermitian, the thermal correlator satisfies the property
\begin{equation}
\left\langle\left[\h{X}(x),\, \h{Y}(s,\, \bm{x}_2) \right] \right\rangle_{T(x)}^*
 = -\left\langle\left[\h{X}(x),\, \h{Y}(s,\, \bm{x}_2) \right] \right\rangle_{T(x)}
\end{equation}
from which follows that the retarded Green's function in eq.~(\ref{eqapp:GRCoord}) is real, that
\begin{equation}
\left\{\lim_{\bm{p}\to 0} G^R_{\h{X}\h{Y}}(\omega,\,\bm{p}) \right\}^* =
    \lim_{\bm{p}\to 0} G^R_{\h{X}\h{Y}}(-\omega,\,\bm{p}),\quad
\left\{ C[\h{X},\,\h{Y}](x;\,\omega)\right\}^* = C[\h{X},\,\h{Y}](x;\,-\omega)
\end{equation}
and that
\begin{equation}
\lim_{\bm{p}\to 0} \text{Re} G^R_{\h{X}\h{Y}}(-\omega,\,\bm{p}) =
    \lim_{\bm{p}\to 0} \text{Re} G^R_{\h{X}\h{Y}}(\omega,\,\bm{p}),\quad
\lim_{\bm{p}\to 0} \text{Im} G^R_{\h{X}\h{Y}}(-\omega,\,\bm{p}) =
    -\lim_{\bm{p}\to 0} \text{Im} G^R_{\h{X}\h{Y}}(\omega,\,\bm{p}).
\end{equation}
Since the real (imaginary) part of the retarded Green’s function is an even (odd) function of $\omega$ we can write
\begin{equation}
\left( \h{X},\, \h{Y}\right)_{\rm D} =-\frac{1}{\beta}
    \left.\frac{\di}{\di\omega}\right|_{\omega=0}\lim_{\bm{p}\to 0}
    \text{Im } G^R_{\h{X}\h{Y}}(\omega,\,\bm{p})
\end{equation}
or in a covariant form
\begin{equation}
\left( \h{X},\, \h{Y}\right)_{\rm D} = 
    -\frac{1}{|\beta(x)|}u^\lambda \left.\frac{\de}{\de p^\lambda}\right|_{u\cdot p=0}
    \lim_{p_\perp \to 0} {\rm Im}\, G^R_{\h{X}\h{Y}}(p) .
\end{equation}
For the dissipative part proportional to the gradients of the axial chemical potential and of the spin potential we also have the operators
\begin{equation}
-\de_\rho\zeta_A(x)\int\di^4 x\, (x_2-x)^\rho\de_\lambda\h{j}^\lambda_A,\quad
\de_\tau\mf{S}_{\rho\sigma}(x)\int\di^4 x\, (x_2-x)^\tau\h{T}^{\rho\sigma}_A(x_2)
\end{equation}
that we write generically as
\begin{equation}
\mcU_{\rho(\alpha_2)}(x)\int\di^4 x\, (x_2-x)^\rho \h{Y}^{(\alpha_2)}.
\end{equation}
Even in these cases, we can use the previous formula with the Green function
\begin{equation}
G^R_{\h{X}\h{Y}^{(\alpha_2)}x^\rho}(x) = -\I\theta(t)
    \left\langle\left[\h{X}(x),\, \h{Y}(0) \right] \right\rangle_{T(x)} x^\rho;
\end{equation}
however, it might be convenient to split the operators from the coordinate $x$ and express the correlator through the Green function:
\begin{equation}
G^R_{\h{X}\h{Y}^{(\alpha_2)}}(x) = -\I\theta(t)
    \left\langle\left[\h{X}(x),\, \h{Y}(0) \right] \right\rangle_{T(x)}.
\end{equation}
It is straightforward to show that
\begin{equation}
G^R_{\h{X}\h{Y}^{(\alpha_2)}x^\rho}(p) = -\I\frac{\de}{\de p_\rho}G^R_{\h{X}\h{Y}^{(\alpha_2)}}(p),
\end{equation}
from which immediately follows that
\begin{equation}
\left( \h{X},\, \int\di^4 x\, (x_2-x)^\rho \h{Y}^{(\alpha_2)}\right)_{\rm D} =
\frac{1}{\beta}\left.\frac{\di}{\di\omega}\right|_{\omega=0}\lim_{\bm{p}\to 0}
    \frac{\de}{\de p_\rho} G^R_{\h{X}\h{Y}^{(\alpha_2)}}(\omega,\,\bm{p})
\end{equation}
and in covariant form
\begin{equation}
\left( \h{X},\, \int\di^4 x\, (x_2-x)^\rho \h{Y}^{(\alpha_2)}\right)_{\rm D} = 
    -\frac{1}{|\beta(x)|}u^\lambda \left.\frac{\de}{\de p^\lambda}\right|_{u\cdot p=0}
    \lim_{p_\perp \to 0} \frac{\de}{\de p_\rho} {\rm Re}\, G^R_{\h{X}\h{Y}^{(\alpha_2)}}(p) .
\end{equation}

Consider now the linear response of a local equilibrium operator in the local rest frame
\begin{equation}
\h{B}_{\mcU}^{(1)}
= b_{\mcU}\,\mcU_{\rho(\alpha_2)}(x)\, \h{\mathcal{B}}_{\mcU,1}^{\rho(\alpha_2)}
= b_{\mcU}\,\mcU_{\rho(\alpha_2)}(x)\int_\Sigma\di\Sigma_\lambda(y)\,(y-x)^\rho
    \, \h{\mathcal{B}}_{\mcU}^{\lambda(\alpha_2)}(y)
\end{equation}
that leads to
\begin{equation}
\Delta_{\mcU,1,\,{\rm LTE}} O(x) = \mcU_{\rho(\alpha_2)}(x)\,b_{\mcU}\,
    \left( \h{O},\, \h{\mathcal{B}}_{\mcU,1}^{\rho(\alpha_2)}\right)_{\rm LTE}
\end{equation}
with
\begin{equation}
\begin{split}
\left( \h{O},\, \h{\mathcal{B}}_{\mcU,1}^{(\alpha)}\right)_{\rm LTE}
    = & \int_0^{|\beta|} \frac{\di\tau}{|\beta(x)|}
        \mean{\h{\mathcal{B}}_{\mcU,1[\tau/|\beta|]}^{(\alpha)}\h{O}(x)}_{T(x),\,{\rm c}} \\
    = & \int_0^{|\beta|} \frac{\di\tau}{|\beta(x)|}\int_\Sigma\di\Sigma_\lambda(y)\,(y-x)^\rho
        \mean{\h{\mathcal{B}}_{\mcU[\tau/|\beta|]}^{\lambda(\alpha_2)}(y)\h{O}(x)}_{T(x),\,{\rm c}}\, .
\end{split}
\end{equation}
Considering that the operator $\h{\mathcal{B}}_{\mcU}^{\lambda(\alpha_2)}$ commutes with $\h{Q}$, such that
\begin{equation}
\h{\mathcal{B}}_{\mcU[z]}^{\lambda(\alpha_2)}(y) = 
\e^{z\left(\beta(x)\cdot\h{P}-\zeta(x)\h{Q}\right)}
   \h{\mathcal{B}}_{\mcU}^{\lambda(\alpha_2)}(y)
   \e^{-z\left(\beta(x)\cdot\h{P}-\zeta(x)\h{Q}\right)} =
   \h{\mathcal{B}}_{\mcU}^{\lambda(\alpha_2)}\left(y - \I z \sqrt{\beta(x)^2}\right),
\end{equation}
and that the correlation between the operator $\h{O}$ at time $t$ and $\h{\mathcal{B}}_{\mcU}^{\lambda(\alpha_2)}$ at the infinitely remote past is vanishing, we obtain
\begin{equation}
\begin{split}
\left( \h{O},\, \h{\mathcal{B}}_{\mcU,1}^{(\alpha)}\right)_{\rm LTE}
=& \frac{\I}{|\beta(x)|} \int_{-\infty}^{0}\di t_2 \int_\Sigma \di\Sigma_\lambda(y) (y-x)^{\rho} 
        \left\langle\left[\h{O}(x),\, \h{\mathcal{B}}_{\mcU}^{\lambda(\alpha_2)}(t_2+y_0,\, \bm{y}) \right] \right\rangle_{T(x)}.
\end{split}
\end{equation}
We now replace the thermal average of the commutator with its spectral function in Fourier space
\begin{equation}
\left\langle\left[\h{O}(x),\, \h{\mathcal{B}}_{\mcU}^{\lambda(\alpha_2)}(y_0+t_2,\, \bm{y}) \right] \right\rangle_{T(x)} =
\int \frac{\di^4 p}{(2\pi)^4} \e^{-\I p_0 t_2} \e^{-\I p\cdot (y-x)} \rho^{\lambda(\alpha_2)}_{\h{O}\h{\mathcal{B}}}(p)
\end{equation}
obtaining
\begin{equation}
\begin{split}
\left( \h{O},\, \h{\mathcal{B}}_{\mcU,1}^{(\alpha)}\right)_{\rm LTE}
=& \frac{\I}{|\beta(x)|} \int_{-\infty}^\infty\di t_2\, \theta(-t_2)\int \frac{\di^4 p}{(2\pi)^4} 
    \rho^{\lambda(\alpha_2)}_{\h{O}\h{\mathcal{B}}}(p) \e^{-\I p_0 t_2}
    \int_\Sigma \di\Sigma_\lambda(y) (y-x)^{\rho}  \e^{-\I p\cdot (y-x)}.
\end{split}
\end{equation}
We replace the Heaviside theta with its integral representation and we integrate in $t_2$:
\begin{equation}
\int_{-\infty}^\infty \di t_2\, \theta(-t_2) \e^{-\I p_0 t_2} =
    \I\int_{-\infty}^\infty\frac{\di \omega}{2\pi}\int_{-\infty}^\infty \di t_2\,
    \frac{\e^{-\I (p_0-\omega) t_2} }{\omega + \I 0^+} =
    \I\int_{-\infty}^\infty\frac{\di \omega}{2\pi} \frac{2\pi\delta(p_0-\omega) }{\omega + \I 0^+},
\end{equation}
obtaining
\begin{equation}
\begin{split}
\left( \h{O},\, \h{\mathcal{B}}_{\mcU,1}^{(\alpha)}\right)_{\rm LTE}
=& -\frac{1}{|\beta(x)|} \int_{-\infty}^\infty\frac{\di \omega}{2\pi}\int \frac{\di^4 p}{(2\pi)^4} 
    \frac{2\pi\delta(p_0-\omega) }{\omega + \I 0^+}
    \rho^{\lambda(\alpha_2)}_{\h{O}\h{\mathcal{B}}}(p) \\ &\times
    \int_\Sigma \di\Sigma_\lambda(y) (y-x)^{\rho}  \e^{-\I p\cdot (y-x)}.
\end{split}
\end{equation}
Then, noticing that
\begin{equation}
(y-x)^\rho \e^{-\I p\cdot (y-x)}= \I \frac{\de}{\de p_\rho}\e^{-\I p\cdot (y-x)} 
\end{equation}
and integrating by parts, we obtain
\begin{equation}
\begin{split}
\left( \h{O},\, \h{\mathcal{B}}_{\mcU,1}^{(\alpha)}\right)_{\rm LTE}
=& \frac{\I}{|\beta(x)|} \int_{-\infty}^\infty\frac{\di \omega}{2\pi}\int \frac{\di^4 p}{(2\pi)^3} 
     \, \frac{1}{\omega + \I 0^+}\frac{\de}{\de p_\rho}
     \left[\rho^{\lambda(\alpha_2)}_{\h{O}\h{\mathcal{B}}}(p)\delta(p_0-\omega)\right]
    \\ &\times\int_\Sigma \di\Sigma_\lambda(y) \e^{-\I p\cdot (y-x)}.
\end{split}
\end{equation}
\begin{figure}[tbhp]
\centering
\includegraphics[width=0.45\textwidth]{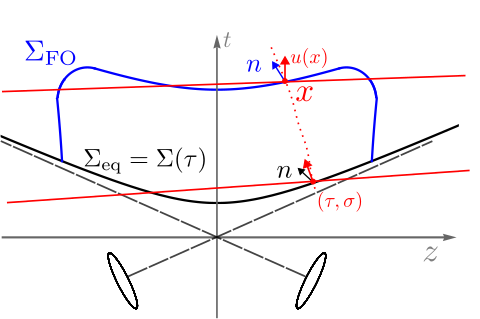}
\includegraphics[width=0.45\textwidth]{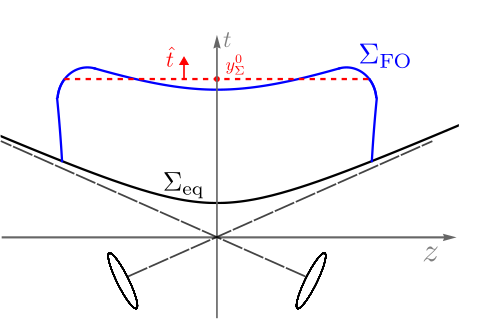}
\caption{Spacetime diagram of a relativistic heavy-ion collision. The hypersurface where LTE is achieved is $\Sigma_{\rm eq}=\Sigma(\tau)$ (black line) and $\Sigma_{\rm FO}$ (blue line) is the hypersurface where the quark-gluon plasma decouples and where spin polarization is evaluated. On the left we approximate the later time hypersurface $\Sigma=\Sigma_{\rm FO}$ with an hyperplane orthogonal to $u$ in $x$ (red line). For the dissipative part where one needs to carry out the integral over the plasma volume $\Omega$, the volume is approximated as the flat volume within the two hyperplanes. On the right we approximate $\Sigma_{\rm FO}$ as the spacelike hyperplane (red dashed line) with normal vector $\hat{t}$, the time direction in the center of mass frame, and at an average time $y^0_\Sigma$.}
\label{fig:Freezeout}
\end{figure}
If $\Sigma$ is the freezeout hypersurface in heavy ion collisions, see figure~\ref{fig:Freezeout}, the integral in $\Sigma$ can be obtained with the Gauss theorem, rewriting the integral in a flat plane at the initial phase of the collision and the difference is given by a volume integral~\cite{Becattini:2021suc,Sheng:2024pbw}. In this way, one finds \cite{Sheng:2024pbw}:
\begin{equation}
\int_\Sigma \di\Sigma_\lambda(y) \e^{-\I p\cdot (y-x)} = 
    (2\pi)^3 \e^{-\I p^0 (y^0_\Sigma - x_0)} \delta^3(\bm{p})\hat{t}_\lambda,
\end{equation}
where $\hat{t}$ is the time direction in the center of mass frame and $y^0_\Sigma$ denotes a constant time, dependent on $p$, which can be considered as an average time for the spacelike part of the hypersurface $\Sigma$, see figure~\ref{fig:Freezeout}.
After integration first in $\bm{p}$ and then in $p_0$ we have
\begin{equation}
\begin{split}
\left( \h{O},\, \h{\mathcal{B}}_{\mcU,1}^{(\alpha)}\right)_{\rm LTE}
=& \frac{\I \hat{t}_\lambda}{|\beta(x)|} \int_{-\infty}^\infty\frac{\di \omega}{2\pi}\int 
    \frac{\di p_0}{\omega + \I 0^+}\e^{-\I p^0 (y^0_\Sigma - x_0)}
     \\ &\times\left[\delta(p_0-\omega)\frac{\de}{\de p_\rho}
      + \eta^\rho_0 \delta'(p_0-\omega)\right]
   \rho^{\lambda(\alpha_2)}_{\h{O}\h{\mathcal{B}}}(p_0,\bm{p}=0)\\
= & \frac{\I \hat{t}_\lambda}{|\beta(x)|} \int_{-\infty}^\infty\frac{\di \omega}{2\pi}\int 
    \frac{\di p_0}{\omega + \I 0^+}\e^{-\I p^0 (y^0_\Sigma - x_0)}
     \delta(p_0-\omega)\\ &\times\left[\frac{\de}{\de p_\rho}
      - \eta^\rho_0 \frac{\de}{\de p_0}+\I (y^0_\Sigma - x_0)\right]
   \rho^{\lambda(\alpha_2)}_{\h{O}\h{\mathcal{B}}}(p_0,\bm{p}=0)\\
= & \frac{\I \hat{t}_\lambda}{|\beta(x)|} \int_{-\infty}^\infty\frac{\di \omega}{2\pi}
    \frac{1}{\omega + \I 0^+}
     \left[\frac{\de}{\de p_\rho}
      - \eta^\rho_0 \frac{\de}{\de p_0}+\I (y^0_\Sigma - x_0)\right]
   \rho^{\lambda(\alpha_2)}_{\h{O}\h{\mathcal{B}}}(\omega,\bm{p}=0).
\end{split}
\end{equation}
Given that $1/(\omega + \I 0^+) =$ PV$(1/\omega)-\I\pi\delta(\omega)$ and that the spectral function is real, we obtain
\begin{equation}
\begin{split}
\left( \h{O},\, \h{\mathcal{B}}_{\mcU,1}^{(\alpha)}\right)_{\rm LTE}
=& \frac{\hat{t}_\lambda}{2|\beta(x)|} \lim_{\omega\to 0} \lim_{\bm{p}\to 0}
     \left[\frac{\de}{\de p_\rho}-\eta^\rho_0\frac{\de}{\de p_0}\right]
     \rho^{\lambda(\alpha_2)}_{\h{O}\h{\mathcal{B}}}(\omega,\bm{p})
\end{split}
\end{equation}
or in covariant form and reminding that the spectral function is twice the imaginary part of the retarded function, we finally have:
\begin{equation}
\label{eqAPP:CorrGRU1FO}
\left( \h{O},\, \h{\mathcal{B}}_{\mcU,1}^{(\alpha)}\right)_{\rm LTE}
= \frac{\hat{t}_\lambda}{|\beta(x)|} \lim_{p\cdot u\to 0} \lim_{p_T\to 0}
     \Delta^\rho_\sigma \frac{\de}{\de p_\sigma}
     {\rm Im}\,G^{R}_{\h{O}\h{\mathcal{B}}^{\lambda(\alpha_2)}_{\mcU}}(p).
\end{equation}
If instead we would have approximated the integral over the hypersurface $\Sigma$ as an integral over the hyperplane with normal $u(x)$, see figure~\ref{fig:Freezeout}, then we could have used the translational invariance of the hyperplane to write:
\begin{equation}
\begin{split}
\left( \h{O},\, \h{\mathcal{B}}_{\mcU,1}^{(\alpha)}\right)_{\rm LTE}
\simeq & \frac{\I u_\lambda}{|\beta(x)|} \int_{-\infty}^\infty\di t_2\, \theta(-t_2)\int \frac{\di^4 p}{(2\pi)^4} 
    \rho^{\lambda(\alpha_2)}_{\h{O}\h{\mathcal{B}}}(p)
    \int_\Sigma \di^3 y\, (y-x)^{\rho}  \e^{-\I p\cdot (y-x)},
\end{split}
\end{equation}
and following the same steps as before we would have obtained:
\begin{equation}
\left( \h{O},\, \h{\mathcal{B}}_{\mcU,1}^{(\alpha)}\right)_{\rm LTE}
= \frac{u_\lambda}{|\beta(x)|} \lim_{p\cdot u\to 0} \lim_{p_T\to 0}
     \frac{\de}{\de p_\rho}
     {\rm Im}\, G^{R}_{\h{O}\h{\mathcal{B}}^{\lambda(\alpha_2)}_{\mcU}}(p).
\end{equation}
The linear response to thermal vorticity is related to the conserved operator $\h{J}$, meaning that the result is independent on the specific form of the hypersurface $\Sigma$. Therefore, we can use the result above to compute its associated correlator.
The other type of correlators that we obtained for local thermal equilibrium is obtained from operators of the form
\begin{equation}
\h{B}_{\mcU,0}
= b_{\mcU}\,\mcU_{(\alpha)}(x)\, \h{\mathcal{B}}_{\mcU,0}^{(\alpha)}
= b_{\mcU}\,\mcU_{(\alpha)}(x)\int_\Sigma\di\Sigma_\lambda(y)
    \, \h{\mathcal{B}}_{\mcU}^{\lambda(\alpha)}(y),
\end{equation}
that lead to
\begin{equation}
\Delta_{\mcU,0,\,{\rm LTE}} O(x) = \mcU_{(\alpha)}(x)\,b_{\mcU}\,
    \left( \h{O},\, \h{\mathcal{B}}_{\mcU,0}^{(\alpha)}\right)_{\rm LTE}
\end{equation}
where
\begin{equation}
\begin{split}
\left( \h{O},\, \h{\mathcal{B}}_{\mcU,0}^{(\alpha)}\right)_{\rm LTE}
    = & \int_0^{|\beta|} \frac{\di\tau}{|\beta(x)|}
        \mean{\h{\mathcal{B}}_{\mcU,0[\tau/|\beta|]}^{(\alpha)}\h{O}(x)}_{T(x),\,{\rm c}} \\
    = & \int_0^{|\beta|} \frac{\di\tau}{|\beta(x)|}\int_\Sigma\di\Sigma_\lambda(y)\,
        \mean{\h{\mathcal{B}}_{\mcU[\tau/|\beta|]}^{\lambda(\alpha)}(y)\h{O}(x)}_{T(x),\,{\rm c}}\, .
\end{split}
\end{equation}
This can be worked out in a similar fashion to what described above, obtaining
\begin{equation}
\label{eqApp:CorrGRU0FO}
\left( \h{O},\, \h{\mathcal{B}}_{\mcU,0}^{(\alpha)}\right)_{\rm LTE}
= \frac{\hat{t}_\lambda}{|\beta(x)|} \lim_{p\cdot u\to 0} \lim_{p_T\to 0}
    {\rm Re}\,G^{R}_{\h{O}\h{\mathcal{B}}^{\lambda(\alpha)}_{\mcU}}(p)
\end{equation}
when approximating $\Sigma$ with an average hyperplane orthogonal to $\hat{t}$, and
\begin{equation}
\begin{split}
\left( \h{O},\, \h{\mathcal{B}}_{\mcU,0}^{(\alpha)}\right)_{\rm LTE}
\simeq& \frac{ u_\lambda}{|\beta(x)|} \lim_{p\cdot u\to 0} \lim_{p_T\to 0}
    {\rm Re}\,G^{R}_{\h{O}\h{\mathcal{B}}^{\lambda(\alpha)}_{\mcU}}(p)
\end{split}
\end{equation}
when approximating the hypersurface with the hyperplane orthogonal to $u$ evaluated in $x$.

%*******************************************************************************************************
%*******************************************************************************************************
%*******************************************************************************************************
\section{Translations and Lorentz transformations of the Wigner operator}
\label{App:WigVectors}
This appendix demonstrates that the axial and vector parts of the Wigner operator transform like a vector under Lorentz transformations. Recall that the Dirac field under translation transform as
\begin{equation}
\h{T}(a) \h{\psi}(x) \h{T}(a)^\dagger = \h{\psi}(x+a),\quad
\h{T}(a) \h{\bar\psi}(x) \h{T}(a)^\dagger = \h{\bar\psi}(x+a),
\end{equation}
and reminding the Wigner operator definition
\begin{equation}
\h{W}(x,k)=-\frac{1}{(2\pi)^4}\int\di^4 y\,\e^{-\I k\cdot y} \h{\psi}(x-y/2) \h{\bar\psi}(x+y/2),
\end{equation}
it follows that the translated Wigner operator is
\begin{equation}
\h{T}(a) \h{W}(x,k) \h{T}(a)^\dagger = \h{W}(x+a,k).
\end{equation}
Similarly, under a Lorentz transformation, the Dirac field becomes
\begin{equation}
\h{\Lambda} \h{\psi}(x) \h{\Lambda}^\dagger = S(\Lambda) \h{\psi}(\Lambda^{-1}x),\quad
\h{\Lambda} \h{\bar\psi}(x) \h{\Lambda}^\dagger = \h{\bar\psi}(\Lambda^{-1} x) S(\Lambda^{-1}),
\end{equation}
then we can derive how the Wigner function transforms:
\begin{equation}
\begin{split}
\h{\Lambda} \h{W}(x,k) \h{\Lambda}^\dagger = & -\frac{1}{(2\pi)^4}\int\di^4 y\,\e^{-\I k\cdot y}
    \h{\Lambda}\h{\psi}(x-y/2)\h{\Lambda}^\dagger \h{\Lambda} \h{\bar\psi}(x+y/2)\h{\Lambda}^\dagger\\
= & -\frac{1}{(2\pi)^4}\int\di^4 y\,\e^{-\I k\cdot y} S(\Lambda)\h{\psi}(\Lambda^{-1}x-\Lambda^{-1}y/2)
    \h{\bar\psi}(\Lambda^{-1}x+\Lambda^{-1}y/2)S(\Lambda^{-1})\\
= & -\frac{1}{(2\pi)^4}\int\di^4 y'\,\e^{-\I k\cdot \Lambda y'} S(\Lambda)\h{\psi}(\Lambda^{-1}x-y'/2)
    \h{\bar\psi}(\Lambda^{-1}x+y'/2)S(\Lambda^{-1})\\
= & -\frac{1}{(2\pi)^4}\int\di^4 y\,\e^{-\I \Lambda^{-1} k\cdot y}
    S(\Lambda)\h{\psi}(\Lambda^{-1}x-y/2) \h{\bar\psi}(\Lambda^{-1}x+y/2)S(\Lambda^{-1})\\
= & S(\Lambda) \h{W}(\Lambda^{-1}x,\,\Lambda^{-1}k) S(\Lambda^{-1}).
\end{split}
\end{equation}

The axial part of the Wigner operator is defined as
\begin{equation}
\hWFA(x,k) = \tr_4\left[\gamma^\mu\gamma^5\h{W}(x,k)\right]
\end{equation}
from which, using the previous relations, it follows that
\begin{equation}
\hWFA^\mu(x,k)=\e^{-\I\h{P}\cdot x} \hWFA^\mu(0,k)\,\e^{\I\h{P}\cdot x}
\end{equation}
and
\begin{equation}
\begin{split}
\h{\Lambda} \hWFA^\mu(x,k) \h{\Lambda}^\dagger = &
    \tr\left[\gamma^\mu\gamma^5\h{\Lambda}\h{W}(x,k)\h{\Lambda}^\dagger\right]
    =\tr\left[\gamma^\mu\gamma^5 S(\Lambda)\h{W}(\Lambda^{-1}x,\Lambda^{-1}k)S(\Lambda^{-1})\right]\\
 = & \tr\left[S(\Lambda^{-1})\gamma^\mu\gamma^5 S(\Lambda)\h{W}(\Lambda^{-1}x,\Lambda^{-1}k)\right]
 = \Lambda^\mu_{\,\nu}\tr\left[\gamma^\nu\gamma^5 \h{W}(\Lambda^{-1}x,\Lambda^{-1}k)\right]\\
 =& \Lambda^\mu_{\,\nu} \hWFA^\nu(\Lambda^{-1}x,\Lambda^{-1}k).
\end{split}
\end{equation}
In the same fashion, one can show that
\begin{equation}
\hWFV^\mu(x,k)=\e^{-\I\h{P}\cdot x} \hWFV^\mu(0,k)\,\e^{\I\h{P}\cdot x},\quad
\h{\Lambda} \hWFV^\mu(x,k) \h{\Lambda}^\dagger =
    \Lambda^\mu_{\,\nu} \hWFV^\nu(\Lambda^{-1}x,\Lambda^{-1}k).
\end{equation}
%

%*******************************************************************************************************
%*******************************************************************************************************
%*******************************************************************************************************
\section{Chiral spin Hall effect conductivities}
\label{App:ChiralSHE}
This appendix calculates the conductivities related to the spin Hall like effects; that is
\begin{align}
\label{eq:appSHEA}
\Delta_{\rm SHE} \mathcal{A}_+^\mu(x,k)=&\epsilon^{\mu\nu\rho\sigma}\frac{k^\perp_\nu u_\sigma}{(k\cdot u)}\left[ a^c_{r\epsilon}(k)\de_\rho\zeta + \mathfrak{a}_{r_A\epsilon}(k)\de_\rho\zeta_A \right],\\
\label{eq:appSHEV}
\Delta_{\rm SHE} \mathcal{V}_+^\mu(x,k)=&\epsilon^{\mu\nu\rho\sigma}\frac{k^\perp_\nu u_\sigma}{(k\cdot u)}\left[ \mathfrak{v}_{r\epsilon}(k)\de_\rho\zeta + v^c_{r_A\epsilon}(k)\de_\rho\zeta_A \right],
\end{align}
for a fluid composed of a non-interacting massless Dirac particle and with chiral imbalance $\zeta_A$. The Kubo formulas for the conductivities are given in eqs.~(\ref{eq:SHEKf}) and (\ref{eq:CSHEKf}) that read
\begin{align}
a_{r \epsilon}^c = & \frac{(k\cdot u)}{2 k_\perp^2} u^\lambda\epsilon_{\lambda\mu\tau\rho}k_\perp^\tau
    \int\di\Sigma_\sigma(y-x)^{\ortu\rho}\corrLTE{\hWFA_+^\mu(x,k)}{\h{j}^\sigma(y)},\\
\mf{v}_{r \epsilon} = & \frac{(k\cdot u)}{2 k_\perp^2} u^\lambda\epsilon_{\lambda\mu\tau\rho}k_\perp^\tau
    \int\di\Sigma_\sigma(y-x)^{\ortu\rho}\corrLTE{\hWFV_+^\mu(x,k)}{\h{j}^\sigma(y)},\\
\mf{a}_{r_A \epsilon} =& \frac{(k\cdot u)}{2 k_\perp^2} u^\lambda\epsilon_{\lambda\mu\tau\rho}
  k_\perp^\tau\int\di\Sigma_\sigma(y-x)^{\ortu\rho}\corrLTE{\hWFA_+^\mu(x,k)}{\h{j}_A^\sigma(y)},\\
v^c_{r_A \epsilon} = & \frac{(k\cdot u)}{2 k_\perp^2} u^\lambda\epsilon_{\lambda\mu\tau\rho}k_\perp^\tau
    \int\di\Sigma_\sigma(y-x)^{\ortu\rho}\corrLTE{\hWFV_+^\mu(x,k)}{\h{j}_A^\sigma(y)},
\end{align}
where, see eq.~(\ref{eq:CorrLTECoord}),
\begin{equation}
\left( \h{X}(x,k),\, \h{Y}(y)\right)_{\rm LTE}
    = \int_0^{1} \di z\, \mean{\h{Y}(y+\I z\beta)\h{X}(x)}_{\beta(x),\,{\rm c}}\, .
\end{equation}
The thermal expectation values are obtained with the statistical operator of equilibrium including the axial chemical potential, that is
\begin{equation}
\label{eq:SOZetaA}
\wrho_{{\rm eq}(x)} = \frac{1}{\mathcal{Z}_{\rm eq}}\exp\left\{-\beta(x)\cdot\wP + \zeta(x)\h{Q} + \zeta_A(x)\h{Q}_A\right\} .
\end{equation}
For a massless field, instead of using the vector and axial currents and the vector and axial parts of the Wigner function, it is more convenient to use the right- and left-handed chiralities:
\begin{align}
\mu = & \mu_R + \mu_L, &
    \h{j}^\mu(y) = & \h{j}_R^\mu(y) + \h{j}_L^\mu(y),&
    \WFV_+^\mu(x,k) = & \mathcal{R}_+^\mu(x,k) + \mathcal{L}_+^\mu(x,k),\\
\mu_A = & \mu_R - \mu_L, &
    \h{j}_A^\mu(y) = & \h{j}_R^\mu(y) - \h{j}_L^\mu(y),&
    \WFA_+^\mu(x,k) = & \mathcal{R}_+^\mu(x,k) - \mathcal{L}_+^\mu(x,k).\\
\end{align}
Using the right and left chiralities, the effects in eqs.~(\ref{eq:appSHEA}) and (\ref{eq:appSHEV}) are translated to
\begin{align}
\Delta_{\rm SHE} \mathcal{R}_+^\mu(x,k) = \Delta_{\rm SHE} \mathcal{V}_{R\,+}^\mu(x,k)
=&\epsilon^{\mu\nu\rho\sigma}\frac{k^\perp_\nu u_\sigma}{(k\cdot u)}\left[ v_{R,R}(k)\de_\rho\zeta_R + v_{R,L}(k)\de_\rho\zeta_L \right],\\
\Delta_{\rm SHE} \mathcal{L}_+^\mu(x,k)= \Delta_{\rm SHE} \mathcal{V}_{L\,+}^\mu(x,k)
=&\epsilon^{\mu\nu\rho\sigma}\frac{k^\perp_\nu u_\sigma}{(k\cdot u)}\left[ v_{L,R}(k)\de_\rho\zeta_r + v_{L,L}(k)\de_\rho\zeta_L \right],
\end{align}
or altogether, denoting with $\chi,\,\chi'=+1,\,-1=R,\,L$ corresponding respectively to right and left chirality,
\begin{equation}
\Delta_{\chi'} \mathcal{V}_{\chi\,+}^\mu(x,k) = \epsilon^{\mu\nu\rho\sigma}\frac{k^\perp_\nu u_\sigma}{(k\cdot u)}
    v_{\chi,\chi'} \de_\rho\zeta_{\chi'}
\end{equation}
with the conductivity obtained with
\begin{equation}
v_{\chi,\chi'} = \frac{(k\cdot u)}{2 k_\perp^2} u^\lambda\epsilon_{\lambda\mu\tau\rho}k_\perp^\tau
    \int\di\Sigma_\sigma(y-x)^{\ortu\rho}\corrLTE{\hWFV_{\chi\,+}^\mu(x,k)}{\h{j}_{\chi'}^\sigma(y)}.
\end{equation}
Replacing the definitions of the current and of the right/left part of Wigner function, the conductivity is
\begin{equation}\begin{split}
v_{\chi,\chi'} =& \frac{(k\cdot u)}{2 k_\perp^2} u^\lambda\epsilon_{\lambda\mu\tau\rho}k_\perp^\tau
    \int\di\Sigma_\sigma(y-x)^{\ortu\rho}\left[\gamma^\mu \frac{1+\chi\gamma^5}{2} \right]_{BA}\\
    &\times\corrLTE{\h{W}^+_{AB}(x,k)}{\bar{\h{\Psi}}(y)\frac{1+\chi'\gamma^5}{2}\gamma^\sigma\h{\Psi}(y)}.
\end{split}\end{equation}

Then, using the normal modes expansion of the free Dirac field:
\begin{equation}
\h{\Psi}_+(x) = \sum_h \frac{1}{(2\pi)^{3/2}}\int\frac{\di^3 p}{2\varepsilon_p}
    u_h(p)\e^{-\I p\cdot x} \h{a}_h(p)
\end{equation}
with $h$ their chirality, we obtain
\begin{equation}
\h{W}^+_{AB}(x,k) = \frac{1}{(2\pi)^3} \sum_{h,h'} \int\frac{\di^3 p}{2\varepsilon_p}
    \int\frac{\di^3 p'}{2\varepsilon_{p'}}\delta\left(k-\frac{p+p'}{2}\right)\e^{-\I x(p'-p)}
    u_{h'}(p')_A \bar{u}_{h}(p)_B  \h{a}^\dagger_h(p) \h{a}_{h'}(p'),
\end{equation}
and
\begin{equation}
\h{j}^\sigma_{\chi'\,+}(y+\I z\beta) = \sum_{\tau,\tau'}
    \int\frac{\di^3 q}{2\varepsilon_q}\int\frac{\di^3 q'}{2\varepsilon_{q'}}
     \frac{\e^{\I y(q'-q)-z\beta\cdot(q'-q)}}{(2\pi)^3}
    \bar{u}_{\tau'}(q')\gamma^\sigma \frac{1+\chi'\gamma^5}{2} u_{\tau}(q)
    \h{a}^\dagger_{\tau'}(q') \h{a}_{\tau}(q),
\end{equation}
where we neglected the antiparticle part of the current because its contribution vanish when correlated with the particle part of the Wigner function. Replacing these expressions in the conductivity, we obtain
\begin{equation}
\begin{split}
v_{\chi,\chi'} =& \frac{(k\cdot u)}{2 k_\perp^2} u^\lambda\epsilon_{\lambda\mu\tau\rho}k_\perp^\tau
    \int\di\Sigma_\sigma(y-x)^{\ortu\rho}\left[\gamma^\mu \frac{1+\chi\gamma^5}{2} \right]_{BA}
    \left[\gamma^\sigma \frac{1+\chi'\gamma^5}{2} \right]_{B'A'}\frac{1}{(2\pi)^6}\\
    &\times  \sum_{h,h',\tau,\tau'}\int_0^1\di z
    \int\frac{\di^3 p}{2\varepsilon_p}\int\frac{\di^3 p'}{2\varepsilon_{p'}}
    \int\frac{\di^3 q}{2\varepsilon_q}\int\frac{\di^3 q'}{2\varepsilon_{q'}}
    \delta\left(k-\frac{p+p'}{2}\right)\e^{-\I x(p'-p)} \e^{(\I y-z\beta)\cdot(q'-q)}\\
    &\times u_{h'}(p')_A \bar{u}_{\tau'}(q')_{B'} u_{\tau}(q)_{A'} \bar{u}_{h}(p)_B
    \mean{\h{a}^\dagger_h(p) \h{a}_{h'}(p')\h{a}^\dagger_{\tau'}(q') \h{a}_{\tau}(q)}_{\beta,\, c}.
\end{split}
\end{equation}
Using $\mean{\h{a}^\dagger_1\, \h{a}_2 \, \h{a}^\dagger_3 \, \h{a}_4}_{\beta,\, c} = \mean{\h{a}^\dagger_1\, \h{a}_4}_\beta \mean{\h{a}_2 \, \h{a}^\dagger_3}_\beta$, and the expectation values with the statistical operator~(\ref{eq:SOZetaA}), that is
\begin{align}
\mean{\h{a}^\dagger_h(p) \h{a}_{\tau}(q)}_{\beta} = & \delta_{h\tau}2\varepsilon_q
    \delta^3(\bm{p}-\bm{q})n^{h}_F(x,p),\\
\mean{\h{a}_{h'}(p')\h{a}^\dagger_{\tau'}(q')}_{\beta} = & \delta_{h'\tau'}2\varepsilon_{q'}
    \delta^3(\bm{p}'-\bm{q}')(1-n^{h'}_F(x,p')),
\end{align}
where
\begin{equation}
n^{\chi}_F(x,k) = \frac{1}{\e^{\beta(x)\cdot k -\zeta(x)-\chi\zeta_A(x)}+1},
\end{equation}
we can immediately integrate over $q$ and $q'$ and sum over $\tau$ and $\tau'$.  We now consider the sums over the chiralities and we construct the trace over the spinoral indices:
\begin{equation}
\begin{split}
\sum_{h,h'}&\tr\left[\bar{u}_{h}(p) \gamma^\mu  \frac{1+\chi\gamma^5}{2} u_{h'}(p')
    \bar{u}_{h'}(p') \gamma^\sigma  \frac{1+\chi'\gamma^5}{2} u_{h}(p)\right]n^{h}_F(x,p)(1-n^{h'}_F(x,p')) \\
= & \sum_{h,h'}\tr\left[\bar{u}_{h}(p) \gamma^\mu  \frac{1+\chi h'}{2} u_{h'}(p')
    \bar{u}_{h'}(p') \gamma^\sigma  \frac{1+\chi' h}{2} u_{h}(p)\right]n^{h}_F(x,p)(1-n^{h'}_F(x,p'))\\
= & \sum_{h,h'}\delta_{h',\chi}\delta_{h,\chi'}\tr\left[\bar{u}_{h}(p) \gamma^\mu   u_{h'}(p')
    \bar{u}_{h'}(p') \gamma^\sigma u_{h}(p)\right]n^{h}_F(x,p)(1-n^{h'}_F(x,p'))\\
= & \tr\left[\bar{u}_{\chi'}(p) \gamma^\mu   u_{\chi}(p')
    \bar{u}_{\chi}(p') \gamma^\sigma u_{\chi'}(p)\right]n^{\chi'}_F(x,p)(1-n^{\chi}_F(x,p')),
\end{split}
\end{equation}
where we used $\gamma^5 u_h(p) = h\, u_h(p)$. Now, using the cyclicity of the trace and $u_{\chi'}(p) \bar{u}_{\chi'}(p)=(1+\chi'\gamma^5)\slashed{p}/2$, we obtain
\begin{equation}
\begin{split}
T^{\mu\sigma}(p,p';\chi,\chi') = & \tr\left[\bar{u}_{\chi'}(p) \gamma^\mu   u_{\chi}(p')
    \bar{u}_{\chi}(p') \gamma^\sigma u_{\chi'}(p)\right]\\
 =&  \tr\left[ \gamma^\mu \frac{1+\chi\gamma^5}{2} \slashed{p}'
    \gamma^\sigma \frac{1+\chi'\gamma^5}{2} \slashed{p}\right]\\
= & (1 +\chi \chi')(p^\mu p^{\prime\sigma} + p^{\prime\mu} p^{\sigma} - (p\cdot p')\eta^{\mu\sigma} )
    + \I\epsilon^{\mu\sigma\alpha\beta}p_\alpha p'_\beta .
\end{split}
\end{equation}
After these steps we obtained
\begin{equation}
\begin{split}
v_{\chi,\chi'} =& \frac{(k\cdot u)}{2 k_\perp^2} u^\lambda\epsilon_{\lambda\mu\tau\rho}k_\perp^\tau
    \frac{1}{(2\pi)^6} \int_0^1\di z \int\frac{\di^3 p}{2\varepsilon_p}
    \int\frac{\di^3 p'}{2\varepsilon_{p'}} \delta\left(k-\frac{p+p'}{2}\right)\\
    &\times T^{\mu\sigma}(p,p';\chi,\chi') n^{\chi'}_F(x,p)(1-n^{\chi}_F(x,p'))
    \e^{-z\beta\cdot(p'-p)}\int\di\Sigma_\sigma(y-x)^{\ortu\rho}\e^{\I (p'-p)(x-y)}.
\end{split}
\end{equation}
The integral over the hypersurface gives
\begin{equation}
\int\di\Sigma_\sigma(y-x)^{\ortu\rho}\e^{\I (p'-p)(x-y)} =
    -\I u_\sigma \Delta^\rho_{\,\kappa} (2\pi)^3 \frac{\de}{\de p'_\kappa}\delta^3(\bm{p}-\bm{p}').
\end{equation}
The derivative of the Dirac delta can be integrated by parts. After that, it is easy to use the deltas to integrate in $p$ and $p'$, resulting in $p=p'=k$. Noticing that
\begin{equation}
\epsilon_{\lambda\mu\tau\rho} u_\lambda u_\sigma k_\perp^\tau
    T^{\mu\sigma}(p=k,p'=k;\chi,\chi') =
    2(1 + \chi\chi')(k\cdot u) \epsilon_{\lambda\mu\tau\rho}  u^\lambda k_\perp^\tau k^\mu =0,
\end{equation}
the only non-vanishing contribution is given by
\begin{equation}
\begin{split}
v_{\chi,\chi'} =& \I\frac{(k\cdot u)}{2 k_\perp^2}\epsilon_{\lambda\mu\tau\rho}u^\lambda k_\perp^\tau
    u_\sigma\int_0^1\frac{\di z}{(2\pi)^3} \int\frac{\di^3 p}{2\varepsilon_p}
    \int\frac{\di^3 p'}{2\varepsilon_{p'}} \e^{z\beta\cdot(p-p')} \delta\left(k-\frac{p+p'}{2}\right)\\
    &\times \delta^3(\bm{p}-\bm{p}') n^{\chi'}_F(x,p)(1-n^{\chi}_F(x,p'))
     \frac{\de}{\de p'_{\ortu\rho}}T^{\mu\sigma}(p,p';\chi,\chi').
\end{split}
\end{equation}
The derivatives in $p'$ must be carried out reminding that $p'$ is on-shell:
\begin{equation}
 \frac{\de p^{\prime\sigma}}{\de p'_{\ortu\rho}} = \Delta^\sigma_{\,\rho}
    -u^\sigma\frac{p^{\prime\ortu\rho}}{\varepsilon_{p'}},
\end{equation}
therefore, after integrating in $p'$, one obtains
\begin{equation}\begin{split}
\left.\frac{\de}{\de p'_{\ortu\rho}}T^{\mu\sigma}(p,p';\chi,\chi')\right|_{p'=p} =&
    (1 +\chi'\chi')\left( p^\mu \frac{\de p^\sigma}{\de p_{\ortu\rho}}
        + \frac{\de p^\mu}{\de p_{\ortu\rho}} p^\sigma
        - p_\alpha \frac{\de p^\alpha}{\de p_{\ortu\rho}}\eta^{\mu\sigma} \right)\\
        &+\I(\chi+\chi')\epsilon^{\mu\sigma\alpha\beta} p_\alpha \frac{\de p_\beta}{\de p_{\ortu\rho}}.
\end{split}\end{equation}
After integration in $p$, that sets $p=k$, the first term in the round bracket gives a vanishing contribution to the conductivity; indeed:
\begin{equation}
\epsilon_{\lambda\mu\tau\rho} k^\tau k^\mu =0.
\end{equation}
The same happens for the second:
\begin{equation}
\epsilon_{\lambda\mu\tau\rho} u^\lambda (\Delta^\mu_{\,\rho}-u^\mu\frac{k^{\ortu\rho}}{\varepsilon_k}) =0,
\end{equation}
and the third:
\begin{equation}
\epsilon_{\lambda\mu\tau\rho} u^\lambda u_\sigma \eta^{\mu\sigma} =0.
\end{equation}
Therefore, only the last term gives non-vanishing contributions and the conductivity reads
\begin{equation}
\begin{split}
v_{\chi,\chi'} =& \I\frac{(k\cdot u)}{2 k_\perp^2}\epsilon_{\lambda\mu\tau\rho}u^\lambda k_\perp^\tau
    u_\sigma\int_0^1\frac{\di z}{(2\pi)^3} \int\frac{\di^3 p}{2\varepsilon_p}
    \int\frac{\di^3 p'}{2\varepsilon_{p'}} \e^{z\beta\cdot(p-p')} \delta\left(k-\frac{p+p'}{2}\right)\\
    &\times\delta^3(\bm{p}-\bm{p}')  n^{\chi'}_F(x,p)(1-n^{\chi}_F(x,p'))
     \I(\chi+\chi')\epsilon^{\mu\sigma\alpha\beta} p_\alpha \frac{\de p'_\beta}{\de p'_{\ortu\rho}}\\
= & -\frac{\chi+\chi'}{(2\pi)^3}\frac{(k\cdot u)}{2 k_\perp^2}\epsilon_{\lambda\mu\tau\rho}
    \epsilon^{\mu\sigma\alpha\beta} u^\lambda k_\perp^\tau k_\alpha \left(  \Delta^\rho_{\,\beta}
    -u_\beta\frac{k^{\ortu\rho}}{\varepsilon_k}\right)n^{\chi'}_F(x,p)(1-n^{\chi}_F(x,p'))\\ & \times
    \int_0^1 \di z\, \frac{\delta(k\cdot u - \varepsilon_k)}{2(k\cdot u)}\frac{1}{2(k\cdot u)}\\
= & -\frac{\chi+\chi'}{2}\frac{(k\cdot u)}{2 k_\perp^2}\epsilon_{\lambda\mu\tau\rho}
    \epsilon^{\mu\sigma\alpha\beta} u^\lambda k_\perp^\tau k_\alpha \left(  \Delta^\rho_{\,\beta}
    -u_\beta\frac{k^{\ortu\rho}}{\varepsilon_k}\right)n^{\chi'}_F(x,p)(1-n^{\chi}_F(x,p'))\\ & \times
    \frac{1}{(2\pi)^3} \frac{\delta(k\cdot u - \varepsilon_k)}{2(k\cdot u)}.
\end{split}
\end{equation}
After carrying out the contractions, one obtains
\begin{equation}
v_{\chi,\chi'} = -\frac{\chi+\chi'}{2}\frac{\delta(k^2)\theta(k\cdot u)}{(2\pi)^3}n^{\chi'}_F(x,p)(1-n^{\chi}_F(x,p')).
\end{equation}
We note that the conductivities are vanishing when $\chi'=-\chi$. To obtain the original conductivities in eqs.~(\ref{eq:appSHEA}) and (\ref{eq:appSHEV}) we simply have (in compact notation):
\begin{align}
a_{r \epsilon}^c =& \left(\h{\mathcal{R}} - \h{\mathcal{L}},\,  \h{j}_R + \h{j}_L \right)
= \left(\h{\mathcal{R}},\,  \h{j}_R\right) + \left(\h{\mathcal{R}},\,\h{j}_L \right)
    -\left( \h{\mathcal{L}},\,  \h{j}_R\right) -\left( \h{\mathcal{L}},\, \h{j}_L \right)\\
=& \left(\h{\mathcal{R}},\,  \h{j}_R\right)  -\left( \h{\mathcal{L}},\, \h{j}_L \right)
= v_{R,R} - v_{L,L}\\
=&-\frac{\delta(k^2)\theta(k\cdot u)}{(2\pi)^3}\left[n_F^R(x,k)(1-n_F^R(x,k)) + n_F^L(x,k)(1-n_F^L(x,k))\right]\\
=& \left(\h{\mathcal{R}} + \h{\mathcal{L}},\,  \h{j}_R - \h{j}_L \right)
= v^c_{r_A \epsilon},
\end{align}
and
\begin{align}
\mf{a}_{r_A \epsilon} =& \left(\h{\mathcal{R}} - \h{\mathcal{L}},\,  \h{j}_R - \h{j}_L \right)
= \left(\h{\mathcal{R}},\,  \h{j}_R\right) - \left(\h{\mathcal{R}},\,\h{j}_L \right)
    -\left( \h{\mathcal{L}},\,  \h{j}_R\right) + \left( \h{\mathcal{L}},\, \h{j}_L \right)\\
=& \left(\h{\mathcal{R}},\,  \h{j}_R\right)  + \left( \h{\mathcal{L}},\, \h{j}_L \right)
= v_{R,R} + v_{L,L} \\
=&-\frac{\delta(k^2)\theta(k\cdot u)}{(2\pi)^3}\left[n_F^R(x,k)(1-n_F^R(x,k)) - n_F^L(x,k)(1-n_F^L(x,k))\right]\\
=& \left(\h{\mathcal{R}} + \h{\mathcal{L}},\,  \h{j}_R + \h{j}_L \right)
= \mf{v}_{r \epsilon}.
\end{align}
These are the values reported in section~\ref{sec:CSHE}.
%\enlargethispage{3\baselineskip}

\providecommand{\href}[2]{#2}\begingroup\raggedright\endgroup


\begin{thebibliography}{100}

\bibitem{Liang:2004ph}
Z.-T.~Liang and X.-N.~Wang, \emph{{Globally polarized quark-gluon plasma in non-central A+A collisions}}, \href{https://doi.org/10.1103/PhysRevLett.94.102301}{\emph{Phys. Rev. Lett.} {\bfseries 94} (2005) 102301} [\href{https://arxiv.org/abs/nucl-th/0410079}{{\ttfamily nucl-th/0410079}}].

\bibitem{Gao:2007bc}
J.-H.~Gao, S.-W.~Chen, W.-t.~Deng, Z.-T.~Liang, Q.~Wang and X.-N.~Wang, \emph{{Global quark polarization in non-central A+A collisions}}, \href{https://doi.org/10.1103/PhysRevC.77.044902}{\emph{Phys. Rev. C} {\bfseries 77} (2008) 044902} [\href{https://arxiv.org/abs/0710.2943}{{\ttfamily 0710.2943}}].

\bibitem{Huang:2011ru}
X.-G.~Huang, P.~Huovinen and X.-N.~Wang, \emph{{Quark Polarization in a Viscous Quark-Gluon Plasma}}, \href{https://doi.org/10.1103/PhysRevC.84.054910}{\emph{Phys. Rev. C} {\bfseries 84} (2011) 054910} [\href{https://arxiv.org/abs/1108.5649}{{\ttfamily 1108.5649}}].

\bibitem{Becattini:2007sr}
F.~Becattini, F.~Piccinini and J.~Rizzo, \emph{{Angular momentum conservation in heavy ion collisions at very high energy}}, \href{https://doi.org/10.1103/PhysRevC.77.024906}{\emph{Phys. Rev. C} {\bfseries 77} (2008) 024906} [\href{https://arxiv.org/abs/0711.1253}{{\ttfamily 0711.1253}}].

\bibitem{Becattini:2013fla}
F.~Becattini, V.~Chandra, L.~Del~Zanna and E.~Grossi, \emph{{Relativistic distribution function for particles with spin at local thermodynamical equilibrium}}, \href{https://doi.org/10.1016/j.aop.2013.07.004}{\emph{Annals Phys.} {\bfseries 338} (2013) 32} [\href{https://arxiv.org/abs/1303.3431}{{\ttfamily 1303.3431}}].

\bibitem{STAR:2017ckg}
{\scshape STAR} collaboration, \emph{{Global $\Lambda$ hyperon polarization in nuclear collisions: evidence for the most vortical fluid}}, \href{https://doi.org/10.1038/nature23004}{\emph{Nature} {\bfseries 548} (2017) 62} [\href{https://arxiv.org/abs/1701.06657}{{\ttfamily 1701.06657}}].

\bibitem{STAR:2018gyt}
{\scshape STAR} collaboration, \emph{{Global polarization of $\Lambda$ hyperons in Au+Au collisions at $\sqrt{s_{_{NN}}}$ = 200 GeV}}, \href{https://doi.org/10.1103/PhysRevC.98.014910}{\emph{Phys. Rev. C} {\bfseries 98} (2018) 014910} [\href{https://arxiv.org/abs/1805.04400}{{\ttfamily 1805.04400}}].

\bibitem{STAR:2021beb}
{\scshape STAR} collaboration, \emph{{Global $\Lambda$-hyperon polarization in Au+Au collisions at $\sqrt {s_{NN}}$=3~GeV}}, \href{https://doi.org/10.1103/PhysRevC.104.L061901}{\emph{Phys. Rev. C} {\bfseries 104} (2021) L061901} [\href{https://arxiv.org/abs/2108.00044}{{\ttfamily 2108.00044}}].

\bibitem{STAR:2020xbm}
{\scshape STAR} collaboration, \emph{{Global Polarization of $\Xi$ and $\Omega$ Hyperons in Au+Au Collisions at $\sqrt {s_{NN}}$ = 200 GeV}}, \href{https://doi.org/10.1103/PhysRevLett.126.162301}{\emph{Phys. Rev. Lett.} {\bfseries 126} (2021) 162301} [\href{https://arxiv.org/abs/2012.13601}{{\ttfamily 2012.13601}}].

\bibitem{ALICE:2019onw}
{\scshape ALICE} collaboration, \emph{{Global polarization of $\Lambda \bar \Lambda$ hyperons in Pb-Pb collisions at $\sqrt {s_{NN}}$ = 2.76 and 5.02 TeV}}, \href{https://doi.org/10.1103/PhysRevC.101.044611}{\emph{Phys. Rev. C} {\bfseries 101} (2020) 044611} [\href{https://arxiv.org/abs/1909.01281}{{\ttfamily 1909.01281}}].

\bibitem{Becattini:2015ska}
F.~Becattini, G.~Inghirami, V.~Rolando, A.~Beraudo, L.~Del~Zanna, A.~De~Pace et~al., \emph{{A study of vorticity formation in high energy nuclear collisions}}, \href{https://doi.org/10.1140/epjc/s10052-015-3624-1}{\emph{Eur. Phys. J. C} {\bfseries 75} (2015) 406} [\href{https://arxiv.org/abs/1501.04468}{{\ttfamily 1501.04468}}].

\bibitem{Karpenko:2016jyx}
I.~Karpenko and F.~Becattini, \emph{{Study of $\Lambda $ polarization in relativistic nuclear collisions at $\sqrt{s_\mathrm {NN}}=7.7$ \textendash{}200 GeV}}, \href{https://doi.org/10.1140/epjc/s10052-017-4765-1}{\emph{Eur. Phys. J. C} {\bfseries 77} (2017) 213} [\href{https://arxiv.org/abs/1610.04717}{{\ttfamily 1610.04717}}].

\bibitem{Xie:2017upb}
Y.~Xie, D.~Wang and L.P.~Csernai, \emph{{Global \ensuremath{\Lambda} polarization in high energy collisions}}, \href{https://doi.org/10.1103/PhysRevC.95.031901}{\emph{Phys. Rev. C} {\bfseries 95} (2017) 031901} [\href{https://arxiv.org/abs/1703.03770}{{\ttfamily 1703.03770}}].

\bibitem{Wu:2019eyi}
H.-Z.~Wu, L.-G.~Pang, X.-G.~Huang and Q.~Wang, \emph{{Local spin polarization in high energy heavy ion collisions}}, \href{https://doi.org/10.1103/PhysRevResearch.1.033058}{\emph{Phys. Rev. Research.} {\bfseries 1} (2019) 033058} [\href{https://arxiv.org/abs/1906.09385}{{\ttfamily 1906.09385}}].

\bibitem{Ivanov:2020udj}
Y.B.~Ivanov, \emph{{Global $\Lambda$ polarization in moderately relativistic nuclear collisions}}, \href{https://doi.org/10.1103/PhysRevC.103.L031903}{\emph{Phys. Rev. C} {\bfseries 103} (2021) L031903} [\href{https://arxiv.org/abs/2012.07597}{{\ttfamily 2012.07597}}].

\bibitem{Fu:2020oxj}
B.~Fu, K.~Xu, X.-G.~Huang and H.~Song, \emph{{Hydrodynamic study of hyperon spin polarization in relativistic heavy ion collisions}}, \href{https://doi.org/10.1103/PhysRevC.103.024903}{\emph{Phys. Rev. C} {\bfseries 103} (2021) 024903} [\href{https://arxiv.org/abs/2011.03740}{{\ttfamily 2011.03740}}].

\bibitem{Niida:2018hfw}
{\scshape STAR} collaboration, \emph{{Global and local polarization of $\Lambda$ hyperons in Au+Au collisions at 200 GeV from STAR}}, \href{https://doi.org/10.1016/j.nuclphysa.2018.08.034}{\emph{Nucl. Phys. A} {\bfseries 982} (2019) 511} [\href{https://arxiv.org/abs/1808.10482}{{\ttfamily 1808.10482}}].

\bibitem{STAR:2019erd}
{\scshape STAR} collaboration, \emph{{Polarization of $\Lambda$ ($\bar{\Lambda}$) hyperons along the beam direction in Au+Au collisions at $\sqrt{s_{_{NN}}}$ = 200 GeV}}, \href{https://doi.org/10.1103/PhysRevLett.123.132301}{\emph{Phys. Rev. Lett.} {\bfseries 123} (2019) 132301} [\href{https://arxiv.org/abs/1905.11917}{{\ttfamily 1905.11917}}].

\bibitem{ALICE:2021pzu}
{\scshape ALICE} collaboration, \emph{{Polarization of $\Lambda$ and $\bar \Lambda$ Hyperons along the Beam Direction in Pb-Pb Collisions at $\sqrt {s_{NN}}$=5.02\,\,TeV}}, \href{https://doi.org/10.1103/PhysRevLett.128.172005}{\emph{Phys. Rev. Lett.} {\bfseries 128} (2022) 172005} [\href{https://arxiv.org/abs/2107.11183}{{\ttfamily 2107.11183}}].

\bibitem{Becattini:2020ngo}
F.~Becattini and M.A.~Lisa, \emph{{Polarization and Vorticity in the Quark\textendash{}Gluon Plasma}}, \href{https://doi.org/10.1146/annurev-nucl-021920-095245}{\emph{Ann. Rev. Nucl. Part. Sci.} {\bfseries 70} (2020) 395} [\href{https://arxiv.org/abs/2003.03640}{{\ttfamily 2003.03640}}].

\bibitem{Liu:2021uhn}
S.Y.F.~Liu and Y.~Yin, \emph{{Spin polarization induced by the hydrodynamic gradients}}, \href{https://doi.org/10.1007/JHEP07(2021)188}{\emph{JHEP} {\bfseries 07} (2021) 188} [\href{https://arxiv.org/abs/2103.09200}{{\ttfamily 2103.09200}}].

\bibitem{Becattini:2021suc}
F.~Becattini, M.~Buzzegoli and A.~Palermo, \emph{{Spin-thermal shear coupling in a relativistic fluid}}, \href{https://doi.org/10.1016/j.physletb.2021.136519}{\emph{Phys. Lett. B} {\bfseries 820} (2021) 136519} [\href{https://arxiv.org/abs/2103.10917}{{\ttfamily 2103.10917}}].

\bibitem{Liu:2020dxg}
S.Y.F.~Liu and Y.~Yin, \emph{{Spin Hall effect in heavy-ion collisions}}, \href{https://doi.org/10.1103/PhysRevD.104.054043}{\emph{Phys. Rev. D} {\bfseries 104} (2021) 054043} [\href{https://arxiv.org/abs/2006.12421}{{\ttfamily 2006.12421}}].

\bibitem{Fu:2021pok}
B.~Fu, S.Y.F.~Liu, L.~Pang, H.~Song and Y.~Yin, \emph{{Shear-Induced Spin Polarization in Heavy-Ion Collisions}}, \href{https://doi.org/10.1103/PhysRevLett.127.142301}{\emph{Phys. Rev. Lett.} {\bfseries 127} (2021) 142301} [\href{https://arxiv.org/abs/2103.10403}{{\ttfamily 2103.10403}}].

\bibitem{Becattini:2021iol}
F.~Becattini, M.~Buzzegoli, G.~Inghirami, I.~Karpenko and A.~Palermo, \emph{{Local Polarization and Isothermal Local Equilibrium in Relativistic Heavy Ion Collisions}}, \href{https://doi.org/10.1103/PhysRevLett.127.272302}{\emph{Phys. Rev. Lett.} {\bfseries 127} (2021) 272302} [\href{https://arxiv.org/abs/2103.14621}{{\ttfamily 2103.14621}}].

\bibitem{Yi:2021ryh}
C.~Yi, S.~Pu and D.-L.~Yang, \emph{{Reexamination of local spin polarization beyond global equilibrium in relativistic heavy ion collisions}}, \href{https://doi.org/10.1103/PhysRevC.104.064901}{\emph{Phys. Rev. C} {\bfseries 104} (2021) 064901} [\href{https://arxiv.org/abs/2106.00238}{{\ttfamily 2106.00238}}].

\bibitem{Wu:2022mkr}
X.-Y.~Wu, C.~Yi, G.-Y.~Qin and S.~Pu, \emph{{Local and global polarization of \ensuremath{\Lambda} hyperons across RHIC-BES energies: The roles of spin hall effect, initial condition, and baryon diffusion}}, \href{https://doi.org/10.1103/PhysRevC.105.064909}{\emph{Phys. Rev. C} {\bfseries 105} (2022) 064909} [\href{https://arxiv.org/abs/2204.02218}{{\ttfamily 2204.02218}}].

\bibitem{Alzhrani:2022dpi}
S.~Alzhrani, S.~Ryu and C.~Shen, \emph{{\ensuremath{\Lambda} spin polarization in event-by-event relativistic heavy-ion collisions}}, \href{https://doi.org/10.1103/PhysRevC.106.014905}{\emph{Phys. Rev. C} {\bfseries 106} (2022) 014905} [\href{https://arxiv.org/abs/2203.15718}{{\ttfamily 2203.15718}}].

\bibitem{Palermo:2024tza}
A.~Palermo, E.~Grossi, I.~Karpenko and F.~Becattini, \emph{{$\Lambda $ polarization in very high energy heavy ion collisions as a probe of the quark\textendash{}gluon plasma formation and properties}}, \href{https://doi.org/10.1140/epjc/s10052-024-13229-z}{\emph{Eur. Phys. J. C} {\bfseries 84} (2024) 920} [\href{https://arxiv.org/abs/2404.14295}{{\ttfamily 2404.14295}}].

\bibitem{Kapusta:2019sad}
J.I.~Kapusta, E.~Rrapaj and S.~Rudaz, \emph{{Relaxation Time for Strange Quark Spin in Rotating Quark-Gluon Plasma}}, \href{https://doi.org/10.1103/PhysRevC.101.024907}{\emph{Phys. Rev. C} {\bfseries 101} (2020) 024907} [\href{https://arxiv.org/abs/1907.10750}{{\ttfamily 1907.10750}}].

\bibitem{Ayala:2019iin}
A.~Ayala, D.~De~La~Cruz, S.~Hern\'andez-Ort\'\i{}z, L.A.~Hern\'andez and J.~Salinas, \emph{{Relaxation time for quark spin and thermal vorticity alignment in heavy-ion collisions}}, \href{https://doi.org/10.1016/j.physletb.2019.135169}{\emph{Phys. Lett. B} {\bfseries 801} (2020) 135169} [\href{https://arxiv.org/abs/1909.00274}{{\ttfamily 1909.00274}}].

\bibitem{Hongo:2022izs}
M.~Hongo, X.-G.~Huang, M.~Kaminski, M.~Stephanov and H.-U.~Yee, \emph{{Spin relaxation rate for heavy quarks in weakly coupled QCD plasma}}, \href{https://doi.org/10.1007/JHEP08(2022)263}{\emph{JHEP} {\bfseries 08} (2022) 263} [\href{https://arxiv.org/abs/2201.12390}{{\ttfamily 2201.12390}}].

\bibitem{Wagner:2024fhf}
D.~Wagner, M.~Shokri and D.H.~Rischke, \emph{{Damping of spin waves}}, \href{https://doi.org/10.1103/PhysRevResearch.6.043103}{\emph{Phys. Rev. Res.} {\bfseries 6} (2024) 043103} [\href{https://arxiv.org/abs/2405.00533}{{\ttfamily 2405.00533}}].

\bibitem{Florkowski:2017ruc}
W.~Florkowski, B.~Friman, A.~Jaiswal and E.~Speranza, \emph{{Relativistic fluid dynamics with spin}}, \href{https://doi.org/10.1103/PhysRevC.97.041901}{\emph{Phys. Rev. C} {\bfseries 97} (2018) 041901} [\href{https://arxiv.org/abs/1705.00587}{{\ttfamily 1705.00587}}].

\bibitem{Weickgenannt:2019dks}
N.~Weickgenannt, X.-L.~Sheng, E.~Speranza, Q.~Wang and D.H.~Rischke, \emph{{Kinetic theory for massive spin-1/2 particles from the Wigner-function formalism}}, \href{https://doi.org/10.1103/PhysRevD.100.056018}{\emph{Phys. Rev. D} {\bfseries 100} (2019) 056018} [\href{https://arxiv.org/abs/1902.06513}{{\ttfamily 1902.06513}}].

\bibitem{STAR:2022fan}
{\scshape STAR} collaboration, \emph{{Pattern of global spin alignment of \ensuremath{\phi} and K$^{*0}$ mesons in heavy-ion collisions}}, \href{https://doi.org/10.1038/s41586-022-05557-5}{\emph{Nature} {\bfseries 614} (2023) 244} [\href{https://arxiv.org/abs/2204.02302}{{\ttfamily 2204.02302}}].

\bibitem{ALICE:2019aid}
{\scshape ALICE} collaboration, \emph{{Evidence of Spin-Orbital Angular Momentum Interactions in Relativistic Heavy-Ion Collisions}}, \href{https://doi.org/10.1103/PhysRevLett.125.012301}{\emph{Phys. Rev. Lett.} {\bfseries 125} (2020) 012301} [\href{https://arxiv.org/abs/1910.14408}{{\ttfamily 1910.14408}}].

\bibitem{Yang:2017sdk}
Y.-G.~Yang, R.-H.~Fang, Q.~Wang and X.-N.~Wang, \emph{{Quark coalescence model for polarized vector mesons and baryons}}, \href{https://doi.org/10.1103/PhysRevC.97.034917}{\emph{Phys. Rev. C} {\bfseries 97} (2018) 034917} [\href{https://arxiv.org/abs/1711.06008}{{\ttfamily 1711.06008}}].

\bibitem{Sheng:2019kmk}
X.-L.~Sheng, L.~Oliva and Q.~Wang, \emph{{What can we learn from the global spin alignment of $\phi$ mesons in heavy-ion collisions?}}, \href{https://doi.org/10.1103/PhysRevD.101.096005}{\emph{Phys. Rev. D} {\bfseries 101} (2020) 096005} [\href{https://arxiv.org/abs/1910.13684}{{\ttfamily 1910.13684}}].

\bibitem{Sheng:2020ghv}
X.-L.~Sheng, Q.~Wang and X.-N.~Wang, \emph{{Improved quark coalescence model for spin alignment and polarization of hadrons}}, \href{https://doi.org/10.1103/PhysRevD.102.056013}{\emph{Phys. Rev. D} {\bfseries 102} (2020) 056013} [\href{https://arxiv.org/abs/2007.05106}{{\ttfamily 2007.05106}}].

\bibitem{Sheng:2022wsy}
X.-L.~Sheng, L.~Oliva, Z.-T.~Liang, Q.~Wang and X.-N.~Wang, \emph{{Spin Alignment of Vector Mesons in Heavy-Ion Collisions}}, \href{https://doi.org/10.1103/PhysRevLett.131.042304}{\emph{Phys. Rev. Lett.} {\bfseries 131} (2023) 042304} [\href{https://arxiv.org/abs/2205.15689}{{\ttfamily 2205.15689}}].

\bibitem{Xia:2020tyd}
X.-L.~Xia, H.~Li, X.-G.~Huang and H.~Zhong~Huang, \emph{{Local spin alignment of vector mesons in relativistic heavy-ion collisions}}, \href{https://doi.org/10.1016/j.physletb.2021.136325}{\emph{Phys. Lett. B} {\bfseries 817} (2021) 136325} [\href{https://arxiv.org/abs/2010.01474}{{\ttfamily 2010.01474}}].

\bibitem{Wei:2023pdf}
M.~Wei and M.~Huang, \emph{{Spin alignment of vector mesons from quark dynamics in a rotating medium*}}, \href{https://doi.org/10.1088/1674-1137/acf036}{\emph{Chin. Phys. C} {\bfseries 47} (2023) 104105} [\href{https://arxiv.org/abs/2303.01897}{{\ttfamily 2303.01897}}].

\bibitem{Kumar:2023ghs}
A.~Kumar, B.~M\"uller and D.-L.~Yang, \emph{{Spin alignment of vector mesons by glasma fields}}, \href{https://doi.org/10.1103/PhysRevD.108.016020}{\emph{Phys. Rev. D} {\bfseries 108} (2023) 016020} [\href{https://arxiv.org/abs/2304.04181}{{\ttfamily 2304.04181}}].

\bibitem{Sheng:2023urn}
X.-L.~Sheng, S.~Pu and Q.~Wang, \emph{{Momentum dependence of the spin alignment of the \ensuremath{\phi} meson}}, \href{https://doi.org/10.1103/PhysRevC.108.054902}{\emph{Phys. Rev. C} {\bfseries 108} (2023) 054902} [\href{https://arxiv.org/abs/2308.14038}{{\ttfamily 2308.14038}}].

\bibitem{Lv:2024uev}
J.-p.~Lv, Z.-h.~Yu, Z.-t.~Liang, Q.~Wang and X.-N.~Wang, \emph{{Global quark spin correlations in relativistic heavy ion collisions}}, \href{https://doi.org/10.1103/PhysRevD.109.114003}{\emph{Phys. Rev. D} {\bfseries 109} (2024) 114003} [\href{https://arxiv.org/abs/2402.13721}{{\ttfamily 2402.13721}}].

\bibitem{Becattini:2024uha}
F.~Becattini, M.~Buzzegoli, T.~Niida, S.~Pu, A.-H.~Tang and Q.~Wang, \emph{{Spin polarization in relativistic heavy-ion collisions}}, \href{https://doi.org/10.1142/S0218301324300066}{\emph{Int. J. Mod. Phys. E} {\bfseries 33} (2024) 2430006} [\href{https://arxiv.org/abs/2402.04540}{{\ttfamily 2402.04540}}].

\bibitem{Niida:2024ntm}
T.~Niida and S.A.~Voloshin, \emph{{Polarization phenomenon in heavy-ion collisions}}, \href{https://doi.org/10.1142/S0218301324300108}{\emph{Int. J. Mod. Phys. E} {\bfseries 33} (2024) 2430010} [\href{https://arxiv.org/abs/2404.11042}{{\ttfamily 2404.11042}}].

\bibitem{Romatschke:2017ejr}
P.~Romatschke and U.~Romatschke, \emph{{Relativistic Fluid Dynamics In and Out of Equilibrium}}, Cambridge Monographs on Mathematical Physics, Cambridge University Press (5, 2019), \href{https://doi.org/10.1017/9781108651998}{10.1017/9781108651998}, [\href{https://arxiv.org/abs/1712.05815}{{\ttfamily 1712.05815}}].

\bibitem{Becattini:2019dxo}
F.~Becattini, M.~Buzzegoli and E.~Grossi, \emph{{Reworking the Zubarev's approach to non-equilibrium quantum statistical mechanics}}, \href{https://doi.org/10.3390/particles2020014}{\emph{Particles} {\bfseries 2} (2019) 197} [\href{https://arxiv.org/abs/1902.01089}{{\ttfamily 1902.01089}}].

\bibitem{Shi:2020htn}
S.~Shi, C.~Gale and S.~Jeon, \emph{{From chiral kinetic theory to relativistic viscous spin hydrodynamics}}, \href{https://doi.org/10.1103/PhysRevC.103.044906}{\emph{Phys. Rev. C} {\bfseries 103} (2021) 044906} [\href{https://arxiv.org/abs/2008.08618}{{\ttfamily 2008.08618}}].

\bibitem{Weickgenannt:2022qvh}
N.~Weickgenannt, D.~Wagner, E.~Speranza and D.H.~Rischke, \emph{{Relativistic dissipative spin hydrodynamics from kinetic theory with a nonlocal collision term}}, \href{https://doi.org/10.1103/PhysRevD.106.L091901}{\emph{Phys. Rev. D} {\bfseries 106} (2022) L091901} [\href{https://arxiv.org/abs/2208.01955}{{\ttfamily 2208.01955}}].

\bibitem{Li:2025pef}
Y.~Li and S.Y.F.~Liu, \emph{{Zubarev response approach to polarization phenomena in local equilibrium}}, {\emph{arXiv preprint} (2025) } [\href{https://arxiv.org/abs/2501.17861}{{\ttfamily 2501.17861}}].

\bibitem{Wagner:2024fry}
D.~Wagner, \emph{{Resummed spin hydrodynamics from quantum kinetic theory}}, \href{https://doi.org/10.1103/PhysRevD.111.016008}{\emph{Phys. Rev. D} {\bfseries 111} (2025) 016008} [\href{https://arxiv.org/abs/2409.07143}{{\ttfamily 2409.07143}}].

\bibitem{Becattini:2018duy}
F.~Becattini, W.~Florkowski and E.~Speranza, \emph{{Spin tensor and its role in non-equilibrium thermodynamics}}, \href{https://doi.org/10.1016/j.physletb.2018.12.016}{\emph{Phys. Lett. B} {\bfseries 789} (2019) 419} [\href{https://arxiv.org/abs/1807.10994}{{\ttfamily 1807.10994}}].

\bibitem{Buzzegoli:2021wlg}
M.~Buzzegoli, \emph{{Pseudogauge dependence of the spin polarization and of the axial vortical effect}}, \href{https://doi.org/10.1103/PhysRevC.105.044907}{\emph{Phys. Rev. C} {\bfseries 105} (2022) 044907} [\href{https://arxiv.org/abs/2109.12084}{{\ttfamily 2109.12084}}].

\bibitem{Zubarev:1979}
D.N.~Zubarev, A.V.~Prozorkevich and S.A.~Smolyanskii, \emph{Derivation of nonlinear generalized equations of quantum relativistic hydrodynamics}, \href{https://doi.org/10.1007/BF01032069}{\emph{Theoretical and Mathematical Physics} {\bfseries 40} (1979) 821}.

\bibitem{vanWeert1982}
C.G.~van Weert, \emph{Maximum entropy principle and relativistic hydrodynamics}, \href{https://doi.org/10.1016/0003-4916(82)90338-4}{\emph{Annals of Physics} {\bfseries 140} (1982) 133}.

\bibitem{Becattini:2014yxa}
F.~Becattini, L.~Bucciantini, E.~Grossi and L.~Tinti, \emph{{Local thermodynamical equilibrium and the beta frame for a quantum relativistic fluid}}, \href{https://doi.org/10.1140/epjc/s10052-015-3384-y}{\emph{Eur. Phys. J. C} {\bfseries 75} (2015) 191} [\href{https://arxiv.org/abs/1403.6265}{{\ttfamily 1403.6265}}].

\bibitem{Hayata:2015lga}
T.~Hayata, Y.~Hidaka, T.~Noumi and M.~Hongo, \emph{{Relativistic hydrodynamics from quantum field theory on the basis of the generalized Gibbs ensemble method}}, \href{https://doi.org/10.1103/PhysRevD.92.065008}{\emph{Phys. Rev. D} {\bfseries 92} (2015) 065008} [\href{https://arxiv.org/abs/1503.04535}{{\ttfamily 1503.04535}}].

\bibitem{Hosoya:1983id}
A.~Hosoya, M.-a.~Sakagami and M.~Takao, \emph{{Nonequilibrium Thermodynamics in Field Theory: Transport Coefficients}}, \href{https://doi.org/10.1016/0003-4916(84)90144-1}{\emph{Annals Phys.} {\bfseries 154} (1984) 229}.

\bibitem{Huang:2011dc}
X.-G.~Huang, A.~Sedrakian and D.H.~Rischke, \emph{{Kubo formulae for relativistic fluids in strong magnetic fields}}, \href{https://doi.org/10.1016/j.aop.2011.08.001}{\emph{Annals Phys.} {\bfseries 326} (2011) 3075} [\href{https://arxiv.org/abs/1108.0602}{{\ttfamily 1108.0602}}].

\bibitem{Harutyunyan:2018cmm}
A.~Harutyunyan, A.~Sedrakian and D.H.~Rischke, \emph{{Relativistic Dissipative Fluid Dynamics from the Non-Equilibrium Statistical Operator}}, \href{https://doi.org/10.3390/particles1010011}{\emph{Particles} {\bfseries 1} (2018) 155} [\href{https://arxiv.org/abs/1804.08267}{{\ttfamily 1804.08267}}].

\bibitem{Harutyunyan:2021rmb}
A.~Harutyunyan, A.~Sedrakian and D.H.~Rischke, \emph{{Relativistic second-order dissipative hydrodynamics from Zubarev\textquoteright{}s non-equilibrium statistical operator}}, \href{https://doi.org/10.1016/j.aop.2022.168755}{\emph{Annals Phys.} {\bfseries 438} (2022) 168755} [\href{https://arxiv.org/abs/2110.04595}{{\ttfamily 2110.04595}}].

\bibitem{Hu:2021lnx}
J.~Hu, \emph{{Kubo formulae for first-order spin hydrodynamics}}, \href{https://doi.org/10.1103/PhysRevD.103.116015}{\emph{Phys. Rev. D} {\bfseries 103} (2021) 116015} [\href{https://arxiv.org/abs/2101.08440}{{\ttfamily 2101.08440}}].

\bibitem{Montenegro:2020paq}
D.~Montenegro and G.~Torrieri, \emph{{Linear response theory and effective action of relativistic hydrodynamics with spin}}, \href{https://doi.org/10.1103/PhysRevD.102.036007}{\emph{Phys. Rev. D} {\bfseries 102} (2020) 036007} [\href{https://arxiv.org/abs/2004.10195}{{\ttfamily 2004.10195}}].

\bibitem{Torrieri:2022ogj}
G.~Torrieri and D.~Montenegro, \emph{{Linear response hydrodynamics of a relativistic dissipative fluid with spin}}, \href{https://doi.org/10.1103/PhysRevD.107.076010}{\emph{Phys. Rev. D} {\bfseries 107} (2023) 076010} [\href{https://arxiv.org/abs/2207.00537}{{\ttfamily 2207.00537}}].

\bibitem{Becattini:2023ouz}
F.~Becattini, A.~Daher and X.-L.~Sheng, \emph{{Entropy current and entropy production in relativistic spin hydrodynamics}}, \href{https://doi.org/10.1016/j.physletb.2024.138533}{\emph{Phys. Lett. B} {\bfseries 850} (2024) 138533} [\href{https://arxiv.org/abs/2309.05789}{{\ttfamily 2309.05789}}].

\bibitem{Tiwari:2024trl}
A.~Tiwari and B.K.~Patra, \emph{{Second-order spin hydrodynamics from Zubarev's nonequilibrium statistical operator formalism}}, {\emph{arXiv preprint} (2024) } [\href{https://arxiv.org/abs/2408.11514}{{\ttfamily 2408.11514}}].

\bibitem{Dey:2024cwo}
S.~Dey and A.~Das, \emph{{Kubo formula for spin hydrodynamics: spin chemical potential as leading order in gradient expansion}}, {\emph{arXiv preprint} (2024) } [\href{https://arxiv.org/abs/2410.04141}{{\ttfamily 2410.04141}}].

\bibitem{She:2024rnx}
D.~She, Y.-W.~Qiu and D.~Hou, \emph{{Relativistic Second-order Spin Hydrodynamics: A Kubo-Type Formulation for the Quark-Gluon Plasma}}, {\emph{arXiv preprint} (2024) } [\href{https://arxiv.org/abs/2410.15142}{{\ttfamily 2410.15142}}].

\bibitem{Buzzegoli:2024mra}
M.~Buzzegoli and A.~Palermo, \emph{{Emergent Canonical Spin Tensor in the Chiral-Symmetric Hot QCD}}, \href{https://doi.org/10.1103/PhysRevLett.133.262301}{\emph{Phys. Rev. Lett.} {\bfseries 133} (2024) 262301} [\href{https://arxiv.org/abs/2407.14345}{{\ttfamily 2407.14345}}].

\bibitem{Buzzegoli:2017cqy}
M.~Buzzegoli, E.~Grossi and F.~Becattini, \emph{{General equilibrium second-order hydrodynamic coefficients for free quantum fields}}, \href{https://doi.org/10.1007/JHEP10(2017)091}{\emph{JHEP} {\bfseries 10} (2017) 091} [\href{https://arxiv.org/abs/1704.02808}{{\ttfamily 1704.02808}}].

\bibitem{Buzzegoli:2018wpy}
M.~Buzzegoli and F.~Becattini, \emph{{General thermodynamic equilibrium with axial chemical potential for the free Dirac field}}, \href{https://doi.org/10.1007/JHEP12(2018)002}{\emph{JHEP} {\bfseries 12} (2018) 002} [\href{https://arxiv.org/abs/1807.02071}{{\ttfamily 1807.02071}}].

\bibitem{Buzzegoli:2020ycf}
M.~Buzzegoli, \emph{{Thermodynamic Equilibrium of Massless Fermions with Vorticity, Chirality and Electromagnetic Field}}, \href{https://doi.org/10.1007/978-3-030-71427-7_3}{\emph{Lect. Notes Phys.} {\bfseries 987} (2021) 59} [\href{https://arxiv.org/abs/2011.09974}{{\ttfamily 2011.09974}}].

\bibitem{Becattini:2020qol}
F.~Becattini, M.~Buzzegoli and A.~Palermo, \emph{{Exact equilibrium distributions in statistical quantum field theory with rotation and acceleration: scalar field}}, \href{https://doi.org/10.1007/JHEP02(2021)101}{\emph{JHEP} {\bfseries 02} (2021) 101} [\href{https://arxiv.org/abs/2007.08249}{{\ttfamily 2007.08249}}].

\bibitem{Palermo:2021hlf}
A.~Palermo, M.~Buzzegoli and F.~Becattini, \emph{{Exact equilibrium distributions in statistical quantum field theory with rotation and acceleration: Dirac field}}, \href{https://doi.org/10.1007/JHEP10(2021)077}{\emph{JHEP} {\bfseries 10} (2021) 077} [\href{https://arxiv.org/abs/2106.08340}{{\ttfamily 2106.08340}}].

\bibitem{Liu:2021nyg}
Y.-C.~Liu and X.-G.~Huang, \emph{{Spin polarization formula for Dirac fermions at local equilibrium}}, \href{https://doi.org/10.1007/s11433-022-1903-8}{\emph{Sci. China Phys. Mech. Astron.} {\bfseries 65} (2022) 272011} [\href{https://arxiv.org/abs/2109.15301}{{\ttfamily 2109.15301}}].

\bibitem{Buzzegoli:2022qrr}
M.~Buzzegoli, \emph{{Spin polarization induced by magnetic field and the relativistic Barnett effect}}, \href{https://doi.org/10.1016/j.nuclphysa.2023.122674}{\emph{Nucl. Phys. A} {\bfseries 1036} (2023) 122674} [\href{https://arxiv.org/abs/2211.04549}{{\ttfamily 2211.04549}}].

\bibitem{Sheng:2024pbw}
X.-L.~Sheng, F.~Becattini, X.-G.~Huang and Z.-H.~Zhang, \emph{{Spin polarization of fermions at local equilibrium: Second-order gradient expansion}}, \href{https://doi.org/10.1103/PhysRevC.110.064908}{\emph{Phys. Rev. C} {\bfseries 110} (2024) 064908} [\href{https://arxiv.org/abs/2407.12130}{{\ttfamily 2407.12130}}].

\bibitem{Li:2022vmb}
F.~Li and S.Y.F.~Liu, \emph{{Tensor Polarization and Spectral Properties of Vector Meson in QCD Medium}}, {\emph{arXiv preprint} (2022) } [\href{https://arxiv.org/abs/2206.11890}{{\ttfamily 2206.11890}}].

\bibitem{Yang:2024fkn}
S.-Z.~Yang, X.-Q.~Xie, S.~Pu, J.-H.~Gao and Q.~Wang, \emph{{Spin alignment of vector mesons in local equilibrium by Zubarev's approach}}, {\emph{arXiv preprint} (2024) } [\href{https://arxiv.org/abs/2412.19400}{{\ttfamily 2412.19400}}].

\bibitem{Zhang:2024mhs}
Z.-H.~Zhang, X.-G.~Huang, F.~Becattini and X.-L.~Sheng, \emph{{Vector and Tensor Spin Polarization for Vector Bosons at Local Equilibrium}}, {\emph{arXiv preprint} (2024) } [\href{https://arxiv.org/abs/2412.19416}{{\ttfamily 2412.19416}}].

\bibitem{Becattini:2020sww}
F.~Becattini, \emph{{Polarization in Relativistic Fluids: A Quantum Field Theoretical Derivation}}, \href{https://doi.org/10.1007/978-3-030-71427-7_2}{\emph{Lect. Notes Phys.} {\bfseries 987} (2021) 15} [\href{https://arxiv.org/abs/2004.04050}{{\ttfamily 2004.04050}}].

\bibitem{Palermo:2023cup}
A.~Palermo and F.~Becattini, \emph{{Exact spin polarization of massive and massless particles in relativistic fluids at global equilibrium}}, \href{https://doi.org/10.1140/epjp/s13360-023-04169-w}{\emph{Eur. Phys. J. Plus} {\bfseries 138} (2023) 547} [\href{https://arxiv.org/abs/2304.02276}{{\ttfamily 2304.02276}}].

\bibitem{Zubarev:1972285}
D.~Zubarev and S.~Tishchenko, \emph{Nonlocal hydrodynamics with memory}, \href{https://doi.org/https://doi.org/10.1016/0031-8914(72)90084-5}{\emph{Physica} {\bfseries 59} (1972) 285}.

\bibitem{Koide:2006ef}
T.~Koide, G.S.~Denicol, P.~Mota and T.~Kodama, \emph{{Relativistic dissipative hydrodynamics: A Minimal causal theory}}, \href{https://doi.org/10.1103/PhysRevC.75.034909}{\emph{Phys. Rev. C} {\bfseries 75} (2007) 034909} [\href{https://arxiv.org/abs/hep-ph/0609117}{{\ttfamily hep-ph/0609117}}].

\bibitem{Koide:2008nw}
T.~Koide and T.~Kodama, \emph{{Transport Coefficients of Non-Newtonian Fluid and Causal Dissipative Hydrodynamics}}, \href{https://doi.org/10.1103/PhysRevE.78.051107}{\emph{Phys. Rev. E} {\bfseries 78} (2008) 051107} [\href{https://arxiv.org/abs/0806.3725}{{\ttfamily 0806.3725}}].

\bibitem{Becattini:2020xbh}
F.~Becattini, M.~Buzzegoli, A.~Palermo and G.~Prokhorov, \emph{{Polarization as a signature of local parity violation in hot QCD matter}}, \href{https://doi.org/10.1016/j.physletb.2021.136706}{\emph{Phys. Lett. B} {\bfseries 822} (2021) 136706} [\href{https://arxiv.org/abs/2009.13449}{{\ttfamily 2009.13449}}].

\bibitem{Du:2008zzb}
F.~Du, L.E.~Finch and J.~Sandweiss, \emph{{Observing spontaneous strong CP violation through hyperon helicity correlations}}, \href{https://doi.org/10.1103/PhysRevC.78.044908}{\emph{Phys. Rev. C} {\bfseries 78} (2008) 044908}.

\bibitem{Maekawa2017-um}
S.~Maekawa, S.O.~Valenzuela, E.~Saitoh and T.~Kimura, eds., \emph{Spin Current}, Series on Semiconductor Science and Technology, Oxford University Press, London, England, 2~ed. (Aug, 2017), \href{https://doi.org/10.1093/oso/9780198787075.001.0001}{10.1093/oso/9780198787075.001.0001}.

\bibitem{AdvancementInChiralSpintronics}
S.~Mishra, A.C.~Jones and C.~Fontanesi, \emph{Recent advancements in chiral spintronics: from molecular-level insights to device applications. a prospect based on the interplay between physical and chemical properties of chiral systems}, \href{https://doi.org/10.1039/D4TC03453H}{\emph{J. Mater. Chem. C} {\bfseries 13} (2025) 2121}.

\bibitem{ChiralInducedSpinSelectivity}
B.P.~Bloom, Y.~Paltiel, R.~Naaman and D.H.~Waldeck, \emph{Chiral induced spin selectivity}, \href{https://doi.org/10.1021/acs.chemrev.3c00661}{\emph{Chemical Reviews} {\bfseries 124} (2024) 1950} [\href{https://arxiv.org/abs/https://doi.org/10.1021/acs.chemrev.3c00661}{{\ttfamily https://doi.org/10.1021/acs.chemrev.3c00661}}].

\bibitem{QSHE}
C.L.~Kane and E.J.~Mele, \emph{Quantum spin hall effect in graphene}, \href{https://doi.org/10.1103/PhysRevLett.95.226801}{\emph{Phys. Rev. Lett.} {\bfseries 95} (2005) 226801}.

\bibitem{Zhang_2018}
Y.~Zhang, J.~Železný, Y.~Sun, J.~van~den Brink and B.~Yan, \emph{Spin hall effect emerging from a noncollinear magnetic lattice without spin–orbit coupling}, \href{https://doi.org/10.1088/1367-2630/aad1eb}{\emph{New Journal of Physics} {\bfseries 20} (2018) 073028}.

\bibitem{THE}
P.~Bruno, V.K.~Dugaev and M.~Taillefumier, \emph{Topological hall effect and berry phase in magnetic nanostructures}, \href{https://doi.org/10.1103/PhysRevLett.93.096806}{\emph{Phys. Rev. Lett.} {\bfseries 93} (2004) 096806}.

\bibitem{Kim2023}
K.~Kim, E.~Vetter, L.~Yan, C.~Yang, Z.~Wang, R.~Sun et~al., \emph{Chiral-phonon-activated spin seebeck effect}, \href{https://doi.org/10.1038/s41563-023-01473-9}{\emph{Nature Materials} {\bfseries 22} (2023) 322–328}.

\bibitem{Konig2007}
M.~K{\"o}nig, S.~Wiedmann, C.~Bru\"une, A.~Roth, H.~Buhmann, L.W.~Molenkamp et~al., \emph{Quantum spin hall insulator state in hgte quantum wells}, \href{https://doi.org/10.1126/science.1148047}{\emph{Science} {\bfseries 318} (2007) 766–770}.

\bibitem{Wu2018}
S.~Wu, V.~Fatemi, Q.D.~Gibson, K.~Watanabe, T.~Taniguchi, R.J.~Cava et~al., \emph{Observation of the quantum spin hall effect up to 100 kelvin in a monolayer crystal}, \href{https://doi.org/10.1126/science.aan6003}{\emph{Science} {\bfseries 359} (2018) 76–79}.

\bibitem{Kolincio2021}
K.K.~Kolincio, M.~Hirschberger, J.~Masell, S.~Gao, A.~Kikkawa, Y.~Taguchi et~al., \emph{Large hall and nernst responses from thermally induced spin chirality in a spin-trimer ferromagnet}, \href{https://doi.org/10.1073/pnas.2023588118}{\emph{Proceedings of the National Academy of Sciences} {\bfseries 118} (2021) }.

\bibitem{Wagner:2022gza}
D.~Wagner, N.~Weickgenannt and E.~Speranza, \emph{{Generating tensor polarization from shear stress}}, \href{https://doi.org/10.1103/PhysRevResearch.5.013187}{\emph{Phys. Rev. Res.} {\bfseries 5} (2023) 013187} [\href{https://arxiv.org/abs/2207.01111}{{\ttfamily 2207.01111}}].

\bibitem{Dong:2023cng}
W.-B.~Dong, Y.-L.~Yin, X.-L.~Sheng, S.-Z.~Yang and Q.~Wang, \emph{{Linear response theory for spin alignment of vector mesons in thermal media}}, \href{https://doi.org/10.1103/PhysRevD.109.056025}{\emph{Phys. Rev. D} {\bfseries 109} (2024) 056025} [\href{https://arxiv.org/abs/2311.18400}{{\ttfamily 2311.18400}}].

\bibitem{Grossi:2024pyh}
E.~Grossi, A.~Palermo and I.~Zahed, \emph{{Polarized \ensuremath{\phi} meson rates with viscous corrections at energies available at the BNL Relativistic Heavy Ion Collider}}, \href{https://doi.org/10.1103/PhysRevC.111.014914}{\emph{Phys. Rev. C} {\bfseries 111} (2025) 014914} [\href{https://arxiv.org/abs/2407.10524}{{\ttfamily 2407.10524}}].

\bibitem{CMS:2025nqr}
{\scshape CMS} collaboration, \emph{{Observation of $\Lambda$ hyperon local polarization in pPb collisions at $\sqrt{s_\mathrm{NN}}$ = 8.16 TeV}}, {\emph{arXiv preprint} (2025) } [\href{https://arxiv.org/abs/2502.07898}{{\ttfamily 2502.07898}}].

\bibitem{Yi:2024kwu}
C.~Yi, X.-Y.~Wu, J.~Zhu, S.~Pu and G.-Y.~Qin, \emph{{Spin polarization of \ensuremath{\Lambda} hyperons along the beam direction in p+Pb collisions at sNN=8.16 TeV using hydrodynamic approaches}}, \href{https://doi.org/10.1103/PhysRevC.111.044901}{\emph{Phys. Rev. C} {\bfseries 111} (2025) 044901} [\href{https://arxiv.org/abs/2408.04296}{{\ttfamily 2408.04296}}].

\end{thebibliography}
\end{document}